\documentclass[superscriptaddress,twocolumn]{revtex4-1}
\usepackage[utf8]{inputenc}
\usepackage{graphicx,color}
\usepackage{amsmath,mathtools}
\usepackage{dsfont}
\usepackage{amssymb}
\usepackage{hyperref}
\usepackage{bbold}
\usepackage{enumitem}

\renewcommand{\Re}{\mathop{\text{Re}}\nolimits}

\newcommand{\Tr}{\mathop{\text{Tr}}\nolimits}
\newcommand{\ket}[1]{|{#1}\rangle}
\newcommand{\bra}[1]{\langle{#1}|}

\newcommand{\bracket}[2]{\langle#1|#2\rangle}

\newcommand{\Up}{{\uparrow}}
\newcommand{\Dn}{{\downarrow}}
\newcommand{\Id}{\mathbb{1}}

\newcommand{\id}{\mathbb{1}}
\newcommand{\hc}{\text{H.c.}}

\newcommand\bwt         {\begin{widetext}}
\newcommand\ewt         {\end{widetext}}

\begin{document}

\title{Dyonic zero-energy modes}

\author{Morten I. K. Munk}
\affiliation{Center for Quantum Devices, Niels Bohr Institute, University of Copenhagen, Juliane Maries Vej 30,
2100 Copenhagen, Denmark.}

\author{Asbjørn Rasmussen}

\affiliation{Center for Atomic-scale Materials Design, Department of Physics, Technical University of Denmark, 2800 Kongens Lyngby, Denmark}

\author{Michele Burrello}
\affiliation{Center for Quantum Devices, Niels Bohr Institute, University of Copenhagen, Juliane Maries Vej 30,
2100 Copenhagen, Denmark.}
\affiliation{Niels Bohr International Academy, Niels Bohr Institute, University of Copenhagen, Juliane Maries Vej 30,
2100 Copenhagen, Denmark.}

\begin{abstract}
One-dimensional systems with topological order are intimately related to the appearance of zero-energy modes localized on their boundaries. The most common example is the Kitaev chain, which displays Majorana zero-energy modes and it is characterized by a two-fold ground state degeneracy related to the global $\mathbb{Z}_2$ symmetry associated with fermionic parity. By extending the symmetry to the $\mathbb{Z}_N$ group, it is possible to engineer systems hosting topological parafermionic modes. In this work, we address one-dimensional systems with a generic discrete symmetry group $G$. We define a ladder model of gauge fluxes that generalizes the Ising and Potts models and displays a symmetry broken phase. Through a non-Abelian Jordan-Wigner transformation, we map this flux ladder into a model of dyonic operators, defined by the group elements and irreducible representations of $G$.
We show that the so-obtained dyonic model has topological order, with zero-energy modes localized at its boundary. These dyonic zero-energy modes are in general weak topological modes, but strong dyonic zero modes appear when suitable position-dependent couplings are considered.

\end{abstract}

\maketitle

\section{Introduction}

With his seminal work \cite{kitaev2001}, Kitaev gave life to the study of one-dimensional models with topological order. These are models displaying degenerate ground states, without any local order parameter able to distinguish them. Their prototypical example is, indeed, the Kitaev chain, a fermionic model characterized by the presence of zero-energy Majorana modes localized at its edges. These modes commute with the Hamiltonian but anticommute with each other, thus enforcing a two-fold degeneracy of the energy spectrum up to exponential corrections in the system size.

The unpaired Majorana modes in Kitaev's model are protected by a global $\mathbb{Z}_2$ symmetry, which corresponds to the conservation of the fermionic parity; once embedded in a two-dimensional system, these zero-energy modes behave like non-Abelian anyons, thus opening an invaluable scenario for topological quantum computation \cite{alicea2012,lejinse2012,beenakker2013}.

In the search for richer kinds of non-Abelian anyons, the Kitaev chain has been generalized to a family of models with global $\mathbb{Z}_N$ symmetries \cite{fendley2012}. These models can be build from a nonlocal representation of the chiral $\mathbb{Z}_N$ Potts model in terms of parafermions, which are a generalization of the Majorana modes to the $\mathbb{Z}_N$ case. Through an iterative procedure, Fendley argued that these $\mathbb{Z}_N$-symmetric chains are characterized by localized zero-energy parafermionic modes \cite{fendley2012} and, consequently, their ground states are $N-$fold degenerate, up to exponential corrections due to finite size effects \cite{jermyn2014,bernevig2016,mazza2017} (see also \cite{Moran2017}). 

Is it possible to generalize further these systems and build one-dimensional topological models characterized by an underlying non-Abelian symmetry group? What are the corresponding zero-energy modes?

These are the questions addressed in this paper. We will define one-dimensional topological models whose Hamiltonian is invariant under the action of a discrete non-Abelian symmetry group $G$ and, based on an iterative expansion, we will show the presence of localized zero-energy modes. These zero-energy modes can be characterized based on their transformation rules under the action of the global symmetry group $G$; similarly to anyons in a two-dimensional quantum double model \cite{kitaev2003}, they will be labeled by both a group element $g$ and an irreducible representation $K$ of $G$. For this reason we call them \textit{dyonic zero-energy modes}. 

Our strategy to build these exotic 1D models with topological order is inspired by the duality between the Ising and Kitaev chains and its generalization to the Potts and parafermionic models: it is known that the Kitaev chain can be described in terms of the Ising model through a Jordan-Wigner transformation mapping spins into fermions; in the same way, the parafermionic chains are equivalent to $\mathbb{Z}_N$ clock models based on a generalized Jordan-Wigner (JW) transformation \cite{fradkin1980,jagannathan}. In both situations the JW transformation maps a bosonic (spin or clock) model, characterized by spontaneous symmetry breaking in an ordered phase, into a model with topological order built from operators (fermionic or parafermionic) which do not commute when spatially separated. The JW transformation is nonlocal and it maps the degeneracy of the ground states in the ordered (ferromagnetic) phase of the bosonic models, into a degeneracy caused by localized zero-energy modes in the topological models.

\begin{table*}[t]
\begin{tabular}{|c|c|c|c|c|}
\hline
\textbf{Global symmetry}  & \textbf{Bosonic model} & \textbf{Mapping} & \textbf{Topological model} & \textbf{Zero modes} \\
\hline
$\mathbb{Z}_2$ & Ising & $\xleftrightarrow{\text{JW}}$ & Kitaev \cite{kitaev2001} & Majorana modes \\
\hline
$\mathbb{Z}_N$ & Chiral Potts & $\xleftrightarrow{{\mathbb{Z}_N} \; \text{JW}}$ & Fendley \cite{fendley2012} &  Parafermionic modes \\
\hline
Non-Abelian $G$ & Chiral gauge flux ladder & $\xleftrightarrow{\text{Non-Abelian JW}}$ & Chiral dyonic model &  Dyonic modes 
 \\
\hline
\end{tabular}
\label{tab}
\caption{The table represents the relation between the topological models by Kitaev and Fendley and their nontopological counterparts given by the Ising and Potts models. The related Jordan-Wigner mapping preserves the corresponding global symmetries. The scope of this paper is to define analogous models with a non-Abelian symmetry and verify the existence of localized zero-energy modes.}
\end{table*}

Our construction will be based on an analogous mapping: we will begin from the ``bosonic'' side and we will first define a $G-$symmetric ladder model, inspired by quantum double models \cite{kitaev2003} and lattice gauge theories. $G$ will be a global non-Abelian gauge symmetry which will be spontaneously broken, thus resulting in an ordered phase. In this ladder model the ground states are $|G|-$fold degenerate, where $|G|$ is the order of the symmetry group, and they can be locally distinguished. We will argue that, for chiral models, the ground-state degeneracy is preserved up to corrections exponentially suppressed in the system size. Then we will proceed by defining a non-Abelian JW transformation, which maps the bosonic ``gauge'' degrees of freedom into dyonic operators labeled by an element $g\in G$ and transforming under the symmetry group $G$ based on its fundamental (standard) irreducible representation $F$. 

Based on both a quasiadiabatic continuation and an iterative construction, we will show that localized dyonic zero-energy modes emerge in the system and we will investigate their fusion rules, which can be understood in terms of the effect of the symmetry transformations and are consistent with the $|G|$-fold degeneracy of the ground state.

Let us summarize the content of this paper. Section \ref{sec:flux} is devoted to the introduction and analysis of the ``bosonic'' gauge-flux ladder model. In Sec. \ref{sec:Ising}, we interpret the Ising and Potts models in terms of flux ladder models to set the stage for the more complicated non-Abelian case; Sec. \ref{sec:op} introduces the building blocks for the non-Abelian flux ladder Hamiltonian, which is built and analyzed in Secs. \ref{sec:Ham} and \ref{sec:order}; Sec. \ref{sec:s3} finally deals with the example provided by the smallest non-Abelian group, $S_3$.
Section \ref{sec:topo1} is dedicated to the construction of the dyonic model; in Sec. \ref{sec:JW}, we introduce the JW transformation for discrete non-Abelian groups and the resulting dyonic operators which allow us to build the dyonic Hamiltonian; in Sec. \ref{sec:topo2}, we define the notion of topological order for one-dimensional systems with a non-Abelian global symmetry. Section \ref{sec:topo} is devoted to the analysis of the zero-energy modes of the dyonic model; in Sec. \ref{sec:weak}, we show that the dyonic model fulfills the criteria for topological order and presents protected weak zero-energy edge modes; in Secs. \ref{sec:strong}-\ref{sec:strong3}, we present the construction of strong topological zero-energy modes and we discuss divergences that hinder their appearance and the conditions the Hamiltonian must fulfill to avoid these divergences; Sec. \ref{sec:fusion} analyzes the fusion properties of the topological modes. Section \ref{sec:aux} discusses further properties of the family of models we introduced and the appearance of additional holographic and local symmetries in the dyonic Hamiltonian. Section \ref{singleflux} presents a numerical analysis of the lowest energy excitations of the model for $G=S_3$ in the single-flux approximation. Finally, in Section \ref{sec:disc}, we summarize our results and Appendices provide additional analyses of some technical aspects.

\section{Non-Abelian gauge flux ladders} \label{sec:flux}

\subsection{Ising and Potts models as gauge-flux ladders} \label{sec:Ising}

\begin{figure}[t]
\includegraphics[width=\columnwidth]{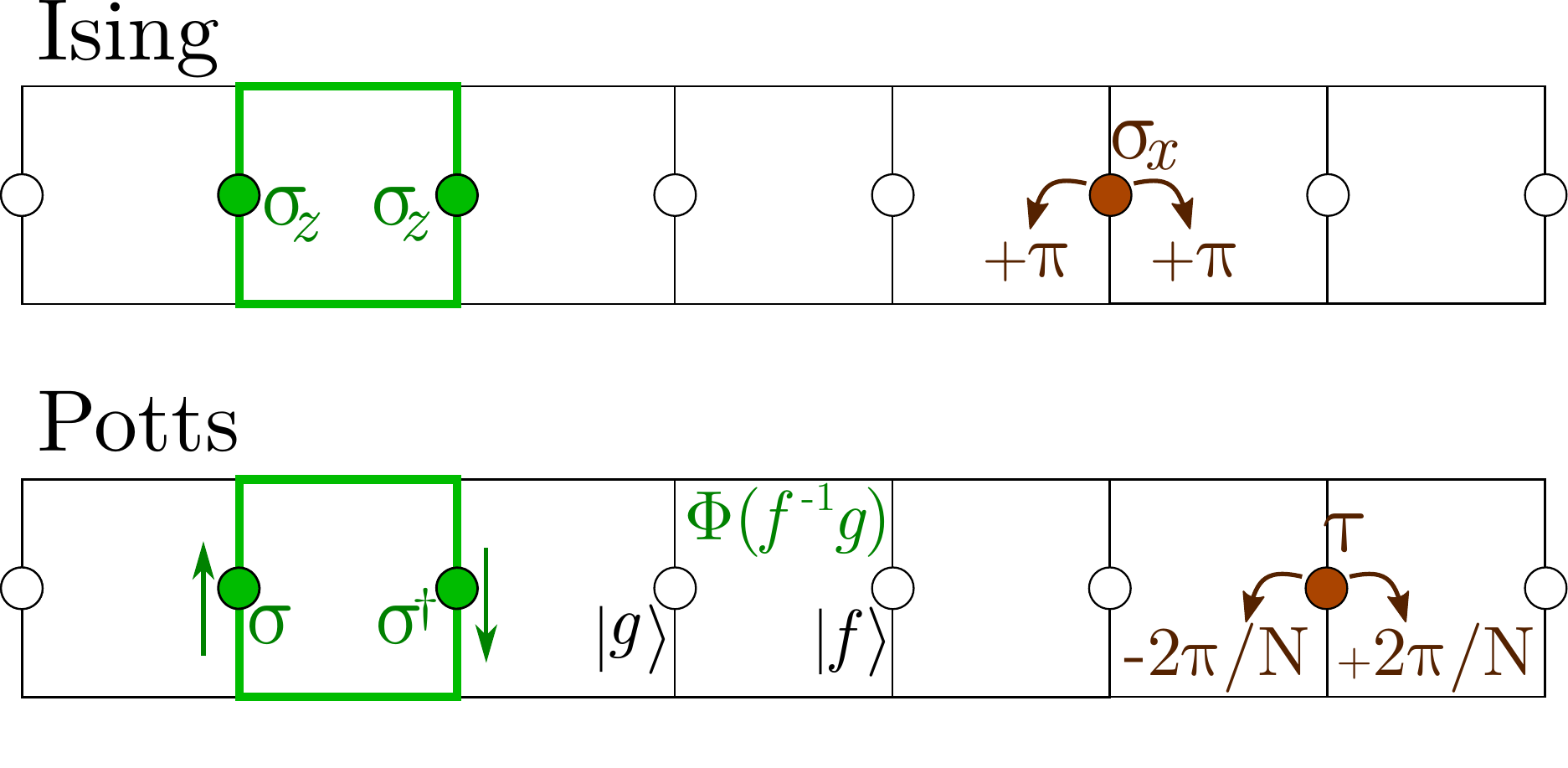}
\caption{The Ising (first row) and Potts (second row) models are interpreted as $\mathbb{Z}_N$ gauge-flux ladders. The nearest-neighbor (green) terms assign a mass to the nontrivial gauge fluxes $\Phi$ and can be interpreted as plaquette operators. The on-site $h$ terms have the effect of adding a pair flux-antiflux to the neighboring plaquettes: $\pi$ fluxes in the Ising case , a pair of $\pm 2\pi/N$ fluxes in the Potts case.}
\label{fig:ising}
\end{figure}

Before beginning the construction of models with non-Abelian symmetries, it is useful to provide a description of the Ising and Potts models in terms of gauge-flux ladders for the Abelian gauge groups and summarize some of their properties. This construction is based on associating each site of the Ising or Potts models with a rung in a ladder and interpreting its states in terms of a gauge degree of freedom related to the $\mathbb{Z}_2$ or $\mathbb{Z}_N$ group.
In particular, let us consider the Ising model:
\begin{equation}
H = -J \sum_{r=1}^{L} \sigma_{z,r} \sigma_{z,r+1} - h \sum_{r=1}^{L} \sigma_{x,r}\,.
\end{equation}
For each site $r$, we can consider the state $\ket{\Up}$ as representing the identity element $e \in \mathbb{Z}_2$ and the state $\ket{\Dn}$ as the nontrivial element $-1 \in \mathbb{Z}_2$. Under this point of view, the term $-J\sigma_{z,r} \sigma_{z,r+1}$ is minimized if the gauge degrees of freedom in neighboring sites are equal. Therefore, by interpreting the ladder as a set of plaquettes in a gauge theory, we can state that this term is minimized if no gauge flux is present in the plaquette $r$, such that a hypothetical particle coupled to this gauge degrees of freedom undergoes a trivial gauge transformation when moving around the plaquette: a gauge flux thus corresponds to a domain wall in the usual ferromagnetic description.
The effect of the $h$ term, instead, is to allow for transitions between the $\ket{\Up}$ and $\ket{\Dn}$ states. This can be interpreted as an electric field term in the $\mathbb{Z}_2$ gauge theory and it amounts to a local gauge transformation acting on a single gauge degree of freedom. 
In this work, we will mostly be interested in the ordered phase $J>h$ of these models. In such a phase, the term $J$ provides a mass for the $\mathbb{Z}_2$ gauge fluxes, whereas the term $h$ nucleates pairs of these fluxes and constitutes their kinetic energy (see Fig. \ref{fig:ising}).
The related global gauge symmetry is given by the string operator $\mathcal{Q}= \prod_r \sigma_{x,r}$. 

An alternative interpretation of the Ising model / $\mathbb{Z}_2$ gauge-flux ladder is provided by the toric code \cite{kitaev2003}. The gauge-flux ladder is a row of the toric code in which all the horizontal degrees of freedom have been frozen into the $\ket{\Up}$ state (corresponding to the identity transformation in $G$) and do not appear in the Hamiltonian. Only the rung degrees of freedom are dynamical and describe the dynamics of the $\mathbb{Z}_2$ magnetic fluxes moving along the ladder.

The same flux-ladder description can be applied to the Potts model:
\begin{equation} \label{potts}
H = -J \sum_{r=1}^{L} \left(e^{i\phi} \sigma_{r+1}^\dag \sigma_{r} + \rm{H.c.}\right) - h \sum_{n=1}^{N-1} \sum_{r=1}^{L} \tau_r^n\,,
\end{equation}
where we introduced the $\mathbb{Z}_N$ clock operators $\sigma$ and $\tau$ obeying the commutation rule $\sigma_r \tau_{r'} = e^{i\frac{2\pi}{N}\delta_{r,r'}}\tau_{r'} \sigma_r$ and the relations $\sigma^N=\tau^N=1$. This model is symmetric under the global $\mathbb{Z}_N$ transformations $\mathcal{Q}_k= \prod_r \tau^k_r$ and can be interpreted as a $\mathbb{Z}_N$ flux-ladder model with the magnetic fluxes taking $N$ different values. 
In the Potts model, we can associate the $N$ eigenstates $\ket{g}$ of the operator $\sigma$, such that $\sigma \ket{g} = e^{i{2\pi n_g}/{N}}\ket{g}$, with the $N$ elements $g$ of the group $\mathbb{Z}_N$; also in this case, we can interpret the states of each site as gauge degrees of freedom lying on the rungs of a ladder. For $\phi=0$, the $J$ term of the Hamiltonian is minimized if the gauge degrees of freedom of neighboring rungs coincide, thus no domain walls are present. This corresponds to a situation in which all the plaquettes host a trivial gauge flux. As in the Ising case, the gauge fluxes correspond to the domain walls of the system and they belong to $N$ inequivalent kinds, one for each element of the group $\mathbb{Z}_N$.
 
Let us consider a single plaquette (see Fig. \ref{fig:ising}). For a generic product state $\ket{g}_r\ket{f}_{r+1}$, the operator $\sigma_{r+1}^\dag \sigma_{r}$ has eigenvalue $e^{i2\pi (n_g-n_f)/N}$. Therefore this state corresponds to a $\mathbb{Z}_N$ gauge flux $\Phi(f^{-1}g) = 2\pi (n_g-n_f)/N$ and the $J$ term of the Hamiltonian returns an energy $-2J\cos[2\pi(n_g-n_f)/N+\phi]$ which determines its mass. By embedding the model in a lattice gauge theory, this gauge flux would correspond to the transformation in $\mathbb{Z}_N$ of a hypothetical matter particle moving clockwise around the ladder	plaquette.

Generalizing the Ising case, the $h$ term in the Hamiltonian corresponds to the sum of the nontrivial local $\mathbb{Z}_N$ gauge transformations that can be applied to each local gauge degree of freedom. In the gauge theory interpretation it is an energy term associated to the electric field in the rung. In particular we have $\tau_r^{n_h}\ket{g}_r = \ket{hg}_r$. The Potts model can thus be interpreted as a ladder of $\mathbb{Z}_N$ magnetic fluxes in the spirit of the $\mathbb{Z}_N$ toric code \cite{bullock2007} (see also \cite{burrello2013} for an analogous stripe model). 

In the case $\phi=0$ the system is invariant under both the time-reversal symmetry $\tau \to \tau^\dag$, $\sigma \to \sigma^\dag$ and the space inversion symmetry $\tau_r \to \tau_{L-r}$, $\sigma_r \to \sigma_{L-r}$, where $L$ is the system size. This implies that the fluxes $\Phi(g)$ and $\Phi(g^{-1})$ have the same mass. When introducing $\phi \neq 0$, both the symmetries are violated and the model becomes chiral. In general, for $\phi \neq 0$, the global $\mathbb{Z}_N$ transformations are the only nonspatial symmetries preserved and it was showed that only in this chiral case zero-energy modes can be stable in the corresponding parafermionic theory \cite{fendley2012}. Therefore, to extend the $\mathbb{Z}_N$ theory to a non-Abelian group, we will adopt a similar approach and consider Hamiltonians violating the time-reversal and space-inversion symmetries.

For both the Ising and Potts models, the phase diagram includes an ordered ferromagnetic phase when $J \gg h$ and a disordered paramagnetic phase for $h \gg J$ (the $\mathbb{Z}_N$ symmetric models include additional gapless phases for $N>4$). The related symmetries are unbroken in the paramagnetic phase and become spontaneously broken for the ferromagnetic phases such that the eigenstates of the models are, in general, not invariant under the gauge group $\mathbb{Z}_N$. The disorder operator introduces a domain wall in the system which corresponds with the gauge-flux in the ladder \cite{fradkin1980}. We define the disorder operators as the product of local gauge symmetries from the left edge of the system to the position of the flux: $\mathcal{L}_g(r) = \prod_{j<r} \tau_j^{n_g}$. These disorder operators are dual to the order operators $\sigma_r$ and, from their product, it is possible to build the Abelian Jordan-Wigner transformations mapping the clock into the parafermionic models \cite{fendley2012}.

\subsection{The rung Hilbert space and operators} \label{sec:op}

The construction of the flux-ladder model is based on lattice gauge theories and quantum double models \cite{kitaev2003} (see also \cite{stern2016}). In particular, we will exploit the formalism adopted for the quantum simulations of lattice gauge theories (see, for example, the reviews \cite{montangero2016,zohar2016}) and we will adopt the notation developed in Refs. \cite{burrello15,burrello16} for their tensor-network study.

Our aim is to define a chiral flux-ladder model invariant under a global gauge group $G$, with $G$ being a discrete group. In analogy with the previous section, we consider degrees of freedom associated with the rungs of the ladder. Each of these rung degrees of freedom spans a local Hilbert space of dimension $|G|$, the order of the group $G$, and a basis for the local states in each rung is given by $\left\{\ket{g}, g\in G\right\}$. This is the group element basis which allows us to easily define the gauge-fluxes populating the plaquettes of the ladder. 

For the construction of our model, we want to generalize both the $\tau$ and the $\sigma$ operators from $\mathbb{Z}_N$ to a generic non-Abelian $G$. These are extended by defining, for each rung: (i) local operators $\theta_g$ and $\tilde{\theta}_g$ that implement left and right local gauge transformations and play the role of the $\tau$ operators; (ii) local matrices $U_{mn}$ of operators which constitute gauge-connection operators and are associated to the fundamental irreducible representation $F$ of $G$; the operators $U$ generalize the $\sigma$ operators in the Potts model.

Based on the group element basis, the previous operators are defined in the following way:
\begin{align}
\theta_g \ket{h} &= \ket{gh} \,, \qquad \quad \qquad \theta^\dag_g \ket{h} = \ket{g^{-1}h},\\
\tilde{\theta}_g \ket{h} &= \ket{hg} \,, \qquad \quad \qquad\tilde{\theta}_g^\dag \ket{h} = \ket{hg^{-1}},\\
U_{mn} \ket{h} &= D_{mn}(h) \ket{h}\,, \quad U_{mn}^\dag \ket{h} = D_{mn}^\dag(h) \ket{h} \label{Udef}
\end{align}
for any $g,h \in G$. In Eq. \eqref{Udef}, the matrix $D_{mn}(h)$ is the unitary matrix which represents the element $h \in G$ in the fundamental representation $F$ of the group. 
More generally, $D^K_{mn}(g)$ will label the $\dim(K) \times \dim(K)$ unitary matrix representing the element $g$ in the representation $K$ of the group; these matrices generalize the Wigner matrices of SU(2). For any irreducible representation $K$, we define operators
\begin{equation}
U_{mn}^K \ket{h} = D^K_{mn}(h) \ket{h}\,, \quad U_{mn}^{K\dag} \ket{h} = D_{mn}^{K\dag}(h) \ket{h} \label{UdefK}\,.
\end{equation}
When the irrep index is not specified, the fundamental representation is assumed.

 We observe that all the connection operators $U$ are diagonal in the group element basis, consistently with our previous description of the $\mathbb{Z}_N$ models; furthermore, we emphasize that $U_{lm}U_{mn}^\dag = \delta_{ln}\Id$, where $\Id$ is the identity operator. Hereafter the Einstein summation convention (summation on repeated indices) is used for the matrix indices.

The operators $\theta_g$ and $\tilde{\theta}_g$ are unitary operators, which transform the state $\ket{h}$ based on the group composition rules. In particular, they fulfill $\theta_g=\theta^\dag_{g^{-1}}$ and  $\tilde{\theta}_g=\tilde{\theta}^\dag_{g^{-1}}$. 

From the previous relations, it is easy to calculate the commutators of these operators:
\begin{align}
U_{mn}\theta_g &= \theta_g [D(g)U]_{mn}\,, \label{Utr1}\\
U_{mn}\tilde{\theta}_g &= \tilde{\theta}_g [UD(g)]_{mn}\,, \label{Utr2} \\
\theta_g \tilde{\theta}_h &= \tilde{\theta}_h \theta_g\,.
\end{align}

Following the convention in Refs. \cite{burrello15}, we finally point out that the matrices $D^K_{mn}(g)$ allow us to define a Fourier transformation that changes the basis for the rung Hilbert space from the group to the irreducible representation basis, and, in particular, from the eigenstates of $U$ to the eigenstates of $\theta$ and $\tilde{\theta}$. This unitary transformation is given by
\begin{equation} \label{reprbases}
\ket{Kmn} = \sum_{g \in G} \sqrt{\frac{\dim K}{|G|}} D^{K}_{mn}(g) \ket{g} \,.
\end{equation}
For the states $\ket{Kmn}$ of this basis, we have
\begin{align}
&\theta_g \ket{Kmn} = D^{K}_{ml}(g^{-1}) \ket{Kln}\,, \label{thetaK} \\
&\tilde{\theta}_g \ket{Kmn} = D^{K}_{ln}(g^{-1}) \ket{Kml}\,.
\end{align}

To describe the flux ladder model, we label the connection operator by $U(r)$ and the gauge transformations acting locally on the rung $r$ by $\theta_g(r)$ and $\tilde{\theta}_g(r)$. In particular, the global left and right gauge transformations assume the form
\begin{equation}
\mathcal{Q}_g = \prod_r \theta_g(r) \,, \qquad 
\tilde{\mathcal{Q}}_g = \prod_r \tilde{\theta}_g(r)\,,
\end{equation}
for any nontrivial group element $g \neq e \in G$, with $e \in G$ labeling the identity element.

Besides the $U$ and $\theta$ operators, we introduce for later convenience the family of  ``dressed'' gauge operators, acting on a single rung:
\begin{equation} \label{thetadr}
\Theta_{g,\sf{ac}}^K = U_{\sf{ab}}^{K\dag} \theta_g U^K_{\sf{bc}} = \theta_g U_{\sf{am}}^{K\dag} D_{\sf{mn}}^{K\dag}(g) U^K_{\sf{nb}}\,.
\end{equation} 
Hereafter we will use different fonts for the matrix indices associated to the dressed gauge operators.
The operators $\Theta$ appear in the study of bond-algebraic dualities for non-Abelian symmetric models developed by Cobanera \emph{et al}. \cite{cobanera2013a}, and obey the same group composition rules of the gauge operators $\theta_g$. In particular, it is easy to verify that
\begin{equation}
\Theta_{g,\sf{ab}}^K\Theta_{h,\sf{bc}}^K=U_{\sf{am}}^{K\dag} \theta_g U_{\sf{mb}}^K U_{\sf{bn}}^{K\dag} \theta_h U^K_{\sf{nc}}  = \Theta^K_{gh,\sf{ac}}\,,
\end{equation} 
for any irreducible representation $K$, and
\begin{equation}
\Theta^K_{g,\sf{ab}}\Theta_{g,\sf{bc}}^{K\dag} = \delta_{\sf{ac}} \Id\,;
\end{equation}
from these relations we get, in particular $\Theta^\dag_g = \Theta_{g^{-1}}$\,. From the definition \eqref{thetadr}, it is easy to derive that the behavior of the $\Theta$ operators under the global left transformations matches the behavior of the gauge operators $\theta$:
\begin{equation}
\mathcal{Q}_h^\dag \Theta^K_g(r) \mathcal{Q}_h = \Theta_{h^{-1}gh}^K(r)\,.
\end{equation}
For Abelian representations $K$, $\Theta_g$ is reduced to $\theta_g D^K(g^{-1})$.

\subsection{The flux Hamiltonian and its symmetries} \label{sec:Ham}

\begin{figure}[t]
\includegraphics[width=\columnwidth]{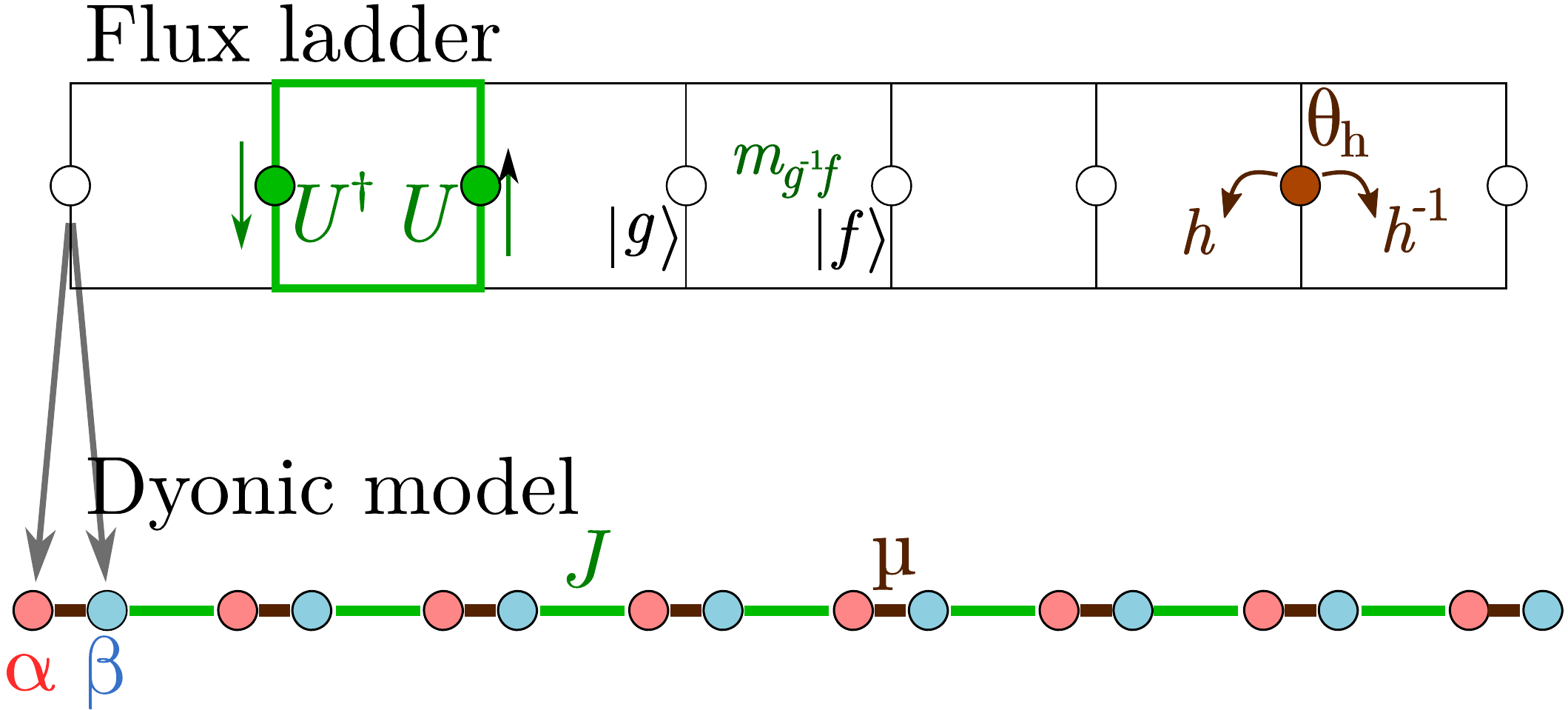}
\caption{(Top) Graphical representation of the operators in the flux-ladder Hamiltonian. The nearest-neighbor (green) terms define $H_J$ and assign a mass to the gauge fluxes: these terms are plaquette operators built from the connection operators $U$. The rung $\theta_h$ terms in $H_\mu$ modify the fluxes in the two neighboring plaquettes. Bottom: the dyonic model is obtained by redefining each rung based on two kinds of operators, $\alpha$ and $\beta$. The Hamiltonians $H_J$ and $H_\mu$ act on different pairs of dyonic operators. }
\label{fig:ladder}
\end{figure}

By exploiting the operators introduced above, we define the flux-ladder model through the Hamiltonian:
\begin{multline}
\label{ham_flux}
H = -J\left(\sum_r \mathrm{Tr}[U(r+1)CU^\dag(r)]+ \mathrm{H.c.} \right) \\
- \mu\sum_{r}\sum_{g \neq e \in G} \chi^A(g^{-1}) \theta_g(r)  \,,
\end{multline}
where $J$ and $\mu$ are real coupling constants and $C$ is a unitary matrix responsible for the chiral nature of the system. In this expression,
\begin{equation}
\chi^A(g^{-1}) = \Tr D^A(g^{-1})\,
\end{equation}
 labels the character of an auxiliary irreducible representation $A$ of the group element $g$. Its role will be important in the definition of the dyonic topological model and it will be discussed in detail in Section \ref{sec:aux}. 

In the following, we label the first term in the Hamiltonian \eqref{ham_flux} by $H_J$ and the second term by $H_\mu$. In this work we are mostly interested in the ordered regime $J \gg \mu$ where $H_J$ dominates and the system presents degenerate ground states in the thermodynamic limit. 

In the following, we discuss the main features of $H_J$ and $H_\mu$, the role of the $C$ matrix and the symmetries of the Hamiltonian $H$. A pictorial representation of the system is provided in Fig. \ref{fig:ladder}.

\subsubsection{$H_J$ and the flux masses}

The first term in the Hamiltonian \eqref{ham_flux} is responsible for the definition of the mass spectrum of the fluxes in the ladder and it generalizes the $J$ term in the chiral Potts model \eqref{potts}. Each operator acts on neighboring degrees of freedom, therefore it is useful to consider the two-rung state $\ket{g_r}_r\ket{g_{r+1}}_{r+1}$: such a state defines a flux $\Phi(r)$ in the $r^{\rm th}$ plaquette which corresponds to the element $g_r^{-1}g_{r+1}$ of the group $G$. In our model, the fluxes are indeed in one-to-one correspondence with the group elements, thus, to define their mass, we exploit the connection operators $U$, which are diagonal in the group element basis. Analogously to the Kogut and Susskind formulation of lattice gauge theories \cite{KogutSusskind}, we consider the trace of these operators as a building block for the masses $m_g$ associated to the fluxes. 
In the simple case of $C=\Id$, the operator $\mathrm{Tr}[U(r+1)U^\dag(r)]$ returns the character $\chi^F(\Phi(r))$ of the group element $\Phi(r)=g_r^{-1}g_{r+1}$ associated to the fundamental representation $F$. The character is maximized by the identity, thus the trivial flux, but it cannot distinguish between group elements in the same conjugacy class, leading to degeneracies in the mass spectrum. To avoid these degeneracies, we introduce the unitary $C$ matrix, of dimension given by $\dim(F) \times \dim(F)$, such that, in general, we can define nondegenerate flux masses:
\begin{equation}\label{mass}
m_g = -J \left(\Tr\left[D(g)C\right] + \Tr\left[C^\dag D^{\dag}(g)\right]\right)\,.
\end{equation}  
For our analysis, it will be important to consider the following conditions on the mass spectrum.
\begin{enumerate}
\renewcommand{\theenumi}{$\mathcal{C}$\arabic{enumi}}
\renewcommand{\labelenumi}{ \theenumi :}
\item \label{condI} For the sake of simplicity we impose that the mass of the trivial flux $e \in G$ is the lowest. This means that the ground states of $H_J$ are states with no fluxes, thus no domain walls in the group element basis. This condition is not necessary for our results, but it simplifies our analysis because it implies that the ordered phase is ferromagnetic-like rather than helical-like. This is analogous to choosing $|\theta| < \pi/3$ in the $\mathbb{Z}_3$ chiral Potts model.
\item \label{condIIa} We impose the mass spectrum to be nondegenerate. As we will discuss in the next sections, this is a necessary but not sufficient requirement for the definition of strong zero-energy modes in the corresponding topological models. This condition implies that we must choose a $C$ matrix such that
\begin{equation}\label{condc}
\Re\left(\mathrm{Tr}[C D(g)]\right) \neq \Re\left(\mathrm{Tr}[C D(h)] \right)\,,
\end{equation}
for any $g \neq h \in G$.
\end{enumerate}

It is now important to define the left and right global gauge transformations of the operators in $H_J$ based on Eqs. \eqref{Utr1} and \eqref{Utr2}:
\begin{align} 
&\mathcal{Q}_g^\dag \mathrm{Tr}[U(r+1)CU^\dag(r)] \mathcal{Q}_g \nonumber \\
&=\Tr[D(g)U(r+1)CU^\dag(r)D^{\dag}(g)] = \mathrm{Tr}[U(r+1)CU^\dag(r)]\,,\\
&\tilde{\mathcal{Q}}_g^\dag \mathrm{Tr}[U(r+1)CU^\dag(r)] \tilde{\mathcal{Q}}_g  \nonumber\\
&=\Tr[U(r+1)D(g)CD^{\dag}(g)U^\dag(r)] \neq \mathrm{Tr}[U(r+1)CU^\dag(r)]\,.
\end{align}
From these equations we see that, in general, $H_J$ is invariant under left global transformation but it is not invariant under right transformations. This is true if $C$ is not a multiple of the identity, since the matrices $D(g)$ are an irreducible representation of the group.
The matrix $C$ breaks the global right gauge symmetry, and this is a manifestation of the chiral nature of the model. We observe that, by exchanging the order of $C$ and $U(r+1)$ in the Hamiltonian, we would get a corresponding model with right rather than left gauge symmetry.

\subsubsection{About the $C$ matrix}

The $C$ matrix is a unitary $\dim(F) \times \dim(F)$ matrix that generalizes the role of the phase $e^{i\theta}$ in the chiral Potts model \eqref{potts} to the non-Abelian case. By expressing the matrix $C= e^{-i \gamma_j T_j}$ as a function of the generators $T_j$ of $U(\dim(F))$, we see that $C$ is a collection of $\dim(F)^2$ parameters. $C$ must be chosen to fulfill the condition \eqref{condc} and, \emph{a priori}, it is not evident that such a matrix exists for all $G$. In the following, we provide a geometrical interpretation of $C$ aimed at showing its existence for groups whose fundamental representation has dimension 2. These include, for example, the group $S_3$, which is the smallest non-Abelian group. In this case, any matrix $D(g)$ can be parametrized as a function of four parameters:
\begin{equation}
D(g)= e^{i \eta_{g,0} \sigma_0 + i \vec{\eta}_g\vec{\sigma}} = e^{i\eta_{g,0}}\left(\cos|\vec{\eta}_g|\sigma_0 + i \sin|\vec{\eta}_g|\hat{\eta}_g \vec{\sigma}  \right)\,,
\end{equation}
where $\sigma_0$ is the $2\times 2$ identity, $\vec{\sigma}$ is the vector of the Pauli matrices, and $\hat{\eta}_g$ is the three-dimensional unit vector in the direction of $\vec{\eta}_g$. A similar decomposition holds for $C=e^{-i\gamma_{0}\sigma_0 -i \vec{\gamma}\vec{\sigma}}$.
We define four-component vectors in the unitary $S^3$ sphere:
\begin{equation}
\mathcal{D}(g) = \begin{pmatrix} \cos|\vec{\eta}_g| \\ \sin|\vec{\eta}_g|\hat{\eta}_g  \end{pmatrix},\qquad \mathcal{C} = \begin{pmatrix} \cos|\vec{\gamma}| \\ \sin|\vec{\gamma}|\hat{\gamma}  \end{pmatrix}.
\end{equation}
Based on this parametrization, the mass of the $g$ flux is
\begin{equation}
m_g = -4J\cos(\eta_{0,g} - \gamma_0) \mathcal{D}(g) \cdot \mathcal{C}\,.
\end{equation}
Hence the condition \eqref{condc}, for any $g \neq h$, becomes
\begin{equation} \label{condc2}
\left[\cos(\eta_{0,g} - \gamma_0) \mathcal{D}(g) - \cos(\eta_{0,h} - \gamma_0) \mathcal{D}(h)\right] \cdot \mathcal{C} \neq 0.
\end{equation}
We fix a value of $\gamma_0$ such that $\cos(\eta_{0,g} - \gamma_0) \neq 0$ for every $g$ and we define a set of rescaled vectors $\mathcal{D}'_g=\cos(\eta_{0,g} - \gamma_0) \mathcal{D}(g)$. In particular, if $F$ is orthogonal (as in the $G=S_3$ case, or any dihedral group), $\eta_{0,g}=0,\pi/2$, and we can choose $\gamma_0=\pi/4$ such that all the cosines become $1/\sqrt{2}$. 
The equations \eqref{condc2} fix $|G|(|G|-1)/2$ conditions that the vector $\mathcal{C}$ must fulfill: the unit vector $\mathcal{C}$ cannot be orthogonal to any of the vectors defined by the differences $\mathcal{D}'_g - \mathcal{D}'_h$ in \eqref{condc2}. Each of these $|G|(|G|-1)/2$ vectors define a great circle on the $S^3$ sphere of orthogonal vector. Therefore we conclude that we can choose any $C$ matrix corresponding to a $\mathcal{C}$ vectors on the $S^3$ sphere that does not belong to any of these great circles. When $\mathcal{C}$ approaches one of these great circles, one of the mass gap closes, thus violating \eqref{condc}. 
A similar geometric interpretation can be build for any irreducible representation in $U(N)$ (see Appendix \ref{app:c}).

\subsubsection{The $H_\mu$ term}

The $H_\mu$ term of the Hamiltonian is meant to provide a dynamics to the fluxes in the ladder, it does not commute with $H_J$ and, differently from $H_J$ is diagonalized in the irreducible representation basis of the rung degrees of freedom, based on Eq. \eqref{thetaK}. 

We observe that, since $\theta_g = \theta_{g^{-1}}^\dag$, $H_\mu$ is Hermitian. Furthermore, for $g=e$, the gauge transformation is just an identity and it provides only an overall energy shift. Therefore we can choose to include or not this term in the Hamiltonian.

$H_\mu$ is meant to generalize the $h$ term in the Potts model \eqref{potts}: for $A$ corresponding to the trivial irreducible representation,  $H_\mu$ is the sum of all the possible gauge transformation operators over all the degrees of freedom and it directly generalizes \eqref{potts}. For a different representation $A$, the resulting Hamiltonian is instead related to a more general form of $\mathbb{Z}_N$ symmetric models studied in \cite{fendley2012}.

The $H_\mu$ term in the Hamiltonian \eqref{ham_flux} corresponds to a projector over the subspace of the states of the rung $r$ corresponding to the irreducible representation $A$. We recall that the projector over a generic irreducible representation $K$ is given by
\begin{equation} \label{Kproj}
\Pi^K = \frac{{\rm dim}(K)}{|G|}\sum_{g \in G} \chi^K(g^{-1}) \theta_g = \sum_{mn} \ket{Kmn}\bra{Kmn}\,.
\end{equation}
Such expression is invariant under both left and right gauge transformations, and $H_\mu$ is thus symmetric under both global transformations. Therefore the (left) set of transformation $\mathcal{Q}_g$ corresponds in general to the global symmetry group for the whole Hamiltonian $H$ when $C\neq \id$. 

The form of $H_\mu$ we have chosen in \eqref{ham_flux} is not the most general preserving such gauge symmetry. We could extend $H_\mu$ to
\begin{equation}
H'_\mu = -\mu\sum_{\substack{r,C_l}} f_{C_l} \sum_{g \in C_l}  \theta_g(r)  = -\mu\sum_{\substack{r,A}} f'_A \Pi^A(r) \,,
\end{equation}   
where $C_l$ runs over the conjugacy classes of $G$, and $A$ runs over the irreducible representations. For the purpose of defining a model with topological order, it is sufficient to consider a single non-Abelian irreducible representation $A$ as in \eqref{ham_flux}.

\subsubsection{The symmetries of the system}

We have already emphasized that the Hamiltonian \eqref{ham_flux} is invariant under the action of the global left gauge transformation for arbitrary $C$, whereas the right transformations do not constitute a symmetry of the system. Analogously to the Potts case, the matrix C breaks also the time-reversal and space-inversion symmetries. The time reversal $T$ transforms the connection and local gauge operators in the following way:
\begin{equation}
T^\dag U(r) T = U^\dag(r)\,,\quad T^\dag \theta_g(r) T = \theta_{g}(r)\,.
\end{equation}
Therefore $H_\mu$ is time-reversal invariant, whereas it is easy to verify that $H_J$ is not for any $C \neq \Id$, due to the representation $F$ being irreducible.
Space inversion $P$ can be defined by
\begin{equation}
P^\dag U(r) P = U(L-r)\,,\quad P^\dag \theta_g(r) P = \theta_g(L-r)\,,
\end{equation}
with $L$ being the system size. $H_\mu$ is invariant also for the inversion transformation, whereas $P^\dag H_J (C) P = H_J(C^\dag)$; therefore $H$ is symmetric under $P$ only if $C$ is Hermitian. For generic unitary $C$ matrices, the system is invariant neither under $P$ and $T$, nor under $PT$. Therefore we do not expect exact degeneracies in the spectrum besides the ones dictated by the global symmetries $\mathcal{Q}_g$.

Concerning the exact degeneracies of the system caused by the global gauge group $G$, each eigenstate of $H$ must transform under $G$ following one of its irreducible representations. Therefore, in general, the spectrum will present exact degeneracies given by the dimensions $\dim(K)$ of the group's irreducible representations.

\subsection{The ordered phase} \label{sec:order}

Let us consider first a system with $\mu=0$. In this case the gauge fluxes have no dynamics and we can associate each state to a collection of fluxes $\left\{\Phi\right\}$. In a ladder of length $L$, the spectrum of $H_J$ is given by the energy levels:
\begin{equation}
E\left(\left\{\Phi\right\}\right) = \sum_{g \in G} n_g m_g \,,
\end{equation}
where $n_g$ counts how many times the flux $g$ appears in the set $\left\{\Phi\right\}$ for a given state of the ladder, and $\sum_g n_g = L-1$. This is analogous to the analysis of the $\mathbb{Z}_N$ symmetric case in \cite{Moran2017}.

When the identity flux is the flux with the lowest mass (condition \ref{condI}), $H_J$ presents $|G|$ ground states corresponding to ferromagnetic states, i.e. without domain walls, in the group element basis. We label these ground states as
\begin{equation}
\ket{\ket{g}} = \bigotimes_{r} \ket{g}_r\,.
\end{equation}
To emphasize the transformation properties of the ground states under the global symmetries $\mathcal{Q}_g$ it is convenient to introduce also a representation basis, analogous to \eqref{reprbases}, for the ground states,
\begin{equation} \label{groundrepr}
\ket{\ket{Kmn}} = \sum_{g \in G} \sqrt{\frac{\dim K}{|G|}} D^{K}_{mn}(g) \, \ket{\ket{g}} \,
\end{equation} 
such that
\begin{equation} \label{gsrepr}
\mathcal{Q}_h^\dag \ket{\ket{Kmn}} = D^{K}_{mm'}(h) \ket{\ket{Km'n}}\,.
\end{equation}

When we introduce a weak $H_\mu$ perturbation, the exact degeneracy of the ground states is split: the $|G|$ ground states are perturbed and separate into a set of families; if $C$ is not a multiple of the identity, the right gauge symmetry is broken and there are $\dim{K}$ families for each irreducible representation $K$. Each of these families has dimension $\dim{K}$. On the other hand, for a trivial $C$ matrix, the right gauge symmetry is restored and there is one family of ground states per irreducible representation, with dimension $(\dim{K})^2$. 

The states within each family maintain their exact degeneracy due to the global symmetry, but, for finite-size systems, small energy gaps are introduced between different ground-state families. Similarly to the $\mathbb{Z}_N$ systems, this splitting of the energies of the ground-state manifold is exponentially suppressed in the system size and it is roughly proportional to $\mu^L/J^{L-1}$. This can be deduced by a perturbative approach: in order for the $H_\mu$ perturbation to turn one ground state into another, it must be applied $L$ times. In this way a flux can be introduced into the system and can propagate from one edge to the other similarly to the domain walls in the $\mathbb{Z}_N$ case \cite{jermyn2014}. Other terms that introduce multiple fluxes are suppressed by their higher energy. Quantitatively, we find that the ground state splitting is given by the effective Hamiltonian:
\begin{multline} \label{perturbativesplitting}
\bra{\bra{gh}} H' \ket{\ket{h}} \\
 = - \left[\frac{\left(\chi^A(g^{-1}) \mu\right)^L}{\left(m_{h^{-1}gh} - m_e\right)^{L-1}} + \frac{\left(\chi^A(g^{-1}) \mu\right)^L}{\left(m_{h^{-1}g^{-1}h} - m_e\right)^{L-1}} \right],
\end{multline}
where the masses $m_g$ are defined in Eq. \eqref{mass} and are proportional to $J$.

The situation is more complicated for the excited states, in which processes of order lower than $L$ can cause transitions between different flux configurations, thus opening gaps that potentially may depend on the specific states involved and break the $|G|$ quasidegeneracy of the spectrum. 

In particular this may happen between degenerate flux configurations, which are states with different flux multiplicities $n_g$ and $n_g'$, but the same energy. In \cite{Moran2017} it has been shown that, in the presence of these resonances among excited states of $H_J$, there may be perturbation processes of low order (namely with an order that does not scale with the system size) which may split these degeneracies in the $\mathbb{Z}_N$ symmetric model \eqref{potts}. Similar processes can imply that the energy splitting of the excited states is not exponentially suppressed with the system size in the non-Abelian model as well.

\subsection{The $S_3$ flux ladder} \label{sec:s3}

To exemplify the flux ladder models in Eq. (\ref{ham_flux}) and verify our analysis of the spectrum of the ordered phase, we consider the smallest non-Abelian group, namely the symmetric group $S_3$ of all the permutations of three elements $(s_1,s_2,s_3)$. $S_3$ has six elements and can also be considered the group of transformations that leave an equilateral triangle invariant. It is generated by two elements, $b$ and $c$ which satisfy the relations $b^2 = c^3 = e$, where $e$ is the identity element, and $bc = c^2b$.

Using the latter relation one can write every element of $S_3$ in ``normal form'': $g = b^nc^m$. In particular, we choose $b$ to permute the first two elements, $b:(s_1,s_2,s_3) \mapsto (s_2,s_1,s_3)$, and $c$ to cyclically permute the three elements, $c:(s_1,s_2,s_3) \mapsto (s_3,s_1,s_2) $.
We denote the representations of this group by $I$ and write the representation matrices as $D^{I}$. There are three irreducible representations of $S_3$: The trivial representation, where each element is represented by the number 1, the parity representation,  where elements $g=b^nc^m$ are represented by $(-1)^n$, and the two-dimensional (fundamental) irreducible representation, which is defined below. We denote these representations by $I=1,-1,2$, respectively.  To construct the Hamiltonian we use the fundamental representation $I=2$ for the definition of the operators $U$. This representation is a subgroup of $O(2)$ and we have
\begin{equation} \label{s3matrix}
D^{(2)}(b) = \begin{pmatrix} 1 && 0 \\ 0 && -1\end{pmatrix}, \;\;
D^{(2)}(c) = \frac{1}{2}\begin{pmatrix} -1 && -\sqrt{3} \\ \sqrt{3} && -1\end{pmatrix}.
\end{equation}
One can think of $D^{(2)}(c)$ as the rotation matrix for a $2\pi/3$ rotation about the $z$-axis, and $D^{(2)}(b)$ as a two-dimensional mirror symmetry about the $x$-axis.

The terms of the Hamiltonian $H_J$ are diagonal in the group element basis. We decide to work in this basis and to use, for each rung, the following ordering of the group elements: $\{\ket{e}, \ket{c}, \ket{c^2}, \ket{b}, \ket{bc}, \ket{bc^2}\}$. 
The states may be conveniently expressed in the tensor product structure $\ket{n}\otimes\ket{m} \equiv \ket{b^nc^m}$, with $n=0,1$ and $m=0,1,2$. From the point of view of the transformations of the equilateral triangle in itself, the states with $n=0$ correspond to the orientation-preserving transformations (rotations), whereas $n=1$ labels the transformation inverting the orientation of the vertices (inversions). In the basis $\ket{n}\otimes\ket{m}$ we may write the local gauge transformations as
\begin{align}
\theta_b &= \begin{pmatrix} 0 & 1 \\ 1 & 0 \end{pmatrix} \otimes \Id, \\
\theta_c &= \begin{pmatrix} 1 & 0 \\ 0 & 0 \end{pmatrix} \otimes \begin{pmatrix} 0 & 0 & 1 \\ 1 & 0 & 0 \\ 0 & 1 & 0\end{pmatrix} + \begin{pmatrix} 0 & 0 \\ 0 & 1 \end{pmatrix} \otimes \begin{pmatrix}0 & 1 & 0 \\ 0 & 0 & 1 \\ 1 & 0 & 0 \end{pmatrix}\,.
\end{align}
All other gauge transformations can be found by compositions of these.

\begin{figure}[t]
\includegraphics[width=\columnwidth]{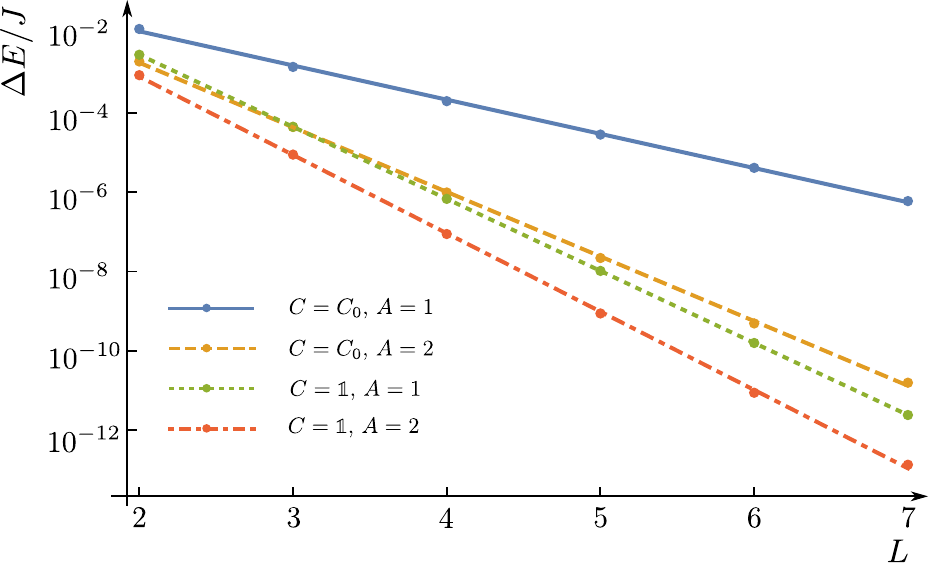}
\caption{
Maximal splitting of the six ground states in units of $J$ of the model with $G=S_3$ and $\mu/J=0.03$, shown on a semilogarithmic plot. There are four different cases, depending on the matrix $C$ (see Eq. \eqref{C0} for the matrix $C_0$) and the irreducible representation $A$.
In all cases the exponential decay of the energy splitting with the system size is evident.}
\label{fig:numerics:groundstates}
\end{figure}

\begin{figure}[t]
\includegraphics[width=\columnwidth]{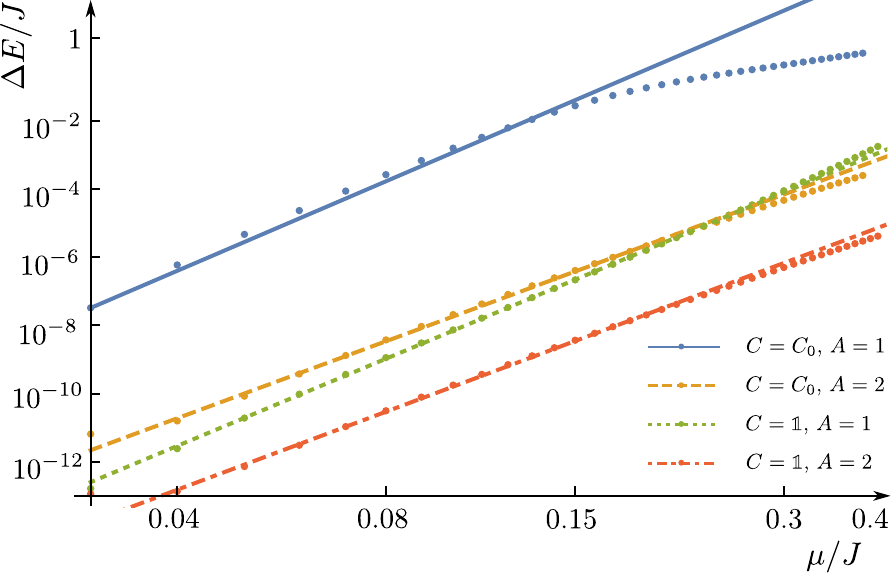}
\caption{
The ground-state splitting in units of $J$, as a function of $\mu/J$ for seven sites, shown on a logarithmic plot, in the same cases as Fig. \ref{fig:numerics:groundstates}. 
The lines are linear fits based on the points with lowest $\mu$, and the change of slope for larger values of $\mu$ is a possible signature of phase transitions.}
\label{fig:numerics:groundstatesfunctionofmu}
\end{figure}

To illustrate the energy features of the ground state manifold, we consider the cases $C = \Id$ and $C = C_0$, with
\begin{equation}  \label{C0}
C_0\equiv \frac{e^{-i\pi/4}}{\sqrt{2}}\left(\Id+ \frac{i}{\sqrt{3}}\sigma_x + \frac{i}{\sqrt{3}}\sigma_y - \frac{i}{\sqrt{3}}\sigma_z\right)\,.
\end{equation}
 The choice $C=\Id$ is the trivial case with fluxes in the same conjugacy class being degenerate, while $C=C_0$ is a choice that satisfies conditions $\mathcal{C}1$ and $\mathcal{C}2$, and presents the following mass spectrum in units of $J$, using the same ordering of the group elements as above: $\{-2, 0,2, 2/\sqrt{3}, 1-1/\sqrt{3}, -1-1/\sqrt{3}\}$.

We calculated the ground-state energies via exact diagonalization as a function of the system size and $\mu$, for $C=\Id,C_0$ and the auxiliary representations $A=1,2$. 
Due to the global symmetries, for generic values of the matrix $C$ and $\mu \ll J$, the six ground states present a degeneracy pattern $1,1,2,2$ corresponding to the nondegenerate states $\ket{\ket{K=1,11}}$, $\ket{\ket{K=-1,11}}$, $\ket{\ket{K=2,j1}}$, and $\ket{\ket{K=2,j2}}$ based on their behavior under the symmetry group expressed in Eq. \eqref{gsrepr}. For $C=\Id$, when the right gauge symmetry is restored, the four ground states with $K=2$ become exactly degenerate.

We define the ground-state splitting $\Delta E$ as the difference between the energies of the highest and lowest state in the ground-state manifold. Based on the perturbative result in Eq. \eqref{perturbativesplitting}, the dominant contribution in this splitting must scale as $\Delta E \propto \mu(\gamma \mu/J)^{L-1}$ for a suitable numerical coefficient $\gamma$. 
The ground-state splitting $\Delta E$ as a function of $L$ is shown in Fig. \ref{fig:numerics:groundstates} for $\mu=0.03J$. For all the analyzed cases, we numerically find the expected exponential suppression of the ground-state splitting with the system size. In Fig. \ref{fig:numerics:groundstatesfunctionofmu}, we illustrate instead the ground-state splittings as a function of $\mu$ for $L = 7$. The power law behavior for small $\mu$ is clearly evident. For all the analyzed cases, the energy splitting approximately behaves like $\delta E \propto \mu^\alpha$ with the exponent $\alpha$ in the range between $7$ and $9$, compatible with the dominant contribution in Eq. \eqref{perturbativesplitting}. For larger values of $\mu$ and $C\neq \Id$, our numerics suggest a change in the exponent, signaling a transition into a different phase.

The study of the full phase diagram as a function of the matrix $C$ and the auxiliary irreducible representation $A$ is an interesting and highly nontrivial problem, which goes beyond the scope of this paper. We observe, however, that for $\mu \to \infty$, $H_\mu$ projects each site on the subspace spanned by the states $\ket{Amn}$. For $A$ Abelian, this implies the existence of a paramagnetic phase for $\mu \gg J$ with a nondegenerate ground state. For $A$ non-Abelian, instead, $H_\mu$ presents a ground-state degeneracy, which grows as $(\dim(A))^{2L}$; these ground states are then split by the introduction of a weak $H_J$. Between the regimes dominated by $H_\mu$ and $H_J$, other phases may be present. For example, in analogy with the $\mathbb{Z}_n$ case, we expect that, for suitable choices of $C$, critical incommensurate phases (see, for instance, Refs. \cite{ostlund1981,burrello2014,hughes2015,sachdev2018}) and phase transitions with dynamical critical exponent $z \neq 1$ \cite{sachdev2018,lukin2018} may appear.

\section{Non-Abelian models with topological order} \label{sec:topo1}

\subsection{The non-Abelian Jordan-Wigner transformation and the dyonic modes} \label{sec:JW}

A model with topological order can be defined by a nonlocal transformation which maps the flux-ladder operators into dyonic operators, characterized by a group element $g$ and by the fundamental representation $F$. These dyonic operators display nontrivial commutation relations even when spatially separated, thus they are nonlocal in the original degrees of freedom of the ladder Hamiltonian. In this respect, they constitute a generalization of the parafermionic operators from $\mathbb{Z}_N$ to non-Abelian groups. In the $\mathbb{Z}_N$ model \cite{fendley2012}, the definition of the parafermionic operators is based on a $\mathbb{Z}_N$ JW transformation that amounts to the multiplication of order and disorder operators \cite{fradkin1980}. The definition of disorder operators, in turn, can be rigorously based on a bond-algebraic duality transformation \cite{cobanera2011}. Inspired by the bond-algebraic dualities for non-Abelian models \cite{cobanera2013a}, we introduce the following disorder operators for the non-Abelian flux-ladder, which is defined in terms of the dressed gauge operators \eqref{thetadr}:
\begin{multline} \label{Ldef}
\mathcal{L}^A_{g,\sf{a}_1\sf{a}_{r+1}}(r) = \Theta^{A\dag}_{g,\sf{a}_1\sf{a}_{2}}(1) \Theta^{A\dag}_{g,\sf{a}_2\sf{a}_3}(2) \ldots \Theta^{A\dag}_{g,\sf{a}_{r}\sf{a}_{r+1}}(r)  \\
= \left[\prod_{x=1}^r \theta_g^\dag(x)\right]  U^\dag(1)D(g)U(1)U^\dag(2) \ldots U^\dag(r)D(g)U(r)\,,
\end{multline}
where we omitted the representation superscript $A$ in the second row.
The string operator $\mathcal{L}$ introduces a flux $g$ in the $r^{\rm th}$ plaquette of the system and returns the matrix $D\left(h_1^{-1}g h_1h_2^{-1} gh_2 \ldots h_r^{-1}g h_r    \right)$ in the auxiliary representation $A$ when applied to any state $\ket{h_1}_1\ldots\ket{h_r}_r$. These operators $\mathcal{L}_g$ fulfill the following properties for any $A$:
\begin{align}
&\mathcal{L}_{g,\sf{ab}}(r) \mathcal{L}^\dag_{g,\sf{bc}}(r) = \mathcal{L}^\dag_{g,\sf{ab}}(r) \mathcal{L}_{g,\sf{bc}}(r)= \delta_{\sf{ac}}\Id  \label{ltr2} \,,\\
&\mathcal{Q}_h^\dag \mathcal{L}_{g}(r) \mathcal{Q}_h =\mathcal{L}_{h^{-1}gh}(r)\,, \label{ltr}\\
&\mathcal{L}_{g,\sf{a}_1\sf{a_2}} \mathcal{L}_{g,\sf{a}_2 \sf{a_3}} \ldots \mathcal{L}_{g,\sf{a}_{|G|}\sf{a}_{|G|+1}} \equiv \left(\mathcal{L}^{|G|}_g\right)_{\sf{a}_1\sf{a}_{|G|+1}} = \delta_{\sf{a}_1\sf{a}_{|G|+1}} \Id\,. \label{ltr3}
\end{align}
The last equation is easily proved by considering that, in the second row of Eq. \eqref{Ldef}, the gauge operator string $\left[\prod_{x=1}^r \theta_g^\dag(x)\right]$ commutes with the string of matrix operators.

We are now ready to define the dyonic operators through a generalized JW transformation obtained by the product of order operators $U^\dag$ and disorder operators $\mathcal{L}$. In full generality, we express the dyonic operators as
\begin{align}
\alpha_{g,mn,\sf{ab}}^{K,A}(2r-1) &=    \mathcal{L}_{g,\sf{ab}}^{A}(r-1)U_{mn}^{K\dagger}(r)\,,\label{JW1}\\
\beta_{g,mn,\sf{ab}}^{K,A}(2r) &= \mathcal{L}_{g,\sf{ab}}^{A}(r)U_{mn}^{K\dagger}(r)\,, \label{JW2}
\end{align}
for every $g\neq e$. These operators carry two pairs of matrix indices, $(mn)$ and $(ab)$, which are associated with the two irreducible representations $K$ and $A$ respectively. If we do not specify otherwise, we will consider $K=A=F$ and we will not specify the irreducible representation superscripts. However, it is necessary to keep the two representation distinguished: we adopt different fonts for their matrix indices and we will label by $\Tr_{K/A}$ the trace over the matrix indices of the two irreducible representations, respectively. 

In analogy with the Kitaev and parafermionic chains, each site $r$ of the flux ladder hosts two kinds of operators, $U$ and $\theta$, and it is decomposed in this dyonic description into a pair of sites, $2r-1$ and $2r$, each hosting the tensors of operators $\alpha$ and $\beta$, living in the odd and even sublattice respectively (see Fig. \ref{fig:ladder}). In the Abelian case, however, all the irreducible representations are one-dimensional, and no tensor structure of this kind appear.

We call these modes \textit{dyonic} because their transformation relations under the global gauge symmetries are similar to the ones of the irreducible representations of the Drienfield quantum double of $G$ \cite{kitaev2003}, as can be derived from Eqs. \eqref{Utr1} and \eqref{ltr}:
\begin{align}
\mathcal{Q}^\dag_h \alpha_{g,mn,\sf{ab}} \mathcal{Q}_h = \alpha_{h^{-1}gh,ml,\sf{ab}} D_{ln}^\dag(h)\,,\\
\mathcal{Q}^\dag_h \beta_{g,mn,\sf{ab}} \mathcal{Q}_h = \beta_{h^{-1}gh,ml,\sf{ab}} D_{ln}^\dag(h)\,,
\end{align}
for any site $r$. These relations are obtained by considering that the disorder operators $\mathcal{L}$ are conjugated by the global symmetry, whereas the operators $U$ transform following the fundamental irreducible representation $F$ (or a different irreducible representation $K$ in the most general case). We also observe that the first operator $\alpha(1) = U^\dag(1)$ does not have a dependence on any group element, differently from all the other operators.

Similarly to parafermionic modes, the following relations hold:
\begin{align}
&\alpha_{g,lm,\sf{ab}}^{\phantom{\dag}} \alpha^\dag_{g,mn,\sf{bc}} = \alpha_{g,lm,\sf{ab}}^\dag \alpha_{g,mn,\sf{bc}}^{\phantom{\dag}}= \delta_{ln} \delta_{\sf ac} \Id \,,\\
&\beta_{g,lm,\sf{ab}}^{\phantom{\dag}} \beta^\dag_{g,mn,\sf{bc}} = \beta_{g,lm,\sf{ab}}^\dag \beta_{g,mn,\sf{bc}}^{\phantom{\dag}}= \delta_{ln} \delta_{\sf ac}\Id\,. 
\end{align}

The commutation relations between $\alpha$ and $\beta$ operators can be obtained from the commutations between $\mathcal{L}(r)$ and $U^\dag(r')$ and the non-Abelian JW transformations, but, for general auxiliary representations $A$, they do not assume a simple form. In the following, we report the results for the special case of Abelian auxiliary representations, which offers the possibility of comparing the dyonic modes to $\mathbb{Z}_N$ parafermionic modes. When $A$ is Abelian, we can omit its trivial indices. Collectively denoting $\alpha(x)$ and $\beta(x)$  by $\gamma(x)$ for odd and even $x$ respectively, we get for $y>x$:
\begin{align} \label{braiding1}
&\gamma_{g,mn}(x) \gamma_{h,pq}(y) = \gamma_{h,pq}(y) \gamma_{hgh^{-1},ml}(x) D_{ln}(h), \\
&\gamma_{g,mn}^{\phantom{\dag}}(x) \gamma^\dag_{h,pq}(y) = \gamma^\dag_{h,pq}(y)  \gamma_{h^{-1}gh,ml}^{\phantom{\dag}}(x) D^\dag_{ln}(h),\\
&\gamma^\dag_{g,mn}(x) \gamma_{h,pq}^{\phantom{\dag}}(y) = \gamma_{h,pq}^{\phantom{\dag}}(y)  D^\dag_{ml}(h)\gamma^\dag_{hgh^{-1},ln}(x) ,\\
&\gamma^\dag_{g,mn}(x) \gamma^\dag_{h,pq}(y) = \gamma^\dag_{h,pq}(y)  D_{ml}^{\phantom{\dag}}(h)\gamma^\dag_{h^{-1}gh,ln}(x), \label{braiding4}
\end{align}
where only the $l$ indices are summed over. The commutation relations for $y<x$ can be derived by conjugation. The relations for $x=y$ and $g\neq h$, instead, differ for $\alpha$ and $\beta$ operators:
\begin{align}
\alpha_{g,mn}(r)\alpha_{h,pq}(r) &= \alpha_{h,pq}(r)\alpha_{hgh^{-1},mn}(r)\nonumber \\ 
&= \alpha_{g^{-1}hg,pq}(r)\alpha_{g,mn}(r)\, ,\\
\beta_{g,mn}(r)\beta_{h,pq}(r) &=  \beta_{h,ps}(r)D_{sq}^\dag(g)\beta_{hgh^{-1},ml}(r)D_{ln}(h) \nonumber \\
&=\beta_{g^{-1}hg,ps}(r)D_{sq}^\dag(g) \beta_{g,ml}(r)D_{ln}(h)\,.
\end{align}
These commutation rules can be seen as a non-Abelian extension of the parafermionic commutation relations. For non-Abelian $A$ representations, the algebra of the dyonic modes is more complicated. Furthermore, differently from their Abelian counterpart, the dyonic operators $\alpha$ and $\beta$ present different algebraic properties. In particular, for any choice of $A$, we observe that
\bwt
\begin{align}
&\alpha_{g,m_1m_2,\sf{a}_1\sf{a}_2} \alpha_{g,m_2m_3,\sf{a}_2\sf{a}_3} \ldots \alpha_{g,m_{|G|}m_{|G|+1},\sf{a}_{|G|}\sf{a}_{|G|+1}}   =   \left(\alpha_{g}^{|G|}\right)_{m_1 m_{|G|+1},\sf{a}_1\sf{a}_{|G|+1}} = \delta_{m_1 m_{|G|+1}} \delta_{\sf{a}_1 \sf{a}_{|G|+1}} \Id \,, \label{alphapower}\\
&\beta_{g,m_1m_2,\sf{a}_1\sf{a}_2} \beta_{g,m_2m_3,\sf{a}_2\sf{a}_3} \ldots \beta_{g,m_{|G|^2}m_{|G|^2+1},\sf{a}_{|G|^2}\sf{a}_{|G|^2+1}}   =   \left(\beta_{g}^{|G|^2}\right)_{m_1 m_{|G|^2+1},\sf{a}_1\sf{a}_{|G|^2+1}} = \delta_{m_1 m_{|G|^2+1}} \delta_{\sf{a}_1 \sf{a}_{|G|^2+1}} \Id \,. \label{betapower}
\end{align}
\ewt
The tensor of operators $\beta^{|G|}$ is not proportional to the identity in general, due to the non trivial commutation relations between $\mathcal{L}(r)$ and $U^\dag(r)$. 

The definitions of the $\alpha$ and $\beta$ modes allow us to express the Hamiltonian $H$ as a local Hamiltonian of the dyonic operators. In particular, the following relation hold for any $h \in G$:
\begin{multline} \label{UUdag}
 \Tr_A \left[\alpha_{h,mn,\sf{ab}}^\dag(2r+1)C_{no}\beta_{h,op,\sf{bc}}(2r)\right]\\
=U_{mn}(r+1)C_{no}U_{op}^\dag(r) \dim(A)\,.
\end{multline}
Here we are tracing only over the indices of the auxiliary representation $A$ characterizing the disorder operators and the effect of this trace is indeed to cancel out the operators $\mathcal{L}$ based on Eq. \eqref{ltr2}. The product with the $C$ matrix instead affects the indices of the $K$ representation. The mapping from the dyonic to the $\theta$ operators instead is based on the following relation:
\begin{multline} 
 \beta_{g,lm,\sf{ab}}^\dag(2r)\alpha_{g,mn,\sf{bc}}(2r-1) 
= U^K_{lm}(r)\Theta^{A}_{g,\sf{ac}}(r)U_{mn}^{K \dag}(r) \\
= U^{A\dag}_{\sf ab}(r) \theta_g(r)  U^{A}_{\sf bc}(r) D^K_{ln}(g) \\
=\theta_g(r) U^{A\dag}_{\sf ab}(r) D^{A \dag}_{\sf bb'}(g)  U^{A}_{\sf b'c}(r) D^{K}_{ln}(g) \,, \label{betaalpha}
\end{multline} 
where we applied \eqref{Utr1}. 
By taking the trace over $A$, we get
\begin{equation}\label{betaalpha2}
\Tr_A \left[\beta_{g,lm}^\dag(2r)\alpha_{g,mn}(2r-1) \right]= \theta_g(r) \chi^{A}(g^{-1}) D^{K}_{ln}(g)\,.
\end{equation}

Therefore, by taking $A=K=F$, we can re-express the Hamiltonian \eqref{ham_flux} as:
\bwt
\begin{equation} \label{ham}
  H= -\frac{J}{\dim(F)}\left(\sum_r \Tr_K\Tr_A\left[ \alpha^{\dagger}_h(2r+1)C\beta_h(2r)\right]+\mathrm{H.c.}\right) 
	- \frac{\mu}{{\rm dim}(F)}\sum_r\sum_{ g\neq e \in G} \mathrm{Tr}_K \Tr_A\left[\beta^{\dagger}_g(2r)\alpha_g(2r-1)D^{K\dag}(g)\right],
\end{equation}
\ewt
where, in the first term, we can choose any $h \in G$ and, in the second, the dimension of $F$ appears because we have chosen to adopt a trace to sum over the matrix indices of the representation $K=F$ in \eqref{betaalpha2}. Both $H_J$ and $H_\mu$ are the sum of local commuting operators in terms of the dyonic modes $\alpha$ and $\beta$. See Fig. \ref{fig:ladder} for a graphical representation of the Hamiltonian.

We observe that Eq. \eqref{betaalpha} implies that the operator $\Theta_g$ is a local operator in the dyonic modes. The operators $\theta_g$, instead, can be obtained as a linear function of $\beta_g^\dag(2r)\alpha_g(2r-1)$ only if $\chi^A(g^{-1}) \neq 0$, as evident from Eq. \eqref{betaalpha2}. Therefore, for a generic choice of the group $G$ and the auxiliary irreducible representation $A$, it is possible that some of the operators $\theta_g$ cannot be defined as local functions of the dyonic modes. We will discuss in detail the role of the auxiliary representation $A$ in Sec. \ref{sec:aux}.

\subsection{Topological order} \label{sec:topo2}

The nonlocal mapping \eqref{JW1} and \eqref{JW2} transforms the quasidegenerate ground states in the spontaneously symmetry-broken phase of the flux ladder Hamiltonian \eqref{ham_flux} into topologically protected ground states of the dyonic Hamiltonian \eqref{ham}. To clarify this point it is useful to introduce a formal definition of topological order for the dyonic system, which is able to generalize the notion of topological order of the Kitaev and parafermionic chains. 
We consider a gapped one-dimensional system defined on an open chain of length $L$, with a set of orthogonal quasidegenerate ground states $\left\{\ket{\psi_q}\right\}$  whose energy splitting decays superpolynomially in the system size. We define the system topologically ordered if it fulfills the following conditions.
\begin{enumerate}
\renewcommand{\theenumi}{$\mathcal{T}$\arabic{enumi}}
\renewcommand{\labelenumi}{ \theenumi :}
\item \label{condt1} For any bounded and local operator $V(r)$, and for any pair of ground states $\ket{\psi_{q_1}},\ket{\psi_{q_2}}$:
\begin{equation}\label{condt1Vexp}
\bracket{\psi_{q_1}}{V(r)\psi_{q_2}} = \bar{V}\delta_{q_1,q_2} + c(r,q_1,q_2),
\end{equation}
where $r$ specifies the position of the support of $V$, the constant $\bar{V}$ does not depend on the ground states, and $c(r,q_1,q_2)$ is a function, which decays superpolynomially with the distance of $r$ from the boundary of the system (thus with the minimum between $r$ and $L-r$).

This condition imposes that no local operator in the bulk of the system can cause transitions between the ground states, up to corrections $c$ that are strongly suppressed with the distance with the boundary. A typical example may be given by considering the Kitaev chain in the topological phase and the annihilation operator of a fermion in the system: if such operator is applied close to the boundary, with a considerable overlap with the zero-energy Majorana modes, then it can cause a transition between the two ground states; if instead it is applied in the bulk, with a negligible overlap with the exponentially localized zero-energy modes, then this transition is exponentially suppressed with the distance with the edges.

\item \label{condt2} Any local observable cannot distinguish the ground states. To formalize this local indistinguishability requirement, we must carefully define what is the set of operators that constitute legitimate observables in the presence of a non-Abelian symmetry. In the case of fermionic systems, the observables are Hermitian operators that commute with the fermionic number; thus they have vanishing matrix elements between states with different fermionic parities. This property is maintained in the parafermionic $\mathbb{Z}_N$ generalization, where the set of observables is restricted to the set of operators commuting with the conserved $\mathbb{Z}_N$ charge and, in general, with the symmetry transformations \cite{bernevig2016}. In the case of a non-Abelian symmetry, the requirement of commuting with the whole symmetry group is very strong, because the group transformations themselves do not fulfill it. Therefore it is useful to weaken this requirement to the purpose of defining a broader set of observables. Instead of considering a set of operators which commute with the conserved charges, we demand that the observables do not allow for transitions between states transforming under different irreducible representations. For our purposes, the irreducible representations play indeed the role of the conserved charges. In particular, we define two distinct sets of operators we label with $\mathcal{C}$ and $\tilde{\mathcal{C}}$.

The set $\mathcal{C}$ includes the rank-2 tensor operators $O_L$ that are block diagonal in the irreducible representation basis and transform under the group symmetry by conjugation, such that
\begin{equation} \label{conj}
\mathcal{Q}_h O_L \mathcal{Q}_h^\dag = \bigoplus_I D^{I}(h) O_L^I D^{I\dag}(h)\,, 
\end{equation}
and 
\begin{equation} \label{conj2}
\tilde{\mathcal{Q}}_h O_L \tilde{\mathcal{Q}}_h^\dag = O_L\,, 
\end{equation}
for a suitable decomposition $O_L=\sum_I O_L^I$ into components $O_L^I = \Pi^I O_L \Pi^I$ where $I$ labels the irreducible representations and the projectors $\Pi^I$ are defined in \eqref{Kproj}.  As a particular case we observe that the elements $\mathcal{Q}_g$ of the symmetry group belong to $\mathcal{C}$ since they fulfill the transformation relations \eqref{conj} and \eqref{conj2}.

The set $\tilde{\mathcal{C}}$ is the right counterpart of $\mathcal{C}$ and it includes the operators transforming as $\tilde{\mathcal{Q}}_g$. Namely,  $\tilde{\mathcal{C}}$ is the set of the rank-2 tensor operators $O_R$ transforming by conjugation as
\begin{equation} \label{conjright}
\tilde{\mathcal{Q}}_h O_R \tilde{\mathcal{Q}}_h^\dag = \bigoplus_I D^{I \dag }(h) O_R^I D^{I}(h)\,, 
\end{equation}
and
\begin{equation} \label{conj2right}
{\mathcal{Q}}_h O_R {\mathcal{Q}}_h^\dag = O_R\,.
\end{equation}

We observe that, for both sets, these operators reduce to the set of observables invariant under the symmetry group in the Abelian case. The non-Abelian structure of the symmetry group provides in this case an additional richness to the system since it is not possible to define a single conserved charge in the $G$-invariant models.

Finally, we can define the following condition for the local indistinguishability of the ground states in systems with a non-Abelian symmetry group: for any local observable $O(r)$, belonging to either $\mathcal{C}$ or $\tilde{\mathcal{C}}$, and any pair of ground states, the following equation must be satisfied:
\begin{equation} \label{indist}
\bracket{\psi_{q_1}}{O(r)\psi_{q_2}} = \bar{O} \delta_{q_1,q_2} + o(L,q_1,q_2),
\end{equation}
where the parameter $\bar{O}$ does not depend on the ground states, and the function $o(L,q_1,q_2)$ decays superpolinomially in the system size $L$. 

This condition properly generalizes the requirement of the local indistinguishability of the ground states under symmetric observables for the Abelian symmetric systems \cite{bernevig2016} to the non-Abelian case.

\end{enumerate}

Both the conditions \ref{condt1} and \ref{condt2} are related to the notion of locality and, for the dyonic model, we will consider an operator local if it can be defined as a function of the $\alpha$ and $\beta$ modes in a small (nonextensive) domain.

In the dyonic model, analogously to the flux ladder model with $J\gg \mu$, we can label the quasidegenerate ground states as $\ket{\ket{Imn}}$ based on their transformations \eqref{gsrepr} under the global symmetry group. This is indeed a property that does not depend on the definition of locality and it is not affected by the nonlocal nature of the JW transformation.
In this basis, the matrix $\bracket{\psi_{q_1}}{\tilde{V}\psi_{q_2}}$ in Eq. \eqref{condt1Vexp} is diagonal for any operator $W$ which preserves the symmetry under $G$, such that $[W,{\mathcal{Q}}_g]=[W,\tilde{\mathcal{Q}}_g]=0$ for any $g\in G$:
\begin{equation}
\bra{\bra{ Imn}}  W \ket{\ket{Rpq}}  \propto W_I \delta_{RI} \delta_{mp} \delta_{nq} \,.
\end{equation}
This is analogous to the effect of operators preserving the fermionic parity in the Kitaev chain and operators preserving the $\mathbb{Z}_N$ symmetry in the parafermionic chains \cite{bernevig2016}. For the same reason, any observable $O$ that is invariant under the action of the symmetry group, presents all the off-diagonal terms in \eqref{indist} equal to zero if we choose the ground-state basis $\left\{\ket{\ket{ Imn}}\right\}$.
For any observable $O$ in the set $\mathcal{C}$ (or in its right counterpart $\tilde{\mathcal{C}}$), instead, the matrix $\left\langle  \left\langle Imn|| O || I'm'n' \right\rangle\right\rangle$ in Eq. \eqref{indist} has vanishing entries for $I \neq I'$ but the elements of $\mathcal{C}$ and $\tilde{\mathcal{C}}$ enable transitions between $m$ and $m'$, and between $n$ and $n'$, respectively. We conclude that, under this point of view, the condition \ref{condt2} can be considered a stronger condition than its Abelian counterpart \cite{bernevig2016}. 

Both the conditions \ref{condt1} and \ref{condt2} are intimately related to the existence of a set of
topologically protected zero-energy modes, localized on
the boundaries (or, more accurately, on the interface between gapped topological and nontopological regions),
which transform nontrivially under the symmetry group
$G$. The transitions between ground states driven by all
the local operators $V$ must be understood in terms of the
overlap with these zero-energy modes, and the local indistinguishability of the ground states is justified by the fact that these states differ only by the application of these boundary modes.

In the next section, we will discuss the properties of these boundary modes and we will show that the dyonic model fulfills the previous criteria for topological order.

\section{The topological zero-energy modes} \label{sec:topo}

\subsection{Weak zero-energy modes} \label{sec:weak}

The condition \ref{condt1} for the system to be topologically ordered is the most immediately related to the existence of zero-energy modes localized on the boundary of the system. In general, it is necessary to distinguish two kinds of topologically protected zero modes and, consequently, two kinds of one-dimensional topological order \cite{jermyn2014,bernevig2016}. A system enjoys weak topological order, and it possesses weak zero-energy modes, if the ground-state manifold is $|G|$-degenerate up to an energy splitting which is exponentially suppressed in the system size, whereas we speak of strong topological order when the whole energy spectrum is $|G|$-degenerate up to exponentially small corrections in the system size.

Therefore the weak topological order is a property only of the ground states. The excited states may present no specific regularity in their energy. In the $\mathbb{Z}_3$ parafermionic model in proximity of the nonchiral point in parameter space, for example, it is known that excited states labeled by different eigenvalues of the symmetries have relevant energy differences which decay only algebraically with the system size \cite{jermyn2014}. The strong topological order is instead a property of the whole spectrum.

The strong or weak kind of topological order are related to the presence of a strong or weak kind of localized zero-energy modes. Both these kind of modes must fulfill the following properties.
\begin{enumerate}
\item To cause transitions between the quasidegenerate ground states, these modes must transform nontrivially under the global symmetries of the Hamiltonian. We denote these modes with $\Gamma$; in the simplest case, they can be associated to a (nontrivial) irreducible representation $K$ of the symmetry group $G$ in such a way that
\begin{equation} \label{QGamma}
\mathcal{Q}_h \Gamma^K \mathcal{Q}^\dag_h =  \Gamma^K D^K(h)\,,
\end{equation}
or more general nontrivial transformation relations.
In the $\mathbb{Z}_N$ Abelian case this requirement reduces to the condition $\mathcal{Q}_1 \Gamma^K = e^{\frac{i 2\pi K}{N}} \Gamma^K \mathcal{Q}_1$, where $K$, for an Abelian group, can be interpreted simply as a power, $\mathcal{Q}_1$ is the $\mathbb{Z}_N$ charge of the system and $D^K = e^{\frac{i 2\pi K}{N}}$ \cite{fendley2012}.
\item The zero-energy modes must be bounded operators, localized on the edge of the system (or at an interface between different gapped phases).
\end{enumerate}
Besides these common requirements, weak and strong zero-energy modes must respectively satisfy the following conditions.
\begin{enumerate}
\item Weak topological modes $\Gamma_W$ must satisfy
\begin{equation} \label{weakcomm}
\left[\Gamma_W , P_0 H P_0\right] \le \gamma e^{-L/\xi}\,,
\end{equation}
where $P_0$ is the projector operator over the ground-state manifold, $\gamma$ is a generic (bounded) operator acting on the ground-state manifold, $L$ is the system size and $\xi$ is a suitable length scale. This requirement imposes that the weak zero modes quasicommute with the Hamiltonian projected on the ground-state manifold. Therefore, when we consider the subspace of the ground states, the projected Hamiltonian commutes with the symmetries $\mathcal{Q}$ and quasicommute with the mode $\Gamma_W$, but $\Gamma_W$ and $\mathcal{Q}$ do not commute with each-other due to the condition \eqref{QGamma}. This implies the quasidegeneracy of the ground-state manifold.
\item Strong topological modes $\Gamma_S$ must satisfy the stronger requirement 
\begin{equation} \label{gammastr}
\left[\Gamma_S , H \right] \le \gamma e^{-L/\xi}\,.
\end{equation}
This requirement, together with \eqref{QGamma}, implies the $|G|$-degeneracy of the whole spectrum up to exponentially suppressed corrections.
\end{enumerate} 

Let us discuss how the notion of topological order and weak zero-energy modes apply to the dyonic system.
The topological order of the model can be easily verified for the Hamiltonian $H_J$: the Hamiltonian $H_J$ is a sum of commuting terms and its $|G|$ ground states $\ket{\ket{Imn}}$ are determined by imposing that
\begin{multline} \label{bulkt}
\Tr_A \left[\alpha_{h,m_2m_3}^\dag(2r+1)\beta_{h,m_1m_2}(2r) \right] \ket{\ket{Imn}} \\
= \delta_{m_1m_3} \dim(A)\ket{\ket{Imn}}\,,
\end{multline}
for every $r$ and $h\neq e$. This implies that the bulk properties of all the ground states are the same. Like in the parafermionic case, the operators $\alpha(1)$ and $\beta_g(2L)$ do not appear in $H_J$ and commute with it: this can be derived by the definitions in \eqref{JW1} and \eqref{JW2}. Therefore $\alpha(1)$ and $\beta_g(2L)$ constitute localized zero-energy modes. Specifically for the case of $H_J$, they satisfy the requirements of strong topological modes, but, analogously to the $\mathbb{Z}_N$ case, their strong behavior is not stable against the addition of a small term $H_\mu$ in the Hamiltonian, and in general they must be considered weak zero modes.
 
Let us first analyze what happens for the unperturbed Hamiltonian $H_J$. The bulk operators by definition are independent of $\alpha(1)$ and $\beta_g(2L)$, and a generic bulk operator therefore is either composed only by terms independent on the operators $\Theta^A_g$, like the ones in Eq. \eqref{bulkt}, or includes terms which are functions of some of the operators $\Theta^A_g$. In the first case, the operator is proportional to the identity when projected on the ground-state manifold; in the second, instead, the operators $\Theta^A_g$ introduce domain walls in the corresponding flux-ladder model, thus completely driving any ground state into excited states. We conclude in both cases that bulk operators do not violate the condition \ref{condt1} for topological order. 

The ground states cannot be distinguished by observables that do not involve either $\alpha(1)$ or the operators $\beta_g(2L)$. 
Taken singularly, $\alpha(1)$ and $\beta_g(2L)$ do not allow us to build nontrivial observables that belong to the set $\mathcal{C}$  (see Eqs. \eqref{conj} and \eqref{conj2}) or to its right counterpart $\tilde{\mathcal{C}}$. Therefore, operators which are a function of $\alpha(1)$ or $\beta_g(2L)$ only, do not violate condition \ref{condt2}.

Hence, the only possible way to build observables in $\mathcal{C}$ or $\tilde{\mathcal{C}}$ that distinguish the ground states is to multiply either $\alpha(1)$ or $\beta_g(2L)$ with suitable bulk dyonic modes. These additional modes, however, necessarily introduce domain walls in the model, as it can be seen from the action of their JW strings in Eqs. \eqref{JW1} and \eqref{JW2} on the ground states of $H_J$. Therefore, under the action of these operators, the ground states are fully transformed in excited states and the expectation values of the kind \eqref{indist} vanish.

The only observables which can distinguish the ground states and belong to $\mathcal{C}$ are the ones build by products of the form $\alpha(1) \beta^\dag(2L)$. In particular, for $\mu=0$, it is convenient to define the operators
\begin{equation} \label{Upsilon0}
\Upsilon_g =  \Tr_K \Tr_A \left[\alpha(1) \beta^\dag_g(2L)\right] = \chi^A(g^L)\mathcal{Q}_g,
\end{equation}
where the last equality can be derived from Eq. \eqref{Ldef}. $\Upsilon_g$ transforms as $\mathcal{Q}_h \Upsilon_g \mathcal{Q}_h^\dag = \Upsilon_{hgh^{-1}}$ and it belongs to $\mathcal{C}$. From these operators it is possible to build observables that generalize the conserved $\mathbb{Z}_N$ charge in the Abelian systems and allow us to distinguish the ground states. All these observables, though, are crucially nonlocal. We conclude therefore that also the condition \ref{condt2} is fulfilled by $H_J$. Hence $H_J$ fulfills the criteria to be topologically ordered.

We additionally remark that in the flux-ladder model the symmetry breaking order parameter is provided by the operators $U(r)$. Such operators are nonlocal in the dyonic model if and only if the auxiliary irreducible representation $A$ is non-Abelian.
 In the following, we restrict to this condition, which is necessary to fulfill the criteria \ref{condt1} and \ref{condt2}, thus to have topological order. We will discuss the nontopological system defined by $A$ being the trivial representation in Sec. \ref{sec:aux}.

The existence of weak zero-energy modes for the full Hamiltonian $H$ for $\mu \ll J$ can be inferred by a quasiadiabatic continuation \cite{hastings2005} by following the same procedure presented in \cite{bernevig2016} for the $\mathbb{Z}_N$ symmetric models. In particular, in the presence of a gap $\Delta(\mu)$ separating the ground-state manifold from the excited states, it is possible to define a quasiadiabatic continuation $\mathcal{V}(\mu)$, which is a unitary mapping preserving locality and symmetry under the group $G$ that maps the ground states of $H_J$ into the ground states of $H$: $\ket{\ket{Imn}}_\mu= V(\mu)\ket{\ket{Imn}}_{\mu=0}$. Therefore the continuation $\mathcal{V}(\mu)$ allows us to map the projector $P(0)$ over the ground states of $H_J$ into the projector $P(\mu)=\mathcal{V}(\mu) P(0) \mathcal{V}^\dag(\mu)$ over the ground-state manifold at finite $\mu$. Through the continuation $\mathcal{V}(\mu)$ it is possible to define the new weak zero-energy modes $\mathcal{V}(\mu) \alpha(1) \mathcal{V}^\dag(\mu)$ and $\mathcal{V}(\mu) \beta_g(2L) \mathcal{V}^\dag(\mu)$ and verify that the conditions for topological order hold also for $H$ as long as the energy gap $\Delta(\mu)$ does not close. The arguments presented in \cite{bernevig2016} extend straightforwardly to the non-Abelian case and show the persistence of topological order for the dyonic mode at finite $\mu$.

By following the approach in \cite{bernevig2016}, we obtain the following first-order expression in $\mu/J$ for the left weak zero-energy modes in the case $C=\Id$:
\begin{align} 
\mathcal{V}(\mu) \alpha(1) \mathcal{V}^\dag(\mu) &= \alpha(1)  + \mu \sum_{h \neq e} \frac{\Tr_{K,A}\left[\beta^\dag_h(2)\alpha_h(1)D^{K\dag}(h)\right]}{m_h-m_e} \nonumber\\
 &\times \alpha(1) \left(\Id-D^{K\dag}(h)\right)  + O\left(\frac{\mu^2}{J^2}\right)\,,\label{leftweak}
\end{align}
and an analogous expression holds for the right edge modes (see Appendix \ref{app:adiab} for more detail).
These weak zero modes depend on the ratio of $\mu$ and the energy gaps $m_h-m_e$ between the ground states and the first excited states at $\mu=0$. For $\mu \ll \min\left[m_h-m_e\right]$, this result suggests that the weak zero modes survive and maintain their localization when introducing the $H_\mu$ perturbation, in analogy with the Abelian models \cite{bernevig2016}. This is consistent with the perturbative result in Eq. \eqref{perturbativesplitting}. 

We notice that the left weak zero-energy mode, originating from $\alpha(1)$, does not carry a group index, differently from the right modes, which originate from $\beta_g(2L)$. This apparent discrepancy is due to the open boundary conditions we are using in the analysis of our system. However, we can generalize our investigation by embedding the topological phase in a larger nontopological system: in this case, also the weak left zero-energy modes would acquire a nontrivial JW string, thus acquiring a full dyonic character like the right modes. In Appendix \ref{app:left} we present the first-order calculation of the left zero-energy mode at the interface between a nontopological and a topological region and we verify that the introduction of this different kind of boundary does not spoil the localization of the mode.

\subsection{Strong zero-energy modes} \label{sec:strong}

So far, we considered only the existence of weak zero-energy modes. In the following we will investigate under which conditions it is possible to define strong zero-energy modes. In particular, inspired by the approach in \cite{fendley2012}, we will present a constructive iterative technique for $\mu \ll J$ to build strong zero modes. Such approach will in general result in unbounded operators that, consequently, do not satisfy the criteria for the definition of topological modes. We will show however that by modifying the Hamiltonian \eqref{ham} and introducing additional constraints, it is possible to find strong topological modes on the edges of the system. 

Our goal is to derive zero modes of the form
\begin{equation}
\Gamma^{(r)} = \Gamma_0 + \Gamma_1 + \ldots + \Gamma_r
\end{equation}
such that
\begin{enumerate}
\item $\Gamma_x$ has support on the first $2x+1$ $\alpha$ and $\beta$ dyonic modes starting from the edge. For the zero modes localized on the left edge, this implies that $\Gamma_x$ is a function of $\alpha(1), \beta(2), \ldots \alpha(2x+1)$. In the right case instead we search for a function of $\beta(2L), \alpha(2L-1), \ldots \beta(2L-2x)$.
\item The mode $\Gamma^{(r)}$ must asymptotically fulfill
\begin{equation} \label{bound}
\left[\Gamma^{(r)},H\right] < \mu \rho^r\,,
\end{equation}
where $\rho<1$ is a suitable parameter obtained in general as a function of $\mu$, $J$, and $C$. In this way the requirement \eqref{gammastr} is satisfied for $r \to L$.
\item The zero modes $\Gamma_{g,mn,\sf{ab}}^{(r)}$ may be characterized by a group element $g$, and,  analogously to $\alpha$ and $\beta$ operators, they are tensors of operators defined by four matrix indices which in general obey dyonic transformation rules with respect to the $K$ irreducible representation:
\begin{equation} \label{Qgamma2}
\phantom{aaaa} \mathcal{Q}_h \Gamma_{g,mn,\sf{ab}}^{(r)} \mathcal{Q}_h^\dag =  \Gamma_{hgh^{-1},mm',\sf{ab}}^{(r)} D^K_{m'n}(h)\,.
\end{equation}
The indices $\sf{ab}$ of the auxiliary representation $\mathcal{A}$ are invariant under transformations of the symmetry group and, in the following, we will omit them.

The requirement \eqref{Qgamma2}, analogously to the condition \eqref{QGamma}, implies for $r\to L$ the quasidegeneracy of the whole energy spectrum. Furthermore, starting from the symmetry invariant ground state $\ket{\ket{000}}$ we obtain
\begin{equation}
\Gamma_{g,mn}^{(L)} \ket{\ket{000}} \in {\rm Span}\left\{\ket{\ket{Kpq}},\; p,q = 1,\ldots, \dim{K}\right\}\,.
\end{equation}
This implies that the zero modes allow for transitions between ground states $\ket{\ket{Rpq}}$ with different irreducible representations $R$.
By applying the zero modes multiple times, the resulting ground states are defined by the Clebsch-Gordan series of the group $G$ \cite{brink} and we will show that it is possible to span the whole ground state manifold, thus extending the behavior of zero-energy Majorana and parafermionic modes to the non-Abelian case.
\end{enumerate}

In the following we will use $\Lambda^{(r)}$ to label the strong zero-energy modes localized on the left boundary of the system, and $\Omega^{(r)}_g$ to label the ones on the right boundary. Analogously to their weak counterpart, only the strong right modes carry a group index. This is again due to the chosen boundary conditions (see Appendix \ref{app:left} for more detail).

The first step of the iterative procedure is to impose the first term to be the zero-energy mode of $H_J$. Therefore we have $\Lambda_0=\alpha(1)$ and $\Omega_{g,0}=\beta_g(2L)$ for the left and right boundary respectively. In this way $[\Lambda_0,H_J]=[\Omega_{g,0},H_J]=0$.

Let us consider the right boundary as example. Following \cite{fendley2012}, we define the commutator
\begin{equation}
C_1(g) \equiv \left[\Omega_{g,0},H\right] = \left[\Omega_{g,0},H_\mu\right]\,.
\end{equation}
$C_1$ is of order $\mu$ and it transforms under the symmetry group as $\Omega_{g,0}= \beta_g(2L)$, from which it inherits the dyonic character:
\begin{multline}
\mathcal{Q}_h C_1(g) \mathcal{Q}_h^\dag = \left[\mathcal{Q}_h\Omega_{g,0}\mathcal{Q}_h^\dag,\mathcal{Q}_hH_\mu\mathcal{Q}_h^\dag\right] \\
 = \left[\Omega_{0,hgh^{-1}}D(h),H_\mu\right] = C_1(hgh^{-1}) D(h)\,.
\end{multline}
The next step is finding an operator $\Omega_{1,g}$ obeying the above conditions such that
\begin{equation}
\left[\Omega_{1,g},H_J\right]= -C_1(g)\,.
\end{equation}
In this way we get
\begin{equation}
\left[\Omega_{0,g} + \Omega_{1,g},H\right]= C_1-C_1 + \left[\Omega_{1,g},H_\mu\right] \equiv C_2.
\end{equation}
In general, $\Omega_{1,g}$ is of order $\mu/J$ and, due to the Hamiltonian being symmetric, it is always possible to define it in such a way that it obeys the same transformation rules of $\Omega_{0,g}$. In general, at each iteration step we evaluate the commutator $C_r(g) = \left[\Omega^{(r-1)}_g,H\right]$ and we construct the corresponding operator $\Omega_{r,g}$ such that
\begin{equation}
\left[\Omega_{r,g},H_J\right] = -C_r = -\left[\Omega_{r-1,g},H_\mu\right]\,.
\end{equation}
The resulting operators $\Omega_{r,g}$ are suppressed by a factor of order $(\mu/J)^r$.

This procedure guarantees the fulfillment of the constraints \eqref{bound} and \eqref{Qgamma2} and, as we will show in the following, of the localization constraint. In the following sections we will express all the zero-energy modes in terms of the operators $\theta_g$ and $U$ to exploit their commutation relations. It is important to stress, however, that the resulting modes $\Lambda$ and $\Omega$ are localized based on the notion of locality obtained by the dyonic operators $\alpha$ and $\beta$.

\subsection{Iterative procedure for strong modes on the left boundary} \label{sec:strong2}
The starting point for the left strong zero mode is $\Lambda_0 =\alpha(1) = U^\dagger(1)$ and we have:
\begin{align}
C_1 &= [\Lambda_0, H_\mu] = \mu \sum_{h_1\neq e} \chi^A(h_1^{-1}) [U^\dagger(1),\theta_{h_1}(1)] \nonumber\\
&= - \mu \sum_{h_1\neq e} \chi^A(h_1^{-1}) \theta_{h_1}(1) U^\dagger(1) ( D^\dagger(h)-\id).
\end{align}
We must identify an operator $\Lambda_1$ with support on $\alpha(1),\beta(2)$ and $\alpha(3)$, such that its commutator with $H_J$ cancels $C_1$. 
We observe that $H_J$ commutes with any function of the operators $U$, therefore we may assume that $\Lambda_1$ inherits a factor $U^\dagger(1) ( D^\dagger(h)-\id)$ from $C_1$. Hence we adopt the following ansatz for $\Lambda_1$:
\begin{equation} \label{lambda1}
\Lambda_1 = \frac{\mu}{J} \sum_{h_1\neq e} F_{1}(h_1) \chi^A(h_1^{-1}) \theta_{h_1}(1) U^\dagger(1) (D^\dagger(h_1) - \id),
\end{equation}
where $F_{1}$ is a function only of the operators $U(1)$, $U(2)$ and the matrices $D(h_1)$, in such a way that $\left[F_1,H_J\right]=0$. 
The commutator $[\Lambda_1,H_J]$ gives
\bwt
\begin{align}
[\Lambda_1,H_J]&=\mu \sum_{h_1\neq e}F_{1}(h_1) \chi^A(h_1^{-1})  \left[\theta_{h_1}(1),\Tr[U(2)CU^\dagger(1)] + \hc\right] U^\dagger(1) (D^\dagger(h_1) - \id) \nonumber\\
&= \mu \sum_{h_1\neq e} F_{1}(h_1) \chi^A(h_1^{-1})  \left(\Tr[U(2)CU^\dagger(1)(D(h_1)-\id)] + \hc\right) \theta_{h_1}(1) U^\dagger(1) (D^\dagger(h_1) - \id),
\end{align}
\ewt
which is equal to the desired value $-C_1$ when we take
\begin{equation}\label{eqnF1}
F_{1}(h_1) = \left(\Tr[U(2)CU^\dagger(1)(D(h_1)-\id)] + \hc \right)^{-1}.
\end{equation}

In the group element basis, the operator $F_1$ always corresponds to the inverse of the difference of two different flux masses \eqref{mass}, since $h_1 \neq e$. Therefore, in order to obtain a bounded operator $\Lambda_1$, it is necessary to choose a matrix $C$ such that all the flux masses in the model are different (condition \eqref{condc}). Hence, similarly to the Abelian case \cite{fendley2012}, it is necessary to break the chiral symmetry in order to have strong zero-energy modes. 

In the second iterative step, the commutator $C_2$ gives
\begin{equation}
C_2 = -\mu\sum_{h_2\neq e}\chi^A(h_2^{-1})[\Lambda_1,\theta_{h_2}(1) + \theta_{h_2}(2)].
\end{equation} 
It is convenient to split this commutator into two pieces, $C_2= C_{{\rm in},2} +C_{{\rm out},2}$, representing the contributions given by the term in $\theta_{h_2}(1)$ and $\theta_{h_2}(2)$ respectively. These two terms of $C_2$ are defined on different supports: $C_{{\rm out},2}$ includes all the dyonic modes up to $\beta(4)$ whereas $C_{{\rm in},2}$ has support only up to $\alpha(3)$.  
 Based on this difference, we can distinguish two contributions also for the operator $\Lambda_2 = \Lambda_{{\rm in},2} + \Lambda_{{\rm out},2}$, such that $\left[\Lambda_{{\rm in/out},2},H_J\right]=-C_{{\rm in/out},2}$.  The operator $\Lambda_{{\rm in},2}$ defines the inner part of $\Lambda_2$, with support up to $\alpha(3)$, thus with the same support of $\Lambda_1$; $\Lambda_{{\rm out},2}$, instead, is the outer part and it includes all the terms of $\Lambda^{(2)}$ that extend its support to $\alpha(5)$.

This distinction between inner and outer contributions can be extended to all the iteration levels and, in general, we have
\begin{align}
C_{{\rm out},n} &= -\mu\sum_{h_n\neq e}\chi^A(h_n^{-1})\left[\Lambda_{n-1}, \theta_{h_n}(n)\right]\,,  \label{Coutn} \\
C_{{\rm in},n} &= -\mu\sum_{i<n} \sum_{h_i\neq e} \chi^A(h_i^{-1})\left[\Lambda_{n-1}, \theta_{h_i}(i)\right]\,.
\end{align}
Correspondingly, we define $\Lambda_n = \Lambda_{{\rm in},n} + \Lambda_{{\rm out},n}$ such that
\begin{equation}
\left[\Lambda_{{\rm in/out},n},H_J\right]=-C_{{\rm in/out},n}\,.
\end{equation}
The operator $\Lambda_{{\rm out},n}$ includes all the \textit{outer terms} with domain extending from $\alpha(1)$ to $\alpha(2n+1)$, whereas $\Lambda_{{\rm in},n}$ includes the \textit{inner terms} with the same domain of $\Lambda_{n-1}$. At the $n^{\rm th}$ level of iteration both $\Lambda_{{\rm out},n}$ and $\Lambda_{{\rm in},n}$ appear to be of order $\left(\mu/J\right)^n$, therefore only the outer modes define the spatial penetration of the zero-energy modes in the bulk.

Let us focus first on the calculation of the outer modes: in the second iteration step, $\Lambda_{{\rm out},2}$ is determined from the commutator $C_{{\rm out},2}$ in Eq. \eqref{Coutn}. The only part of $\Lambda_1$ that doesn't commute with $\theta_{h_2}(2)$ is $F_{1}$ (see Eq. \eqref{lambda1}), and we denote $[F_{1},\theta_{h_2}(2)] = \tilde{F}_{1} \theta_{h_2}(2)$. Concretely,
\bwt
\begin{align}
\tilde{F}_{1}(h_1,h_2) &= \left( \Tr[U(2)CU^\dagger(1)(D(h_1)-\id) + \hc]\right)^{-1} - \left( \Tr[U(2)CU^\dagger(1)(D(h_1)-\id )D^\dagger(h_2) + \hc]\right)^{-1},
\end{align}
which implies 
\begin{equation}
C_{{\rm out},2} = -\frac{\mu^2}{J}\sum_{h_1,h_2\neq e}\chi^A(h_1^{-1}) \chi^A(h_2^{-1})\tilde{F}_{1}(h_1,h_2) \theta_{h_1}(1)\theta_{h_2}(2) U^\dagger(1) (D^\dagger(h_1)-\id)\,.
\end{equation}
Similarly to the first step, we assume that the outer mode $\Lambda_{{\rm out},2}$ takes the form
\begin{equation}\label{Gamma2Out}
\Lambda_{{\rm out},2} = \left(\frac{\mu}{J}\right)^2\sum_{h_1,h_2\neq e}\chi^A(h_1^{-1}) \chi^A(h_2^{-1})\tilde{F}_{1}(h_1,h_2) F_{2}(h_1,h_2,h_3)\theta_{h_1}(1)\theta_{h_2}(2) U^\dagger(1) (D^\dagger(h_1)-\id)\,,
\end{equation}
where we introduced a new function of the $U$ operators $F_{2}(h_1,h_2,h_3)$. By taking
\begin{equation}\label{IterativeF2}
F_{2}=\left(\Tr[U(2) C U^\dagger(1)(D(h_1h_2^{-1})-\id) + \hc] + \Tr[U(3) C U^\dagger(2)(D(h_2)-\id)+\hc] \right)^{-1}\,,
\end{equation}
\ewt
we ensure that $[\Lambda_{{\rm out},2},H_J]=-C_{{\rm out},2}$. 

From this expression we deduce that the condition \eqref{condc} on $C$ is not strong enough to guarantee the existence of the strong zero-energy modes. This condition only ensures that each term in \eqref{IterativeF2} do not cancel individually, but they may still cross cancel. This happens when the action of $h_1$ and $h_2$ results in a swap of the gauge fluxes in the first two plaquettes of the ladder model.
For instance, $F_{2}\vert g_1,g_2,g_3,\ldots\rangle$ is singular when $h_2=g_2g_1^{-1}g_2g_3^{-1}$ and $h_1h_2^{-1}=g_1g_2^{-1} g_3 g_2^{-1}$. For a given group $G$, these two equations will be compatible with the requirement $h_1,h_2\neq e$ for some state, thus causing a divergence of the operators $F_2$ and $\Lambda_2$. To avoid this problem, we can introduce a suitable position dependence in either the parameters $J$ or $C$; we will discuss the problem of the possible divergences of the zero-energy modes in Sec. \ref{diverge}, based on the final result for $\Lambda_{{\rm out},n}$.

When calculating $C_{{\rm out},3}$ by computing $[\Lambda_2,\theta_{h_3}(3)]$, only $F_{2}$ is modified by the action of $\theta_{h_3}(3)$, and we define a new function $\tilde{F}_{2}$ analogously to the previous term. In general, all the outer modes follow the same pattern and, at the $n^{\rm th}$ iteration step, we can define
\bwt
\begin{equation}\label{GammaOutn}
\Lambda_{{\rm out},n} = \left(\frac{\mu}{J}\right)^n \sum_{h_1,\ldots,h_n \neq e}\chi^A(h_1^{-1})\ldots\chi^A(h_n^{-1}) \tilde{F}_{1}  \ldots  \tilde{F}_{n-1} F_{n} \theta_{h_1}(1)\ldots \theta_{h_n}(n) U^\dagger(1) (D^\dagger(h_1)-\id),
\end{equation}
where
\ewt
\begin{align}
&F_{n}(h_1, \ldots, h_n)  \nonumber\\
&\equiv \left(\frac{1}{J}[H_J,\theta_{h_1}(1)\ldots\theta_{h_n}(n)]\theta_{h_1}^\dag(1)\ldots \theta_{h_n}^\dag(n)\right)^{-1} \nonumber \\
 &=\left(\sum_{r=1}^n \Tr[U(r+1)CU^\dag(r)(D(h_rh_{r+1}^{-1})-\id) + \hc]\right)^{-1},
	\label{eqn::Fn}
\end{align}
with the constraint $h_{n+1}=e$. The function $\tilde{F}$ is defined in turn as
\begin{equation}
\tilde{F}_{n-1}(h_1, \ldots,h_{n}) = F_{n-1} - \theta_{h_{n}}(n) F_{n-1} \theta_{h_{n}}^\dag(n) \,.
\end{equation}

From the following expression, it is easy to verify that the operator $\Lambda_{{\rm out},n}$ is a function of the dyonic modes from $\alpha(1)$ to $\alpha(2n+1)$ based on the relations (\ref{UUdag},\ref{betaalpha2}), which map all the operators of the flux-ladder Hamiltonian into local combinations of the dyonic modes. A similar result is obtained for the inner modes (see Appendix \ref{app:inner}) which display similar terms with suitable modifications of the $F$ and $\tilde{F}$ functions.

\subsection{Divergences of the strong modes and space-dependent Hamiltonians} \label{diverge}

The previous expressions we derived for the strong zero-energy modes are ill-defined at all the iteration orders after the first. There are two kinds of divergences that affect the operators $F_n$ and $\tilde{F}_n$ entering in the definition of $\Lambda_{{\rm out},n}$. Let us analyze for simplicity the case of $F_n$ defined in Eq. \eqref{eqn::Fn}, since $\tilde{F}_n$ is given by the difference of two analogous operators, and the same conclusions hold for both. For ease of notation we adopt $J=1$ and $\mu \ll 1$ in the following analysis.

Given a state of the flux ladder $\ket{\psi}=\ket{h_1 \ldots h_n}$, the denominator of $F_n$ returns the difference of the $H_J$ eigenenergies of $\ket{\psi}$ and $\ket{\psi'}=\prod_{r=1}^{n-1} \theta_{h_r}^\dag(r) \ket{\psi}$. This denominator can become zero in two different cases: (i) $\psi$ and $\psi'$ are characterized by different sets of gauge fluxes $\{\Phi\}$ and $\{\Phi'\}$ but their energy is the same; (ii) $\psi$ and $\psi'$ are defined by two different permutations of the same gauge fluxes, thus $\{\Phi\}=\{\Phi'\}$.

The case (i) corresponds to resonances of the kind
\begin{equation} \label{resonance}
\sum_g n_g m_g = \sum_g n_g' m_g \,.
\end{equation} 
with $\{n_g\} \neq \{n_g'\}$. This kind of resonance corresponds to the same divergences met in the Abelian $\mathbb{Z}_3$ model analyzed in \cite{Moran2017} and, in general, it hinders the formation of strong modes for large system sizes, although their effects is usually relevant only at large energies. To avoid this kind of resonance, in principle, we could strengthen our requirement \ref{condIIa} on the $C$ matrix by imposing that the $C$ matrix must be such that all the flux masses $m_g$ are incommensurate with each other. In this case the condition \eqref{resonance} can never be fulfilled, although the difference between the energies of the two fluxes configurations can be arbitrary small for sufficiently long systems. In particular, we can estimate that the energy splitting becomes smaller than a quantity $\epsilon$ at order $O(1/\epsilon f(|G|))$ of the iteration process, where $f$ is a suitable function of the group order only. This kind of splitting implies that the norm of the strong mode contribution $\Lambda_n$ behaves like $\sim n f(|G|)\left[(|G|-1)\mu/J\right]^n $, thus displaying an exponential decay for large $n$. Therefore we conclude that, under the previous incommensurability assumption for the flux masses, strong zero-energy modes are, in general, not critically affected by this kind of resonance.

The case (ii) is characteristic of the non-Abelian groups only. For the Abelian models, the requirements $h_1 \neq e$ and $h_{n+1} = e$ in Eq. \eqref{eqn::Fn} would imply that the sets of fluxes defining $\ket{\psi}$ and $\ket{\psi'}$ cannot be the same. This does not hold for non-Abelian groups because, by changing the order of the fluxes in the ladder, it is possible to modify the total flux $\Phi_{\rm tot}= g_1^{-1}g_{n+1}$. Therefore there can be choices of $h_1, \ldots, h_{n}$ and of the state $\psi$ such that $\psi$ and $\psi'$ share exactly the same fluxes, $\{\Phi\}=\{\Phi'\}$. We emphasize, however, that the resonances of kind (ii) require that $\psi$ and $\psi'$ present at least two nontrivial fluxes. If we assume that $\psi$ and $\psi'$ are both states with a single nontrivial flux of the kind $\Phi(g)$, a divergence would entail that $\Phi_{\rm tot}=\Phi_{\rm tot}'=\Phi(g)$, but this is impossible since $\Phi_{\rm tot}$ and $\Phi_{\rm tot}'$ differ by an overall multiplication of the nontrivial group element $h_1$. 
We conclude that, similarly to the ground states, also the single-flux states are protected against this kind of divergence.

For multi-flux states, the resonances of the case (ii) are unavoidable in uniform systems. To obtain well-defined strong zero-energy modes is thus necessary to consider adding a position dependence to the Hamiltonian parameters. We decide, in particular, to focus on the case of a space dependent $J$ of the form $J_r = (1+\eta_r)$ with $|\eta_r| \ll \min\left[|m_{g} - m_{h}|\right]$ for $g,h \in G$. 
To show that strong zero-energy modes can indeed exist in such a situation, we consider the fine-tuned case $\eta_r = \eta_0/2^{r}$. In this situation the maximum value of $F_n$ is 
\begin{equation} \label{maxF}
\max\left[F_n\right] = \frac{2^n}{2\Delta\eta_0}\, ,
\end{equation} 
where we labeled the minimum of the absolute values of the differences between two flux masses with $\Delta$. This value is reached when all the group elements $h_k$ are the same for $k<n-1$, such that the first $n-2$ terms in Eq. \eqref{eqn::Fn} cancel, whereas $h_{n-1}$ and $h_n$ are chosen to exchange the last two fluxes. In a similar configuration, it is possible to check that all the denominators assumed by the operators $\tilde{F_r}$ with $r<n$ are out of resonance, thus bounded by $|\tilde{F_r}| < 2/\Delta$ without any dependence on the $\eta$ coefficients. 
We conclude that
\begin{equation}
\sum_{h_1\ldots h_n \neq e} |\tilde{F}_1|\ldots |\tilde{F}_{n-1}||F_n| < \frac{1}{4\eta_0}\left(\frac{4}{\Delta}\right)^n \,.
\end{equation}
Therefore, for $\mu/\Delta < \left(4\left(|G|-1\right)\right)^{-1}$, the strong zero-energy mode is exponentially suppressed in the bulk of the system.

This result is achieved through an exponential fine-tuning of the coupling constants, however, we expect that the zero-energy modes exist also for disordered setups, in which the parameters $\eta_r$ become random variables with a suitable distribution. This corresponds to assigning a small random contribution to the flux masses which depends on the plaquettes of the model, thus avoiding the possibility of resonances of the second kind. 

The inner terms of the strong zero-energy modes do not introduce additional resonances and, therefore, do not qualitatively modify the general decay behavior of the modes we discussed (see Appendix \ref{app:inner}).

\subsection{Iterative procedure for strong modes on the right boundary} \label{sec:strong3}

The construction of the strong zero-energy mode $\Omega_g$ localized on the right boundary of the system is very similar to the left modes, except for the fact that it carries a JW string $\mathcal{L}^A_g$ and, consequently, a group index. 

The starting point is $\Omega_{g,0}=\beta_g(2L) = \mathcal{L}^A_{g}(L) U^{K\dagger}(L)$. It is important to notice that the full JW string $\mathcal{L}^A_{g}(L)$ commutes with all terms in the Hamiltonian: it is easy to prove that $\left[\mathcal{L}^A_{g}(L),H_J\right]=0$; concerning the commutator with $H_\mu$, instead, it is useful to rewrite $H_\mu$ as a sum of projectors $\Pi^A(r)$  over the auxiliary representation (see Eq. \eqref{Kproj}) and exploit the relation $\left[\Theta_g(r),\Pi^A(r)\right]=0$. Therefore $\mathcal{L}^A_{g}(L)$ is a symmetry of the system, and the iterative definition of the right modes can proceed in the same way of the left modes. We define the commutators
\begin{multline}
C_{1}(g) = [\Omega_{g,0},H_\mu]  \\
 =-\mu \mathcal{L}_g^A(L)U^\dagger(L)\sum_{h_1\neq e} \chi^A(h_1^{-1}) \theta_{h_1}(L)  (\id-D(h_1)),
\end{multline}
and we build the first-order correction of the strong mode:
\begin{align}
&\Omega_{g,1} = \nonumber \\
& -\frac{\mu}{J} \mathcal{L}_g^A(L) U^\dagger(L) \sum_{h_1\neq e} \chi^A(h_1^{-1}) P_1 \theta_{h_1}(L) (\id-D(h_1)) \,, \label{Omegag1}
\end{align}
with
\begin{equation}
P_1 =\left(\Tr\left[U(L)CU^\dagger(L-1) \left( D^\dagger(h_1)-\id \right) \right] + \hc\right)^{-1},\label{eqnP1}
\end{equation}
such that $[\Omega_{g,1},H_J]=-C_{1}(g)$.

Also, in this case, it is convenient to distinguish inner and outer contributions of the operators, where the outer contributions are the ones defining the decay in the bulk of the system:
\begin{equation}
C_{2}(G)= [\Omega_{g,1},H_{\mu} ] = C_{{\rm in},2}(g) + C_{{\rm out},2}(g)
\end{equation}
with
\bwt
\begin{equation}
C_{{\rm out},2}(g) = -\mu \left[\Omega_{g,1}, \sum_{h_2} \chi^A( h_2) \theta_{h_2}(L-1)\right] =-\frac{\mu^2}{J} \mathcal{L}^A_g(L)U^\dagger(L) \sum_{h_1,h_2}\chi^A(h_1)\chi^A(h_2) \tilde{P}_1 \theta_{h_1}(L) \theta_{h_2}(L-1) (\id - D(h_1)),
\end{equation}
where
\begin{multline}
\tilde{P}_1(h_1,h_2) = [P_1, \theta_{h_2}(L-1)]\theta_{h_2}^\dag(L-1) =\\
 \left(\Tr\left[U(L)CU^\dagger(L-1) \left( D^\dagger(h_1)-\id \right) \right] + \hc \right)^{-1} - \left(\Tr\left[U(L)CU^\dagger(L-1) D(h_2) \left( D^\dagger(h_1)-\id \right) \right] + \hc \right)^{-1},
\end{multline}
and the corresponding outermost term at second order is
\begin{equation}
\Omega_{g,{\rm out},2} = -\frac{\mu^2}{J^2} \mathcal{L}^A_g(L) U^\dagger(L) \sum_{h_1,h_2}\chi^A(h_1)\chi^A(h_2) \tilde{P}_1 P_2 \theta_{h_1}(L) \theta_{h_2}(L-1) (\id-D(h_1)).
\end{equation}
The general construction of all the iterative terms in the right modes follows from the one for left modes with a suitable substitution of the functions $F$ and $\tilde{F}$ with their right counterparts $P$ and $\tilde{P}$:
\begin{equation}\label{OmegaOutn}
\Omega_{g,{\rm out},n} = \left(\frac{\mu}{J}\right)^n \mathcal{L}^A_g(L) U^\dagger(L) \sum_{h_1,\ldots,h_n \neq e}\chi^A(h_1^{-1})\ldots\chi^A(h_n^{-1}) \tilde{P}_{1} \ldots  \tilde{P}_{n-1} P_{n} \theta_{h_1}(L)\ldots \theta_{h_n}(L-n+1)  (\id-D(h_1)),
\end{equation}
where
\begin{equation}
P_{n}(h_1, \ldots, h_n) 
\equiv J\left(H_J-\theta_{h_1}(L)\ldots\theta_{h_n}(L-n+1) H_J \theta_{h_1}^\dag(L)\ldots \theta_{h_n}^\dag(L-n+1)\right)^{-1} \,,
\end{equation}
\ewt
and
\begin{equation}
\tilde{P}_n(h_1, \ldots, h_{n+1}) = P_n - \theta_{h_{n+1}}(L-n) P_n \theta^\dag_{h_{n+1}}(L-n)\,.
\end{equation}
It is easy to observe that these operators are local in the dyonic modes: they all result proportional to $\beta_g(2L)$ and all the terms in the sum in Eq. \eqref{OmegaOutn} can be expressed as products of dyonic operators through Eqs. \eqref{UUdag} and \eqref{betaalpha2}.
The operators $P$ and $\tilde{P}$ are subject to the same kind of divergences of their left counterparts and an analogous space dependence of the coupling constant $J$ can be adopted to achieve the exponential suppression of the right modes in the bulk.

\subsection{Properties of the dyonic zero-energy modes} \label{sec:fusion}

The strong zero-energy dyonic modes are characterized in general by the irreducible representation $K$, which determines the transformation relation \eqref{Qgamma2} through the matrices $D^K(h)$, and by the group index $g$ which appears in the right modes through the operator $\mathcal{L}^A_g$ in \eqref{OmegaOutn}. A group index characterizes also the left modes at the interfaces with nontopological regions of the system (see Appendix \ref{app:left}), however, for simplicity, we will restrict our analysis to the uniform case with open boundaries.

The commutation relation between left and right modes is given by
\begin{equation} \label{commstrong}
\Lambda_{m_1m_2} \Omega_{g,m_3m_4} = \Omega_{g,m_3m_4}\Lambda_{m_1m_2'} D^{K\dag}_{m_2'm_2}(g) \,,
\end{equation}
up to corrections exponentially suppressed in the system size.
Here and in the following we will explicitly write only the indices related to the representation $K$, since the auxiliary representation indices are left invariant under this commutation.
The commutation relation \eqref{commstrong} corresponds to the commutation relations between $\alpha(1)$ and $\beta_g(2L)$ and it generalizes the commutation relations of Majorana and parafermionic zero-energy modes to the non-Abelian case. It can be derived by observing that all the contributions of $\Lambda$ and $\Omega$ are proportional to $\alpha(1)$ and $\beta_g(2L)$ respectively; thus, Eq. \eqref{commstrong} results from the commutation between the factor $\alpha(1)$ and the JW string in the factor $\beta_g(2L)$. Other corrections may appear in the commutation relation due to the overlap of the zero modes for $\mu \neq 0$, but they are all of order $(\mu/J)^L$.

It is important to observe that the zero-energy modes $\Lambda_{m_1m_2}$ and $\Omega_{g,m_1m_2}$ do not exhaust all the possible localized zero modes of the model. Different localized zero-energy modes are generated by multiplying left or right modes with each other. This additional modes are associated, in general, with irreducible representations of the group $G$ different from $K$, therefore, in the following, we will label left and right modes by $\Lambda_{m_1m_2}(I)$ and $\Omega_{g,m_1m_2}(I)$ with $I$ belonging to the irreducible representations of $G$. The zero modes built in the previous section correspond to the case $I=K$.

The analogy with Majorana and parafermionic modes suggests that also the dyonic modes can be considered as extrinsic topological defects with projective non-Abelian anyonic statistics \cite{wen2012,barkeshli2013} and their algebra provides information about the corresponding fusion rules. Let us consider first the products obtained by multiplying different left modes:
\begin{equation}
\Lambda_{m_1m_2}(K) \Lambda_{m_3m_4}(K)\,;
\end{equation}
this is the product of two rank-2 operators which transforms following the irreducible representation $K$ under global gauge symmetries:
\begin{multline}
\mathcal{Q}_h \Lambda_{m_1m_2}(K) \Lambda_{m_3m_4}(K) \mathcal{Q}_h^\dag \\
=\Lambda_{m_1m_2'}(K) \Lambda_{m_3m_4'}(K) D^{K}_{m_2'm_2}(h) D^{K}_{m_4'm_4}(h)  \,.
\end{multline} 
To understand the nature of this operator, we exploit the Clebsch-Gordan series relation \cite{brink}:
\begin{multline} \label{CBseries}
D^{I_1}_{m_2' m_2}(h) D^{I_2}_{m_4' m_4}(h) \\
=\sum_{I,n,n'} \bracket{I_1 m_2' I_2 m_4'}{I n'} \bracket{I n}{I_1 m_2 I_2 m_4} D^{I}_{n'n}(h)\,.
\end{multline}
Here we introduced the notation $\bracket{I_1 m_2' I_2 m_4'}{I n'}$ and $\bracket{I n}{I_1 m_2 I_2 m_4}$ for the Clebsch-Gordan coefficients of the group and their conjugate respectively.
By combining the previous two equations we get:
\begin{align}
\mathcal{Q}_h \Lambda_{m_1m_2}&(K) \Lambda_{m_3m_4}(K) \mathcal{Q}_h^\dag  \nonumber \\
&=\sum_{I,n,n',m_2',m_4'} \Lambda_{m_1m_2'}(K) \Lambda_{m_3m_4'}(K) \nonumber \\
&\times  \bracket{K m_2' K m_4'}{I n'} \bracket{I n}{K m_2 K m_4} D^{I}_{n'n}(h)  \,.
\end{align}
This demonstrates that the product of two zero-energy modes $\Lambda$ is a linear superposition of operators transforming according to the irreducible representations $I$ allowed by the Clebsch-Gordan series. Therefore, in general, we must define a family of zero-energy operators localized on the left edge, $\Lambda_{n_1n_2}(I)$, such that
\begin{multline}
\Lambda_{m_1m_2}(I_1) \Lambda_{m_3m_4}(I_2) \\
= \sum_{I,n_1,n_2} \bracket{I_1 m_1 I_2 m_3}{I n_1} \bracket{I n_2}{I_1 m_2 I_2 m_4} \Lambda_{n_1n_2}(I)\,,
\end{multline}
and
\begin{equation} \label{lambdatransf}
\mathcal{Q}_h \Lambda_{mn}(I) \mathcal{Q}_h^\dag = \Lambda_{mn'}(I)D^{I}_{n'n}(h)\,.
\end{equation}
Based on this transformation relation, we obtain that, starting from the gauge-invariant ground state $\ket{\ket{000}}$, the ground state $\Lambda_{mn}^\dag(I)\ket{\ket{000}} = \ket{\ket{Imn}}$ will transform as $\mathcal{Q}_h \ket{\ket{Imn}} = D^{I\dag}_{mm'}(h)\ket{\ket{Im'n}}$.

From Eq. \eqref{lambdatransf} it is also easy to show that $\Lambda^{|G|}(I)$ is invariant under the symmetries $\mathcal{Q}_h$. Therefore, for any irreducible representation $I$ and any ground state $\ket{\ket{Rmn}}$ we obtain $\Lambda^{|G|}(I)\ket{\ket{Rmn}} \propto \ket{\ket{Rmn}}$. This suggests that the operators $\Lambda_{n_1n_2}(I)$ behave like the dyonic operator $\alpha^{K=I}(1)$.

The situation is more complicated for the right modes: also in this case we can consider modes associated with any irreducible representation $I$, but, with respect to the left modes, we must account also for the group element conjugation in \eqref{Qgamma2} and the indices of the irreducible representation $A$:
\begin{multline}
\mathcal{Q}_h{\Omega}_{g,m_1m_2,{\sf ab}}(I_1) {\Omega}_{k,m_3m_4,{\sf cd}}(I_2)\mathcal{Q}_h^\dag  \\
 =\sum_{I,n,n',m_2',m_4'} \Omega_{hgh^{-1},m_1m_2',{\sf ab}}(I_1) \Omega_{hkh^{-1},m_3m_4',{\sf cd}}(I_2)  \\
 \times\bracket{I_1 m_2' I_2 m_4'}{I n'} \bracket{I n}{I_1 m_2 I_2 m_4}  D^I_{n'n}(h)  \,.
\end{multline}
From this relation we deduce that ${\Omega}_{g}(I_1) {\Omega}_{k}(I_2)$ is indeed proportional to $\prod_r \theta^\dag_{kg}(r)$ and can be decomposed into a linear superposition of dyonic operators associated to the irreducible representations $I$. For non-Abelian auxiliary representations, however, the set  ${\Omega}_{g}(I)$ does not exhaust all the possible right zero-energy modes due to the nontrivial composition of the disorder operators $\mathcal{L}^A$. Moreover, given the previous composition rule for $g=k$, it is possible to show that the modes ${\Omega}_{g}(I)$ behave like the operators $\beta_g^{K=I}(2L)$, and, in particular ${\Omega}_{g}^{|G|^2}(I) \propto \Id_I \Id_A$ is a symmetric operator, similarly to Eq. \eqref{betapower}.

The previous rules dictate how left modes fuse with left modes, and right modes with right modes. Concerning the fusion of a left with a right mode, it is convenient to introduce the operator
\begin{equation}
 \Upsilon(g) \equiv \Tr_K \left[\Lambda(K) \Omega_g^\dag(K)\right]
\end{equation}
where the indices of the auxiliary representation do not play any fundamental role. These operators generalize \eqref{Upsilon0} to the general case with $\mu \neq 0$. Their transformation under the symmetry group results in
\begin{align} 
\mathcal{Q}_h&\Upsilon(g)\mathcal{Q}_h^\dag= \mathcal{Q}_h\Tr_K \left[\Lambda(K) \Omega_g^\dag(K)\right]\mathcal{Q}_h^\dag\nonumber \\ 
 &=\Tr_K \left[\Lambda(K) \Omega_{hgh^{-1}}^\dag(K)\right] =  \Upsilon(hgh^{-1})\,.\label{Uptransf}
\end{align}
The operators $\Upsilon(g)$ extend the usual idea of $\mathbb{Z}_N$ parity from the Abelian to the non-Abelian case: in analogy with the gauge transformations $\mathcal{Q}_g$ themselves, they transform under conjugation and they belong to the class of operators $\mathcal{C}$ characterizing the condition \ref{condt2} for topological order. In particular, the operators $\Upsilon_g$ are block diagonal in the irreducible representation basis and can be decomposed in the following way:
\begin{equation} \label{Yudecomp}
\Upsilon(g) = \sum_{I,m,n} D^{I*}_{mn}(g) \tilde{\Upsilon}(I)_{m,n}\,,
\end{equation}
with $\tilde{\Upsilon}(I)_{m,n} = \upsilon_I \sum_l \ket{Iml}\bra{Inl}$ (where $\upsilon_I$ are suitable constants) and
\begin{equation}
\mathcal{Q}_h \tilde{\Upsilon}(I)\mathcal{Q}_h^\dag = D^I(h^{-1}) \tilde{\Upsilon}(I) D^I(h)\,.
\end{equation}

The decomposition \eqref{Yudecomp} can be considered the fusion rule for left and right zero modes: $\Upsilon(g)$, which plays the role of their operator product, results in a set of \textit{fusion channels} in one-to-one correspondence with the irreducible representations $I$ of the group, which can be schematically represented as
\begin{equation} \label{fusion}
\Lambda \times \Omega = \oplus_I \, \Xi_I\,.
\end{equation}
Each channel $\Xi_I$ has a quantum dimension given by $\dim(I)^2$, such that, in total, we can attribute the quantum dimension $\sqrt{|G|}$ to the zero-energy mode $\Lambda(K)$ and $\Omega(K)$. This is analogous to the case of Majorana and parafermionic zero modes.

We observe that the decomposition \eqref{Yudecomp} holds true independently of our choice of the irreducible representation of the zero modes $\Lambda(I)$ and $\Omega(I)$: our definition of $\Upsilon(g)$ can indeed be extended to the operators $\Upsilon(I,g) \equiv \Tr_I\left[\Lambda(I) \Omega_g^\dag(I)\right]$. These operators behave under gauge transformations in the same way, and can be decomposed in terms of the same operators $\tilde{\Upsilon}(R)$.

It is possible to extend our analysis also to the case of a topological region embedded in a nontopological environment (see Appendix \ref{app:left}). In this situation, the left modes acquire a group index too, and the operators $\Upsilon(g)$ must be defined by contracting $\Lambda_g$ and $\Omega^{\dag}_g$ taken with the same group index. In this way, the JW strings $\mathcal{L}^A_g$ cancel outside the topological region, and all the previous observations still hold.

This situation is analogous to the study of twist defects in symmetry-enriched phases with topological order \cite{barkeshli2014,fradkin2015,teo2016}. Majorana and parafermionic modes behave like twist defects in the $\mathbb{Z}_2$ and $\mathbb{Z}_N$ toric codes, respectively \cite{barkeshli2014,teo2016}; this suggests that the dyonic modes in the system \eqref{ham} may be interpreted as twist defects in a suitable two-dimensional topological system. The requirement of combining $\Lambda_g$ and $\Omega^{\dag}_g$ corresponds to having two twist defects with opposite flux which identify a $g$-defect branch line \cite{barkeshli2014}, and, in this scenario, the study of the topological and braiding properties of the dyonic zero-energy modes must be framed in a $G$-crossed braided tensor category theory \cite{barkeshli2014}.

\section{The role of the auxiliary representation} \label{sec:aux}

The analysis of the topological models in Eq. \eqref{ham} crucially relies on the choice of the group $G$ and of the auxiliary irreducible representation $A$. The auxiliary representation $A$ enters the definition of the disorder operators $\mathcal{L}^A$, which, in turn, define the dyonic modes (\ref{JW1},\ref{JW2}). Because we define locality through the dyonic modes $\alpha$ and $\beta$, the selection of $A$ directly determines which operators are local in the dyonic model.

The connection operators $U(r)$ constitute order parameters able to distinguish the ground states of the flux-ladder Hamiltonian \eqref{ham_flux} in its ferromagnetic phase. Importantly, these operators are nonlocal in the dyonic modes if and only if the irreducible representation $A$ is non-Abelian. This implies that, in case of an Abelian representation $A$, the topological order of the system \eqref{ham} is lost.

The operators $\Theta_g(r)$, instead, are always local in terms of the dyonic modes (see Eq. \eqref{betaalpha}). Furthermore, from Eq. \eqref{betaalpha2}, we obtain that also the operators $\theta_g(r)$ are local, provided that $\chi^A(g^{-1}) \neq 0$. For $\chi^A(g^{-1}) = 0$, instead, $\theta_g$ may be local or nonlocal depending on the group properties. This is related to certain additional symmetries which may appear in the flux-ladder Hamiltonian \eqref{ham_flux} for particular combinations of $G$ and $A$, as for example, the choice $G=S_3$ with its non-Abelian irreducible representation $A=2$.

In the following we will first examine the features of the systems with a trivial auxiliary representation $A$, which exemplifies what happens for all the Abelian auxiliary representations, then we will consider in more detail the case of non-Abelian irreducible representations $A$ with elements with vanishing character $\chi^A(g^{-1})=0$.

\subsection{Trivial auxiliary representations: absence of topological order and appearance of holographic symmetries} \label{subsec:A}

In the case of an Abelian auxiliary representation $A$, the Hamiltonian \eqref{ham} loses its topological order. This is due to the properties of the Jordan-Wigner strings $\mathcal{L}^A$. For $A$ Abelian and irreducible, the matrices $D^A$ in \eqref{Ldef} become just phases. The composition rules of the disorder operators then simplify, $\mathcal{L}^A_{g_1}\mathcal{L}^A_{g_2}=\mathcal{L}^A_{g_2g_1}$, thus we obtain
\begin{equation} \label{Uwrong}
U^\dag(r) = \beta_{hg^{-1}}(2r) \beta^\dag_{h}(2r) \beta_{g}(2r)\,.
\end{equation}
This relation is fulfilled because the Abelian JW strings in the $\beta$ modes annihilate. Eq. \eqref{Uwrong} proves that the operators $U^\dag(r)$ are local in the dyonic operators, and, from these operators, it is possible to build local operators and observables that violate both the conditions \ref{condt1} and \ref{condt2} for topological order.

On the contrary, when $A$ is non-Abelian, the only combinations of JW strings which allow for their annihilation are given by Eqs. (\ref{ltr2},\ref{ltr3}) and it is impossible to find operators local in the dyonic modes that return $U^\dag(r)$.

Let us focus on the trivial case $A=1$ such that $\mathcal{L}^A_g(r) = \prod_{x=1}^r \theta_g^\dag(x)$, without additional indices related to the auxiliary representation. In this case we obtain the apparent inconsistency:
\begin{equation} \label{paradox}
\beta_{kg^{-1}}(2r) \beta^\dag_{k}(2r) = \prod_{j=1}^r \alpha_g(2j-1)\beta^{\dagger}_g(2j)\,;
\end{equation}
this relation is paradoxical because the left-hand-side is a local operator, expressed as a function of $\beta$'s only, but it is equivalent to a nonlocal string operator when expressed in terms of both $\beta$'s and $\alpha$'s. This contradiction is solved by taking into account that, for $A=1$, the operators $\alpha$ and $\beta$ are not independent from each other. In particular, it is possible to express any operator $\alpha$ as a function of the operators $\beta$:
\begin{align}
&\alpha(1) = \beta_{\tilde{k}\tilde{h}^{-1}}(2) \beta^\dag_{\tilde{k}}(2) \beta_{\tilde{h}}(2)  \label{a1} \,,\\
&\alpha_g(2r-1)  \nonumber \\
&=\underbrace{\beta_k(2r-2)\beta^\dag_{kg^{-1}}(2r-2)}_{\mathcal{L}_g(r-1)}\underbrace{\beta_{hg^{-1}}(2r)\beta^\dag_h(2r)\beta_g(2r)}_{U^\dag(r)} \label{ar} \,,
\end{align}
for $r>1$ and any arbitrary choice of $\tilde{h},\tilde{k},h,k \neq e$ such that $\tilde{h}\neq \tilde{k}$ and $k,h \neq g$.

Eqs. \eqref{a1} and \eqref{ar} allow us to solve the apparent inconsistency of Eq. \eqref{paradox}: for the sake of simplicity we can take $k=h=\tilde{k}$ and $\tilde{h}=g$; in this case it is easy to see that the right-hand-side of Eq. \eqref{paradox} reduces telescopically to the left-hand side, thus verifying its local nature in terms of the $\beta$ operators. 

We conclude that, for the case $A=1$, the notion of locality must be based on the $\beta$ operators only: the $\alpha$ operators can be expressed as local combination of the $\beta$ operators and all the Hamiltonian terms are local in turn.
Based on this notion of locality, also the symmetry operators $\mathcal{Q}_g$ become localized:
\begin{equation} \label{holozero}
\mathcal{Q}_g = \beta_{kg^{-1}}(2L)\beta_k^\dag(2L)\,,
\end{equation}
for an arbitrary $k \neq g,e$.
This relation establishes a mapping from the global (thus nonlocal) gauge symmetry in the flux-ladder Hamiltonian \eqref{ham_flux}, to a set of symmetry operators localized on the last site of the system \eqref{ham}. This is an example of holographic symmetry \cite{cobanera2013}. 

As a result, all the operators of the form \eqref{holozero} are localized and exact zero-energy modes of the Hamiltonian \eqref{ham}, independently on the values of $\mu,J$ or $C$. Therefore, it is possible to identify the behavior of any eigenstate of the system under the symmetry group $G$ just by considering expectation values of suitable observables localized on the last site, thanks to Eq. \eqref{holozero}. This also implies that any local perturbation of the form $\mathcal{Q}_g(2L)$ can split the ground state degeneracy of the system in the $J$-dominated phase. For example, by exploiting the projector \eqref{Kproj}, we can build the following symmetry-invariant operator, which separates in energy the gauge-invariant ground state $\ket{\ket{000}}$ from the others:
\begin{equation} \label{pi1}
\Pi^{(1)}_{\text{tot}}  = -\sum_{g\in G} \mathcal{Q}_g\,.
\end{equation}
This perturbation splits the ground-state degeneracy, despite preserving the group symmetry. We observe, however, that the holographic zero-energy modes can be used to build observables that determine only the global behavior under the symmetry transformation (as in the case of the total fermionic parity in the Kitaev chain); when considering a nonuniform system with alternating $\mu-$dominated and $J-$dominated segments, the number of degenerate ground states scales with the number of interfaces and the holographic modes cannot distinguish all the ground states. 

\subsection{Non-Abelian auxiliary representations and additional symmetries}

For a non-Abelian group $G$ and a non-Abelian auxiliary representation $A$, in general, there will be a set of conjugacy classes such that the character $\chi^A$ vanishes for their elements. Let $G_0$ denote the set of group elements $g$ with vanishing character $\chi^A(g^{-1})$:
\begin{equation}
G_0 = \{ g \in G \quad \text{s. t.} \quad \chi^{A}(g^{-1})=0 \}\,,
\end{equation}
and by $G_0^c$ its complement:
\begin{equation}
G_0^c = \{ g \in G \quad \text{s. t.} \quad \chi^{A}(g^{-1}) \neq 0 \}\,.
\end{equation}

For all the elements $\tilde{g} \in G_0$, $\theta_{\tilde{g}}$ does not appear in the gauge-flux Hamiltonian \eqref{ham_flux}. Furthermore, $\theta_{\tilde{g}}(r)$ cannot be expressed simply in terms of the trace over $A$ of $\beta^\dag_{\tilde{g}}(2r)\alpha_{\tilde{g}}(2r-1)$, because the right-hand side of Eq. \eqref{betaalpha2} vanishes. 

Depending on the choice of $G$ and $A$, we must distinguish two cases: (i) $G_0^c$ is not a proper subgroup of $G$; (ii) $G_0^c$ is a proper subgroup of $G$.

An example of the kind (i) is the $S_4$ group, corresponding to the 24 orientation-preserving symmetries of the cube, associated with its fundamental representation $A=3$ of dimension 3. When $G_0^c$ is not a proper subgroup, the elements of $G_0$ can be generated by the products of elements of $G_0^c$. Therefore, in case (i), all the operators $\theta_g(r)$ can be expressed in a local form in terms of the dyonic modes: for $g \in G_0^c$, it is enough to apply Eq. \eqref{betaalpha}; for $\tilde{g} \in G_0$, instead, we can express $\tilde{g}= g_{1} \ldots g_{l}$ with all the $g_i$'s belonging to $G_0^c$; in this way $\theta_{\tilde{g}}(r)=\theta_{g_1}(r)\ldots \theta_{g_l}(r)$ results from the product of the local terms $\theta_{g_i}$ and it is local in turn.

The case (ii) can be exemplified by the group $S_3$ with its fundamental representation $A=2$ (and analogously by all the groups $D_n$). In this case, the operators $\theta_{\tilde{g}}(r)$ with $\tilde{g} \in G_0$ cannot be obtained in this way because $G_0^c$ is closed under composition. This implies that the operators $\theta_{\tilde{g}}(r)$ are not local operators as a function of the dyonic modes. Therefore, adding to the Hamiltonian small perturbations that include the operators $\theta_{\tilde{g}}(r)$ may in general destroy the topological order. 

Furthermore, in case (ii), the system acquires additional local symmetries. To examine the appearance of these symmetries, it is useful to consider the flux-ladder Hamiltonian \eqref{ham_flux}. The operators $\theta_{\tilde{g}}(r)$ (with $\tilde{g} \in G_0$) do not appear in the Hamiltonian and cannot be obtained as products of the other operators $\theta_g$. Let us consider the unitary operator
\begin{equation}
V(r) = \exp\left[i \sum_{\tilde{g} \in G_0} \alpha(r) \ket{\tilde{g}(r)} \bra{\tilde{g}(r)} \right]\,.
\end{equation}
This is a $U(1)$ local transformation that multiplies the wavefunction by a phase $e^{i\alpha(r)}$ if the $r^{\rm th}$ rung is in a state belonging to $G_0$. It is easy to see that $V^\dag(r) H V(r) = H$: $V(r)$ is diagonal in the group element basis, it trivially commutes with $H_J$ and, in case (ii), there are no terms in the Hamiltonian mixing the states in $G_0$ and $G_0^c$ due to $G_0^c$ being closed under composition.
Therefore there is an extensive set of conserved quantities $Q(r)=\sum_{\tilde{g} \in G_0} \alpha(r) \ket{\tilde{g}(r)} \bra{\tilde{g}(r)}$ which split the Hilbert space in $2^L$ subspaces. In each of these subspaces the Hamiltonian has a reduced global symmetry group $G_0^c$ rather than the full symmetry group $G$.

In the case $G=S_3$ and $A=2$, for example, the degrees of freedom $\ket{m}$ and $\ket{n}$ introduced in Sec. \ref{sec:s3} decouple: the conserved charges $Q(r)$ correspond to the $n=0,1$ degrees of freedom and the dynamics in each subspace is characterized by an Abelian $\mathbb{Z}_3$ symmetry generated by the global $c$ transformations only. The global $b$ transformations, instead, map a subspace into its complementary with charges $1-Q(r)$.

In this case (in a system with open boundary conditions) the left zero-energy modes $\Lambda$ and their weak counterpart do not include any of the operators  $\theta_{\tilde{h}}(r)$ with $\tilde{h} \in G_0$ and act only within a single subspace. Their role becomes analogous to the $\mathbb{Z}_3$ parafermionic zero modes. The right zero modes $\Omega_{\tilde{g}}$ and their weak counterparts, instead, map a subspace into its complementary through the JW string in Eq. \eqref{OmegaOutn}. In case (ii), therefore, it is possible to decompose the dyonic modes into the product of $\mathbb{Z}_3$ parafermionic zero modes with $\mathbb{Z}_2$ operators.
An analogous situation is verified for any group $D_n$ with $A=2$. We conclude, therefore, that the groups $D_n$ are unsuitable to study the genuine non-Abelian nature of the zero-energy dyonic modes. 

The groups with non-Abelian irreducible auxiliary representations of the kind (i), instead, offer the suitable playground to study the topological ordered phases of the dyonic models in their full extent.

\section{Analysis of the single-flux subspace for the group $S_3$} \label{singleflux}

In this section we numerically investigate some of the features of the system for the specific case of the $S_3$ flux ladder introduced in Section \ref{sec:s3}: we discuss the roles of the matrix $C$ and the auxiliary irreducible representation $A$ in the spectrum of the lowest excited states and in the definition of the strong zero-energy modes.

We follow the approach presented in \cite{jermyn2014} for Abelian symmetries, and we restrict our analysis to the subspace of the states with a single-flux excitation in the ladder. This is a strong limitation in the study of the overall system, but, despite that, it is useful to verify some of the analytical results of the previous sections and to investigate the onset of resonances in the first step of the iterative definition of the strong zero-energy modes in Eqs. (\ref{lambda1},\ref{eqnF1}). 

For small values of $\mu/J$, the energy spectrum of the single-flux excitations presents $|G|-1$ energy bands, each associated with one of the nontrivial fluxes $g\in G$ of the model. Each energy band includes $(L-1) \times 6$ states, corresponding to the choice of the plaquette $r$ of the flux $g$ and the background group element $h$, namely the state of the last rung of the ladder. We can represent a basis of the single-flux states based on the domain-wall picture:
\begin{equation} \label{sfbasis}
\ket{g,h,r} = \ket{hg}_1 \ldots \ket{hg}_r \ket{h}_{r+1} \ldots \ket{h}_L\,,
\end{equation}
with $g \neq e$.

The flux-ladder Hamiltonian, projected into the single-flux subspace, includes three contributions related to the masses of the fluxes \eqref{mass}, their kinetic energy, and the boundary terms of the system. We label these contributions by $M,K$ and $B$ respectively, such that
\begin{equation}\label{eqn:SingleFluxHam}
H_{\rm sf} = M + K + B\,,
\end{equation}
with
\begin{equation}
\bracket{g_1, h_1, r_1| M}{g_2, h_2, r_2} = \delta_{g_1,g_2}\delta_{h_1, h_2}\delta_{r_1, r_2} m_{g_2^{-1}},
\end{equation}
\begin{multline}
\bracket{g_1, h_1, r_1| K}{g_2, h_2, r_2} \\
= \delta_{g_1,g_2}\delta_{h_1,h_2} \delta_{r_1\pm 1, r_2} \left[ -\mu \chi^A(g_2^{\pm 1})\right],
\end{multline}
\begin{multline}\label{boundary1}
\bracket{g_1, h_1, 1| B}{g_2, h_2, 1}  \\
=\delta_{h_1, h_2}(1-\delta_{g_1, g_2}) \left[-\mu \chi^A(g_2g_1^{-1})\right],
\end{multline}
\begin{multline} \label{boundary2}
\bracket{g_1, h_1, L-1| B}{g_2, h_2, L-1} \\
= \delta_{h_1g_1, h_2g_2}(1-\delta_{g_1, g_2})\left[-\mu\chi^A(g_1g_2^{-1})\right].
\end{multline}

The resulting spectrum is characterized by three different energy scales. The largest energy scale is determined by the differences of the masses $m_g$ in Eq. \eqref{mass}, which establish the gaps among the energy bands in the limit $\mu \to 0$. The second energy scale is related to the kinetic energy of the fluxes and is approximately proportional to $\mu/L$; it defines the typical energy gaps appearing within each band in finite size system as effect of the dispersion of the fluxes. 
Finally, the smallest energy scale is given by the splitting of the quasidegenerate states corresponding to the same fluxes but different backgrounds and it is determined by the effect of the boundary terms. 

The scaling of the smallest energy splitting is related to the onset of resonances that hinder the formation of the strong zero-energy modes. In a system with well-defined strong zero-energy modes, all the states must be $|G|$-fold degenerate up to exponentially suppressed corrections in the system size. If the splitting among quasidegenerate states decays in a slower way with $L$, therefore, no strong zero-energy modes can be present in the system.

Analogously to the Abelian case \cite{jermyn2014}, we expect in general a large splitting of the $|G|$-plets of quasidegenerate states in regions of the spectrum in which at least two different bands overlap. The most common scenario is that the related splitting may decay algebraically in the system size, as in the case of the nonchiral $\mathbb{Z}_3$ model \cite{jermyn2014}. 
 This is due to the effect of the boundary terms: the term \eqref{boundary2} allows for transitions between states with different fluxes and different backgrounds, whereas the term \eqref{boundary1} allows for transitions between states with different fluxes and the same background. The combined action of the both of them, therefore, couples states with the same flux and different backgrounds, thus splitting the $|G|$-plets. This effect, though, is exponentially suppressed in the system size if there is an energy gap between the bands of different fluxes (as it can be derived through perturbation theory) and it becomes relevant only when two energy bands overlap. Stronger modifications of the spectrum may also occur in the presence of more overlapping band.

In the following, we analyze the case $G=S_3$ and we verify that, indeed, in the presence of overlapping bands, the splitting of the 6-plets of quasidegenerate single-flux states does not decay exponentially with the system size. On the contrary, for well-separated bands, such splitting decays exponentially.
We observe that the exponential decay of the single-flux splitting is certainly not sufficient to assess the presence of strong zero-energy modes: It is only related to the absence of resonances between states with a single flux. This implies, for example, that the first order of the iterative procedure \eqref{lambda1} is well-defined, but it does not provide information about the presence of resonances at higher orders.

We analyze the spectrum of the single-flux Hamiltonian \eqref{eqn:SingleFluxHam} for different two different choices of matrix $C$ and the auxiliary representation $A$.

\begin{figure}[tb]
\includegraphics[width=\columnwidth]{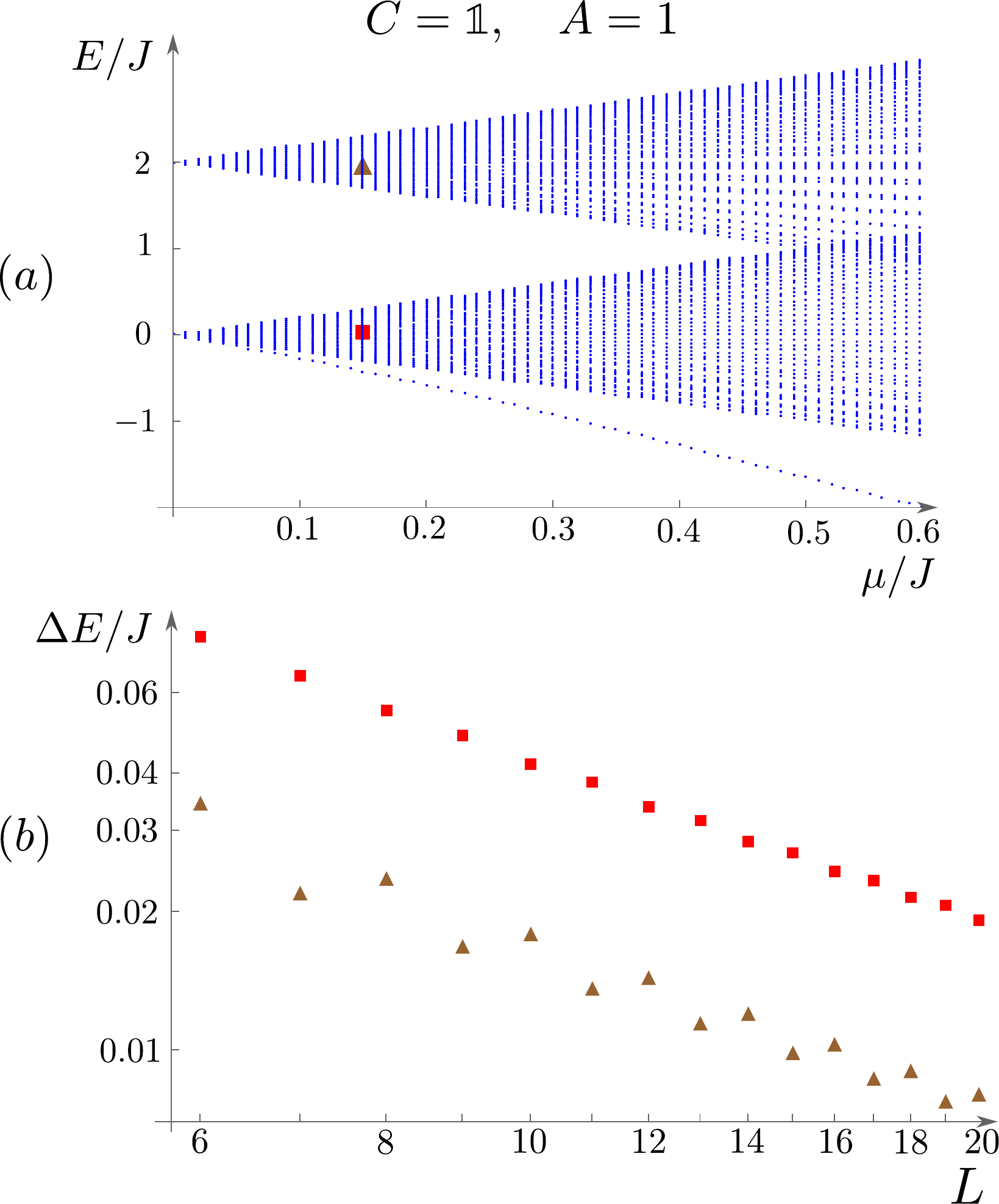}
\caption{(a): Spectrum of the single-flux Hamiltonian for $C=\id$ and $A=1$ for 19 sites. The bottom band consists of the group elements containing inversions, and the top band consists of the rotation fluxes. The branch separating from the lower band consists of 12 exactly degenerate states. (b): Energy splittings in the middle of the two bands for varying system sizes, shown on a logarithmic plot. The red squares indicate splittings between the last set of eight and two degenerate states in the 18-plet in the middle of the lower band, and the brown triangles indicates splittings between the last four and two degenerate states in the 12-plet in the middle of the upper band. In both cases the energy splitting decays roughly as $1/L$. The splitting between other sets of adjacent degenerate states behave similarly throughout the band.}
\label{fig:sfTCA1}
\end{figure}

\subsection{Case $A=1$}

We begin by analyzing the single-flux Hamiltonian in the case of trivial auxiliary representation $A=1$. This case is nontopological, as discussed in Sec. \ref{subsec:A}, but it provides an example of the general behavior of the single-flux energy bands.

For $C=\Id$, the $S_3$ model displays only two single-flux energy bands due to the degeneracy of the masses of the fluxes corresponding to the rotation ($c$ and $c^2$) and inversion ($b,bc$ and $bc^2$) elements of the groups. The doubly degenerate rotations have mass $m_c= 2J$, whereas the three-fold degenerate inversions have mass $m_b=0$ [see the definition \eqref{mass} and the matrices \eqref{s3matrix}]. Both the bands acquire a bandwidth proportional to $\mu$ due to the kinetic energy $K$. 

The spectrum for $C=\Id$ is represented in Fig. \ref{fig:sfTCA1} (a). The lowest (inversion) band includes $18(L-1)$ states corresponding to the six different backgrounds $h$ in \eqref{sfbasis} and the three degenerate fluxes at mass 0. Some of these states are localized at the edges of the system and they include, for instance, the separate branch at the bottom of the band with a 12-fold degeneracy. The remaining states, instead, can be distinguished into families of 18 states with a degeneracy pattern 8-8-2, except for the 24 state closest to the upper edge of the band, which are instead orgainzed in the degeneracy pattern 8-8-8. For $C=\Id$, indeed, the first order resonances in \eqref{lambda1} hinder the formation of strong modes, and the states in the lowest band are not arranged in the typical 6-plets. Our numerical analysis shows that both the splitting of the energies within and between the 18-plets of states decay algebraically and approximately  as $1/L$ in the system size [see Fig. \ref{fig:sfTCA1} (b)].

The upper band is constituted by the two degenerate rotation fluxes. In this case, the spectrum displays families of 12 states with a typical degeneracy pattern 2-4-4-2, and again all energy differences inside and between these 12-plet families decay algebraically [see  Fig. \ref{fig:sfTCA1} (b)].

To split the degeneracies of these the single-flux energy bands for small values of $\mu/J$ we introduce a $C$ matrix that fulfills conditions \ref{condI} and \ref{condIIa}. In particular, we choose
\begin{equation} 
C_1\equiv \frac{e^{-i\pi/4}}{\sqrt{2}}\left(\Id- \frac{i}{\sqrt{3}}\sigma_x + \frac{i}{\sqrt{3}}\sigma_y + \frac{i}{\sqrt{3}}\sigma_z\right)\,.
\end{equation}
The corresponding masses (in ascending order) are $\{-2, -2/\sqrt{3}, -1 + 1/\sqrt{3}, 0, 1 + 1/\sqrt{3}, 2\}$, and we have chosen this matrix in such a way that the gap between the trivial and the first excited fluxes is larger than the gap between the first and second excited fluxes. In this way the predictions of the single-flux Hamiltonian are more accurate for what concerns the lowest band since the transitions with the ground-state manifold and the two-flux states are less relevant than the boundary-term mixing between the first two bands.

\begin{figure}[tb!]
\includegraphics[width=\columnwidth]{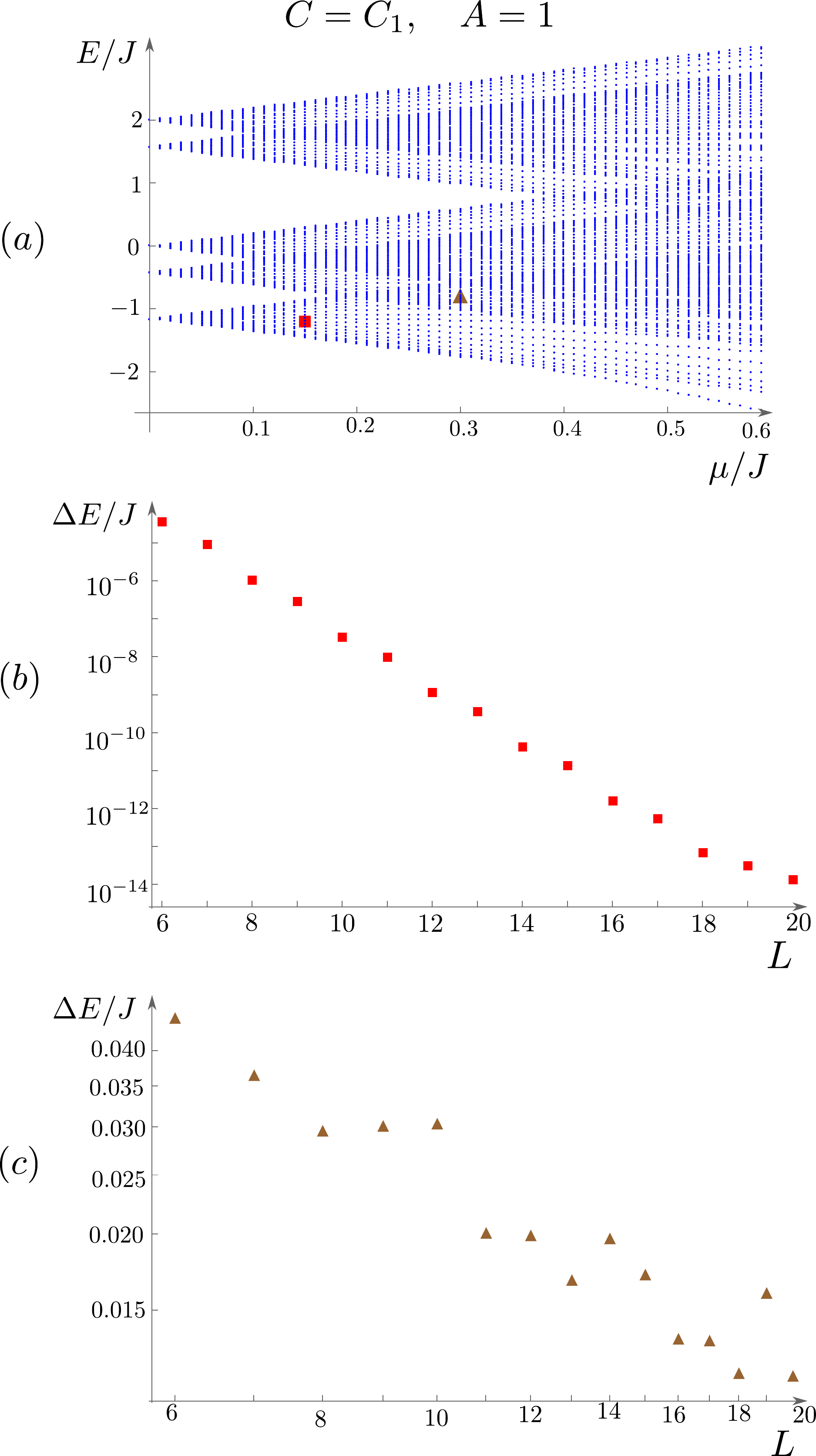}
\caption{(a): Spectrum of the single-flux Hamiltonian for $C=C_1$ and $A=1$ as a function of $\mu/J$ for 19 sites. The bands are nondegenerate by construction of $C_1$. (b): Energy splitting $\Delta E$ in units of $J$ as a function of system size $L$, shown on a semilogarithmic scale; its exponential decay is evident. The splitting is taken between the six quasidegenerate states in the middle of the bottom band at $\mu=0.15$ [red square in (a)]. (c): Splitting as a function of system size $L$ shown on a logarithmic plot and taken within a region of overlap between the bottom and next-lowest band at $\mu=0.3$ [brown triangle in (a)]. The splitting decays approximately algebraically, and we conclude that the zero-energy modes are weak.}
\label{fig:sfCPA1}
\end{figure}

The five resulting single-flux bands are well separated for small $\mu$ [see Fig. \ref{fig:sfCPA1} (a)] and all the states are now organized into $6-$plets separated by gaps scaling as $\mu/L$ due to the kinetic energy. \\
For small $\mu$, in the regions where the bands do not overlap, we observe an exponential decay of the splitting of the 6-plets with the system size [see Fig. \ref{fig:sfCPA1} (b)]. The $C_1$ matrix removes the resonance at the first level of iteration in the definition of the strong-zero energy modes and, consequently, the single-flux spectrum behaves as in the presence of strong modes (whereas states with more than one flux are subject to higher-order resonances).\\
When we consider larger values of $\mu$ and we study the spectrum of the states in a region with two overlapping bands, however, a weaker decay reappears [see Fig. \ref{fig:sfCPA1} (c), which approximately shows an algebraic decay] and the division into 6-plets is no longer precise.

\subsection{Case $A=2$}

\begin{figure}[tb]
\includegraphics[width=\columnwidth]{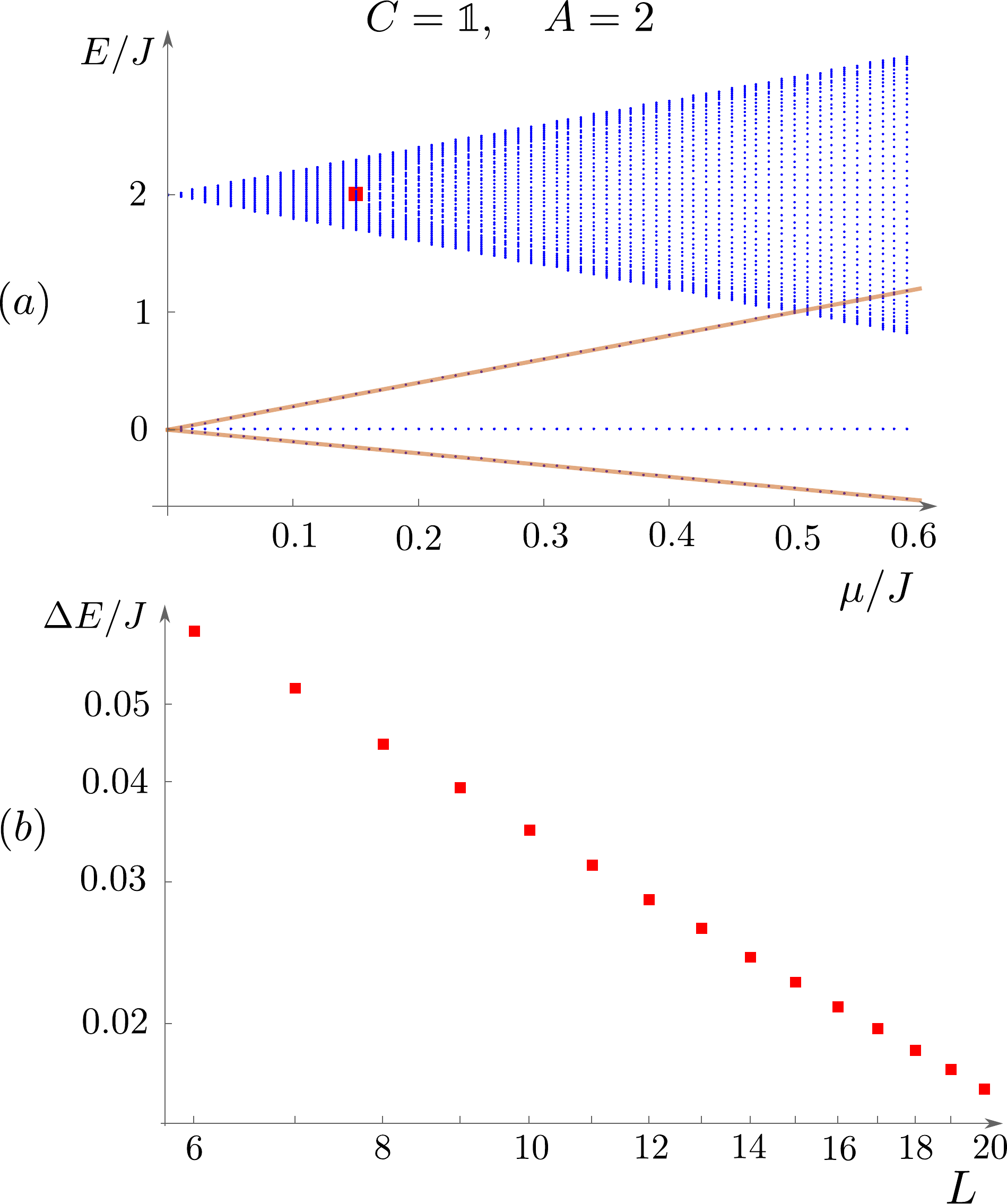}
\caption{(a): Spectrum of the single-flux Hamiltonian for $C=\id$ and $A=2$ for 19 sites. At $\mu=0$, the states associated with the inversions have zero energy (lowest band). These fluxes have no dynamics in the bulk. At the boundary, however, they mix and the boundary states acquire a finite energy when $\mu$ is increased. The energies of these edge states are given by the eigenvalues of the boundary terms \eqref{boundary1} and \eqref{boundary2}: six states acquire the energy $E=2\mu$ and 12 states the energy $-\mu$. These values are indicated by the straight orange lines. (b): Energy splitting $\Delta E$ in units of $J$ shown on a logarithmic plot. The splitting is taken between two sets of 4-fold degenerate states in the middle of the top band at $\mu=0.15$ (red square in the top panel). The splitting is algebraically suppressed in the system size. We conclude that the zero-energy modes are weak.}
\label{fig:sfTCA2}
\end{figure}

\begin{figure}[tb]
\includegraphics[width=\columnwidth]{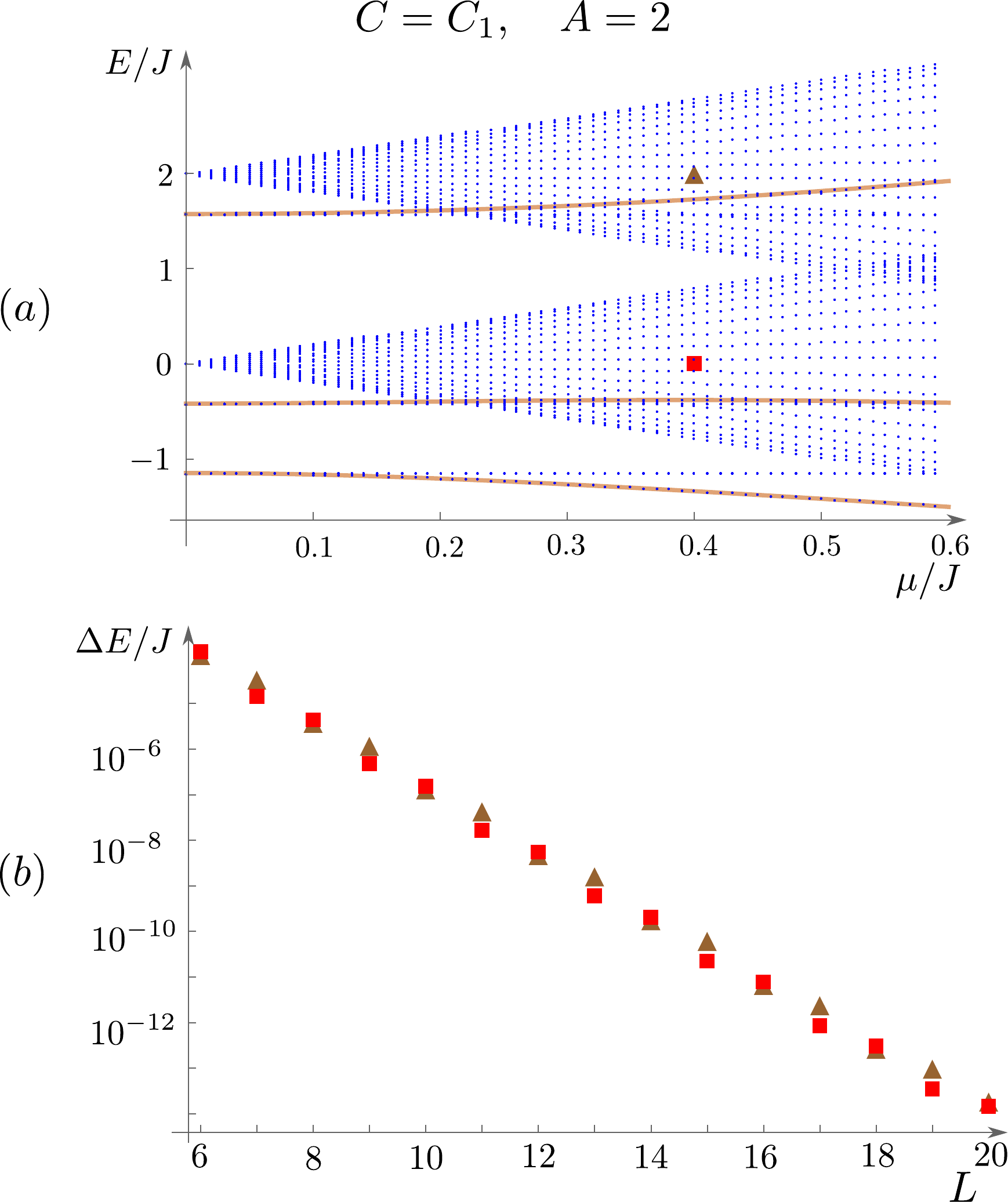}
\caption{(a): Spectrum of the single-flux Hamiltonian for $C=C_1$ and $A=2$ for 19 sites in units of $J$. The flat bands correspond to the inversions, and the branches separating from these bands correspond to boundary states. In analogy to the situation in Fig. \ref{fig:sfTCA2}, these boundary states have energies derived from \eqref{boundary1} and \eqref{boundary2} (orange curves), which are not linear in $\mu$ in this case. (b): Energy splitting $\Delta E$ of two 6-plets in the middle of the two rotation bands at $\mu=0.4$ (red square and brown triangle in the top panel) shown on a logarithmic plot. $\Delta E$ is exponentially suppressed in the system size in both cases.}
\label{fig:sfCPA2}
\end{figure}

True topological order is expected to arise when $A$ is non-Abelian, and therefore we consider the case $A=2$ (the case $A=-1$ is analogous to $A=1$). For the group $S_3$, though, the choice $A=2$ implies that no operator $\theta_{g_b}$ corresponding to the inversion group elements appears in the Hamiltonian, since they have vanishing character in that representation. Consequently, the single-flux inversion bands become flat. This can be seen for both $C=\Id$ (Fig. \ref{fig:sfTCA2}) and $C=C_1$ (Fig. \ref{fig:sfCPA2}).

For $C=\Id$, the system displays a dispersing band corresponding to the degenerate rotation fluxes, and a flat band corresponding to the zero-energy fluxes. Two sets of edge modes branch from the inversion band, with energy $2\mu$ and $-\mu$, as an effect of the boundary terms $T$. 

The degeneracy structure of the rotation band is slightly different from the $A=1$ and $C=\Id$, as the states in this band are 4-fold degenerate with the exception of the states at the edges of the band displaying a 2-fold degeneracy. The splitting between the 4-fold degenerate states is algebraic in the system size [see Fig. \ref{fig:sfTCA2} (b)].

For $C=C_1$, instead, the five bands are well separated. The bands corresponding to the inversions are still dispersionless and, also in this case, branches of edge modes depart from them [see Fig. \ref{fig:sfCPA2} (a)]. The behavior of the rotation bands is analogous to the case $A=1$; the states are arranged in 6-plets and, for values of $\mu$ such that these two bands do not overlap, their splitting is exponentially suppressed in the system size.

\section{Conclusions} \label{sec:disc}

In this work, we defined two models with a global non-Abelian group symmetry. The first is the chiral ladder model for gauge fluxes in Eq. \eqref{ham_flux}. Based on our assumptions on its parameters, this model displays a ferromagnetic symmetry broken phase with $|G|$ degenerate ground states. The second is the model \eqref{ham} built through dyonic operators whose properties are determined by the symmetry group. The two models are unitarily equivalent through a nonlocal Jordan-Wigner transformation based on the non-Abelian group $G$. Such transformation maps the ferromagnetic phase of the ladder model into a phase of the dyonic model that displays topological order and weak zero-energy dyonic modes localized on the boundary of the system. This is analogous to the topological one-dimensional chains of Majorana \cite{kitaev2001} and parafermionic \cite{fendley2012} modes and our construction generalizes these systems and defines a new kind of one-dimensional topological order based on discrete non-Abelian symmetry groups.

To examine the properties of the dyonic model, we extended the definition of one-dimensional topological order (see, for example, Ref. \cite{bernevig2016}) to systems with non-Abelian symmetries. The appearance of topological order in the dyonic model crucially relies on the notion of locality determined by the dyonic modes. For this purpose, the Jordan-Wigner transformation adopted for the definition of the dyonic modes must rely on an auxiliary irreducible representation $A$, which must be non-Abelian. In case of Abelian auxiliary representations, the dyonic model displays holographic symmetries.

We examined the weak localized dyonic topological modes appearing in the system through a quasiadiabatic continuation technique and we presented a constructive approach to investigate the appearance of strong zero-energy modes. We showed that the definition of strong modes is in general flawed by divergences originating from two kinds of resonances between excited states: besides the resonances appearing in the study of the Abelian models \cite{jermyn2014,Moran2017}, the non-Abelian dyonic and ladder models suffer from the degeneracy of states characterized by different permutations of the same set of gauge fluxes. This hinders the formation of strong zero-energy modes unless these degeneracies are removed through the introduction of coupling constants with a weak position dependence.

The gauge-flux ladder models have been inspired by lattice gauge theories and quantum double models. They may display, in general, very rich phase diagrams and it is possible to envision schemes for their quantum simulation in ultracold atom setups based on the protocols developed for the quantum simulation of lattice gauge theories \cite{montangero2016,zohar2016} (see, for example, the proposal \cite{zohar2018} for the simulation of systems with $S_3$ symmetry). The realization of the dyonic model, instead, must rely on topological systems in higher dimensions with one-dimensional edge states with the required $G$ symmetry. Based on matrix-product-state results in Refs. \cite{turner11,fidkowski11,verstraete17}, it is indeed possible to show that there cannot exists a purely one-dimensional realization of these gapped topological phases of matter.

The systems we built are based on discrete symmetry groups. We observe, however, that the flux-ladder model can be extended to truncated Lie groups through suitable modifications of the operators $U$ in the Hamiltonian \eqref{ham_flux} \cite{burrello15} (see Ref. \cite{burrello2016} for the specific SU(2) case), and we can envision extensions to quantum groups as well. The generalization of the dyonic models to these scenarios is an interesting open problem which may connect our model to different systems of interacting anyons.

Finally, we point out that the dyonic modes we defined constitute a particular one-dimensional realization of the extrinsic anyonic twist defects studied in the context of two-dimensional symmetry-enriched systems with topological order \cite{barkeshli2014,fradkin2015,teo2016}. Based on the analogy with quantum double models, we suppose that their projective non-Abelian braiding statistics is universal for a suitable choice of the symmetry group. The braiding of dyonic modes can be studied by embedding the dyonic models in appropriate tri-junction geometries or two-dimensional systems, thus extending the known results for parafermionic modes \cite{shtengel2012,stern2012}.

\section*{Acknowledgments}

We warmly thank A. C. Balram, L. Mazza, G. Ortiz and J. Slingerland for fruitful discussions. M.M. acknowledges support by the Danish National Research Foundation. M.B. acknowledges support from the Villum Foundation.

\appendix

\section{The $C$-matrix in high-dimension representations} \label{app:c}
In the main text we proved the existence of a unitary matrix $C$ satisying Eq. \eqref{condc} when the representation matrices $D^F(g)$ of $G$ belong to $U(2)$. In this section we will extend the proof to the case with $D^F(g)\in U(N)$. We will exploit the decomposition $U(N)=U(1)\times SU(N)$, implying that any matrix $U\in U(N)$ is generated by a phase and the generators of $SU(N)$. $SU(N)$ in turn is generated by $N^2-1$ traceless, Hermitian matrices $T_a$ satisfying $[T_a,T_b] = i \sum_c f_{abc} T_c$, where $f_{abc}$ are the structure constants of $SU(N)$. These matrices $T_a$ satisfy 
\begin{align}
T_a T_b &=  \delta_{ab} \id + \frac{1}{2}\sum_{c=1}^{N^2-1}(i f_{abc} + d_{abc})T_c \,, \label{TProduct}
\end{align}
such that we can write
\begin{align}
D^F(g) &= e^{i\eta_{g,0}\id}e^{it \vec{\eta}_g\cdot \vec{T}}= e^{i\eta_{g,0}\id}\Big( d_{g,0} \id + \vec{d}_g\cdot \vec{T} \Big).
\end{align}
Here $d_{g_0}$ and $\vec{d}_g$ are in general complicated functions of $\vec{\eta}_g$ and the structure constants. For simplicity, let us consider the case $C \in SU(N)$. We can write
\begin{equation}
C =v_{0} \id + \vec{v} \cdot \vec{T}.
\end{equation}
From \eqref{TProduct} we see that 
\begin{align}
K_g &= \Tr\big( C D^F(g) \big) = e^{i\eta_{g,0}} \Big(Nv_{0} d_{g,0} + N\sum_i v_i d_{g,i} \Big)\nonumber\\
&= N {\mathcal{D}}(g)\cdot{\mathcal{C}},
\end{align}
where the $N^2$ dimensional vectors are defined in analogy with the two-dimensional case:
\begin{equation}
{\mathcal{D}}(g)= e^{i\alpha_{g0}}\begin{pmatrix} d_{g,0}\\ \vec{d}_g\end{pmatrix},\quad
{\mathcal{C}}  =\begin{pmatrix} v_{0}\\ \vec{v}\end{pmatrix}.
\end{equation}
Since
\begin{align}
1 &= \frac{1}{N}\Tr[C^\dagger C] = \vert v_0\vert^2 + \sum_i \vert v_i\vert^2 = \vert\vert \mathcal{C} \vert \vert^2,
\end{align}
the vector $\mathcal{C}$ lies on the $(N^2-1)$-sphere. The condition $K_g \neq K_h$ amounts to 
\begin{equation}\label{eqn::CCondUNasVectors}
 \big({\mathcal{D}}(g)- {\mathcal{D}}(h) \big)\cdot {\mathcal{C}}\neq 0,
\end{equation}
and the demand that this holds for all $g\neq h$ gives at most $n = {\vert{G}\vert(\vert{G}\vert-1)}/{2}$ vectors which $\mathcal{C}$ cannot be orthogonal to, or in other words, there are $n$ great circles on the $(N^2-1)$-sphere which $\mathcal{C}$ cannot lie on. For all the vectors $\mathcal{C}$ that do not belong to these great circles, the corresponding matrix $C$ satisfies the condition \eqref{condc}. If we include a general overall phase to the matrix $C$, this does not affect $v_0$ and $\vec{v}$, hence the conditions \eqref{eqn::CCondUNasVectors} are unaffected and the extension to $C \in U(N)$ is straightforward.

\onecolumngrid

\section{Quasiadiabatic continuation of the weak zero-energy modes at first order} \label{app:adiab}

By applying the quasiadiabatic continuation technique \cite{bernevig2016,hastings2005,hastings2010}, we evaluate the first order correction of the weak zero-energy modes of $H_J$ after the introduction of a small perturbation $H_\mu$ such that $\mu \ll J$. We consider for simplicity the case $C=\Id$.

For the left edge, the unperturbed zero energy mode is $\alpha(1)$. We will calculate $\mathcal{V}(\mu) \alpha(1) \mathcal{V}^\dag(\mu)$ where the unitary operator $\mathcal{V}(\mu)$ is defined as the path ordered evolution
\begin{equation} \label{Vmu}
\mathcal{V}(\mu) = \textrm{Texp}\left[i\int_0^\mu \mathcal{D}(\mu') \,{\rm d}\mu'\right]\,
\end{equation}
generated by the operator
\begin{equation}
\mathcal{D}(\mu)=-i\int_{-\infty}^{+\infty} {\rm d}t \, e^{i H t} \mathcal{F}\left(\partial_\mu H\right) e^{-iHt}\,.
\end{equation}
In the previous relation, $H=H_J +H_\mu$ and the function $\mathcal{F}$ is meant to introduce suitable filter functions \cite{hastings2010}, depending on the different kinds of excitations of the ground states, to cut off the time the time evolution of $\partial_\mu H$ for large $|t|$.
In particular we adopt
\begin{equation}\label{filter1}
\mathcal{F}\left(\partial_\mu H\right) = - \sum_r \sum_{h \neq e} F[(m_h-m_e) t] \chi^A(h^{-1}) \theta_h(r)\,, 
\end{equation}
where $m_g$ labels the flux masses \eqref{mass} and $F(t)$ is an imaginary, odd and analytical filter function such that its Fourier transform results in
\begin{equation} \label{fFourier}
\tilde{F}(\omega) = \int_{-\infty}^{+\infty} {\rm d}t\, e^{i\omega t} F(t) \approx -\frac{1}{\omega} \quad \text{for} \quad |\omega| \ge 1\,,
\end{equation}
and $\tilde{F}(0)=0$ \cite{hastings2010}. From Eq. \eqref{Vmu} we get
\begin{equation} \label{firstorder}
\mathcal{V}(\mu) \alpha(1) \mathcal{V}^\dag(\mu) = \alpha(1) +i \mu \left[\mathcal{D}(0),\alpha(1)\right] + \ldots .
\end{equation}

The commutator results in
\begin{multline}
\left[\mathcal{D}(\mu=0),\alpha(1)\right] 
= i\int_{-\infty}^{+\infty}  {\rm d}t\, \left[e^{iH_Jt}\sum_{h \neq e} F[(m_h-m_e) t] \chi^A(h^{-1}) \theta_h(1) e^{-iH_Jt}, U^\dag(1)\right] \\
= i\int_{-\infty}^{+\infty}  {\rm d}t\, e^{iH_Jt} \left[\sum_{h \neq e} F[(m_h-m_e) t] \chi^A(h^{-1}) \theta_h(1) U^\dag(1) \left(\Id - D^\dag(h)\right)\right]e^{-iH_Jt} \\
= i  \sum_{h \neq e}  \chi^A(h^{-1}) \theta_h(1) U^\dag(1) \left(\Id - D^\dag(h)\right) \int_{-\infty}^{+\infty}  {\rm d}t\, F[(m_h-m_e) t] e^{-i J \left(\Tr\left[U(2)CU^\dag(1)\left(D^\dag(h)-\Id\right) + {\rm H.c.} \right] \right) t}
\,.
\end{multline}
We expressed all the terms in the previous relations as a function of the flux operators $U^\dag(1) = \alpha(1)$ and $\chi^A(h^{-1}) \theta_h(1) = \Tr_A\left[\beta^\dag_h(2) \alpha_h(1)\right] D^\dag(h)$. The weak zero-energy modes are defined based on their commutation relation \eqref{weakcomm} with the Hamiltonian projected on the ground-state manifold. Therefore we can specialize the previous expressions by considering their effect on the ground states of $H$ only. To the purpose of evaluating the first order correction in \eqref{firstorder}, we can consider in turn the effect of the commutator on the ground states of $H_J$, since dealing with the eigenstates of $H$ would imply the introduction of a further correction of order $\mu/J$ based on the relation $P(\mu) \approx P(0) + i \mu \left[\mathcal{D}(0),P(0)\right]$, where $P(\mu)$ is the projection operator onto the ground state manifold for finite $\mu$. Under this assumption, in the case $C=\Id$, we obtain
\begin{equation}
\left[\mathcal{D}(\mu=0),\alpha(1)\right] P(\mu) \approx i \sum_{h \neq e}  \chi^A(h^{-1})\theta_h(1) U^\dag(1) \left(\Id - D^\dag(h)\right) P(0) \int_{-\infty}^{+\infty}  {\rm d}t\,  F[(m_h-m_e) t] e^{i\left(m_h-m_e\right)t} + O(\mu/J) \,.
\end{equation}
After considering this ground-state restriction, by applying Eq. \eqref{fFourier} and considering that $\tilde{F}(1)\approx-1$, we finally obtain
\begin{equation} \label{leftweak2}
\mathcal{V}(\mu) \alpha(1) \mathcal{V}^\dag(\mu) = U^\dag(1) + \sum_{h \neq e} \frac{\mu}{m_h-m_e} \chi^A(h^{-1})\theta_h(1) U^\dag(1) \left(\Id - D^\dag(h)\right)\, + O\left(\frac{\mu^2}{J^2}\right) \,.
\end{equation}
This relation corresponds to Eq. \eqref{leftweak} once we express the $\theta$ and $U^\dag$ operators in terms of the dyonic modes. We also observe that this first-order correction coincides with the first-order term $\Lambda^{(1)}$ in Eq. \eqref{lambda1} when we apply the strong zero-energy mode to the ground-state manifold of $H_J$ in the limit $C \to \Id$. The case with a general $C$ matrix in the Hamiltonian can be investigated with the same approach. The final result indeed matches $\Lambda^{(1)}$ in Eq. \eqref{lambda1}.

For $C=\Id$, a similar calculation can be performed for the right edge modes. For this purpose, it is necessary to generalize the functional $\mathcal{F}(t)$. Instead of considering the set of functions $F(m_h-m_e) t$ in Eq. \eqref{filter1}, we define $\mathcal{F}$ based on a set of operators $f_J$:
\begin{equation}\label{filter2}
\mathcal{F}\left(\partial_\mu H\right) = - \sum_r \sum_{h \neq e} F(f_J(h,r) t)  \chi^A(h^{-1}) \theta_h(r)\,, 
\end{equation}
where
\begin{equation}\label{filter3}
f_J(h,r) = H_J - \theta_h(r)H_J\theta_h^\dag(r)\,.
\end{equation}
The role of the operators $f_J$ is to extract the correct spectral gap of the unperturbed Hamiltonian $H_J$ to be associated with each term of $\partial_\mu H$. 

The key property in the definition \eqref{filter2} is that both $\mathcal{F}\left(\partial_\mu H\right)$ and the resulting $\mathcal{D}(0)$ commute with the string operator $\mathcal{L}_g(L)$ appearing in $\beta_g(2L)$. By exploiting this property and $\left[\beta_g(2L),H_J\right]=0$, we get
\begin{align}
\left[\mathcal{D}(\mu=0),\beta_g(2L)\right] P(\mu) &\approx i\int_{-\infty}^{+\infty}  {\rm d}t\, \left[e^{iH_Jt}\sum_{r,h \neq e} F[f_J(h,r) t] \chi^A(h^{-1}) \theta_h(r) e^{-iH_Jt}, \mathcal{L}_g(L) U^\dag(L) \right] P(0)  \nonumber \\
&=i\int_{-\infty}^{+\infty}  {\rm d}t\, \left[\sum_{r,h \neq e} e^{if_J(h,r) t} F[f_J(h,r) t] \chi^A(h^{-1}) \theta_h(r), \mathcal{L}_g(L) U^\dag(L) \right] P(0) \nonumber\\
&= i \mathcal{L}_g(L) \sum_{h \neq e} \int_{-\infty}^{+\infty}  {\rm d}t\,  e^{if_J(h,L) t} F[f_J(h,L) t] \chi^A(h^{-1}) \left[\theta_h(L), U^\dag(L)\right] P(0) \nonumber \\
&=-i\sum_{h \neq e} \mathcal{L}_g(L) U^\dag(L)\left[D(h)-\Id \right]\int_{-\infty}^{+\infty}  {\rm d}t\,  e^{if_J(h,L) t} F[f_J(h,L) t] \chi^A(h^{-1}) \theta_h(L) P(0) \nonumber\\
&\approx i\sum_{h \neq e} \mathcal{L}_g(L) U^\dag(L)\left[D(h)-\Id \right]  \chi^A(h^{-1}) \theta_h(L) \frac{1}{m_h-m_e} P(0)
 \,.
\end{align}

Thanks to the definitions \eqref{filter2} and \eqref{filter3}, the last line holds also for $C \neq \Id$ and can be derived by commuting $f_J$ with $\theta_h$ and applying it to the projector $P(0)$. We conclude, in general:
\begin{equation}
\mathcal{V}(\mu) \beta_g(2L) \mathcal{V}^\dag(\mu) = \beta_g(2L) + \beta_g(2L) \sum_{h \neq e} \frac{\mu}{m_h-m_e} \left(D(h)-\Id\right)\Tr_K\Tr_A\left[\beta^\dag_h(2)\alpha_h(1)D^{K\dag}(h)\right] + O\left(\frac{\mu^2}{J^2}\right)\,,
\end{equation}
which is also consistent with the form of the right zero-energy strong-mode \eqref{Omegag1} applied to the unperturbed ground states.

\vspace{0.5cm}

\twocolumngrid

\section{The weak modes at the interface between nontopological and topological regions} \label{app:left}

The analysis in Sec. \ref{sec:topo} assumes a finite and uniform chain in its topological phase with $\mu \ll J$. For $\mu=0$, the left zero-energy mode is $\alpha(1)$ which, based on the definition \eqref{JW1}, does not carry a Jordan-Wigner string, and, consequently, a group element index. This property is inherited by all the left weak zero-energy modes defined by adiabatic continuation in Appendix \ref{app:adiab} and it holds also for the calculation of the strong zero-energy modes in Sec. \ref{sec:topo}.

\begin{figure}[tb]
\includegraphics[width=\columnwidth]{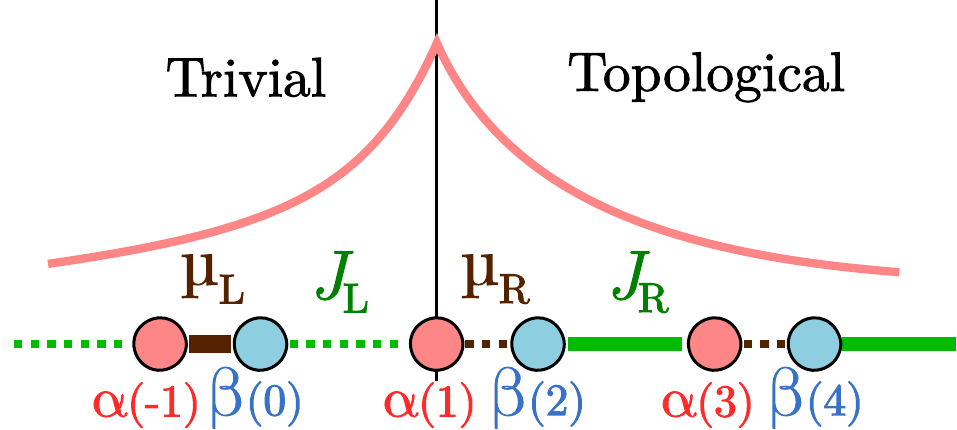}
\caption{Schematic representation of the interface between trivial $(r<1)$ and topological $(r\ge 1)$ interface. The dotted/full lines represent weak/strong couplings and the resulting weak zero-energy modes is localized around $\alpha_g(1)$.}
\label{fig:interface}
\end{figure}

In this Appendix we analyze what happens when we consider a boundary between a nontopological region, located at $r<1$ and a topological region at $r \ge 1$. In this case, the system is infinitely extended in both directions and the Jordan-Wigner strings must be redefined by extending them to $r=-\infty$: $\mathcal{L}_g(r) = \prod_{x=-\infty}^r \Theta_g(x)$ where the product is an ordered product generalizing Eq. \eqref{Ldef}.

We model the system through the Hamiltonian
\begin{equation}
H = H_L(\mu_L,J_L) + H_R(\mu_R,J_R)
\end{equation}
where the left Hamiltonian $H_L$ is defined for $r<1$ and is in the trivial regime $\mu_L \gg J_L$, whereas the right Hamiltonian $H_R$ is defined in the topological region $r\ge 1$ with $\mu_R \ll J_R$ (see Fig. \ref{fig:interface}). For $\mu_R=J_L=0$, the operators $\alpha_g(1)$ do not appear in $H$ and constitute zero-energy modes. In the following we will discuss how these zero-energy modes evolve quasiadiabatically, at first order, when introducing perturbations given by $J_L$ and $\mu_R$.

The unperturbed Hamiltonians $H_L(\mu_L,0)$ and $H_R(0,J_R)$ commute, since they are defined in nonoverlapping domains. This makes it possible to evaluate the two first-order contributions resulting in Eq. \eqref{firstorder} separately. The contribution given by $\mu_R$ coincides with the result in Eq. \eqref{leftweak2}. Therefore we focus on the introduction of $J_L$ only. For ease of notation we drop the subscript $L$ referring to the domain $r<1$. The operator $\mathcal{D}(J=0)$ is defined as
\begin{multline}
\mathcal{D}(J=0)= i \int_{-\infty}^{+\infty} dt e^{iH_\mu t} F(\Delta t)  \\
\times  \left[\sum_{r<1} \left(\Tr\left[U(r+1)CU^\dag(r)\right] + {\rm H.c.} \right)\right] e^{-iH_\mu t}
\end{multline}
Since we are interested in the weak modes, the operator $\Delta$ represents the gap caused by the application of the plaquette operators over the ground states of $H_\mu$. By using the projectors \eqref{Kproj} we can rewrite
\begin{equation}
H_\mu = - \frac{\mu |G|}{\dim A}\sum_{r<1} \Pi^A(r)\,,
\end{equation}
therefore the ground states of $H_\mu$ corresponds to states in which all the sites in the ladder model are in an arbitrary state $\ket{Aab}$. 
We conclude that the gap operator $\Delta$ can be defined as
\begin{equation}
\Delta = \frac{\mu |G|}{\dim A}\sum_{r<1} \left(\Id - \Pi^A(r)\right)\,.
\end{equation} 
We observe that the projector over the ground states of $H_\mu$ is $P(J=0)= \prod_{r<1} \Pi^A(r)$ and it commutes with $\alpha_g(1)$. Therefore, by following the approach in Appendix \ref{app:adiab}, we obtain
\begin{widetext}
\begin{multline}
\left[\mathcal{D}(J=0),\alpha_g(1)\right] P(J) \approx i \int_{-\infty}^{+\infty} dt F(\Delta t) e^{i\Delta t} \left[\left(\Tr\left[U(1)CU^\dag(0)\right] + {\rm H.c.}\right), \mathcal{L}_g(0)\right]U^\dag(1) P(0) \\
= -i \frac{\dim A}{\mu |G|} \left(\Id - \Pi^A(0)\right)\left(\Tr\left[U(1)CU^\dag(0)\left(\Id - D^\dag(g)\right)\right] + {\rm H.c.}\right) \alpha_g(1) P(0)
\end{multline}
where we exploited that $\tilde{F}(0)=0$. The first-order correction to $\alpha_g(1)$ on the trivial region results in
\begin{equation}
\mathcal{V}(J_L)\alpha_g(1)\mathcal{V^\dag}(J_L) = \alpha_g(1) + \frac{J_L \dim A}{\mu_L |G|} \left(\Id - \Pi^A(0)\right) \left(\Tr\left[U(1)CU^\dag(0)\left(\Id - D^\dag(g)\right)\right] + {\rm H.c.}\right) \alpha_g(1)  + O\left(\frac{J_L^2}{\mu_L^2}\right)\,.
\end{equation}
\end{widetext}
This relation can be fully recast in a local form as a function of the operators $\alpha_g(1) $, $\alpha_g(-1)$ and $\beta_g(0)$ and it suggests that, under quasiadiabatic evolution, the weak zero-energy modes at the interfaces between topological and nontopological regions maintain their locality.
A similar approach can be applied to estimate the strong-zero energy modes at such interface. Also, in this case, the left modes acquire a group index $g$ and the result is fully dyonic.

\section{Inner term of the zero modes} \label{app:inner}

\begin{figure}[tb]
\includegraphics[width=\columnwidth]{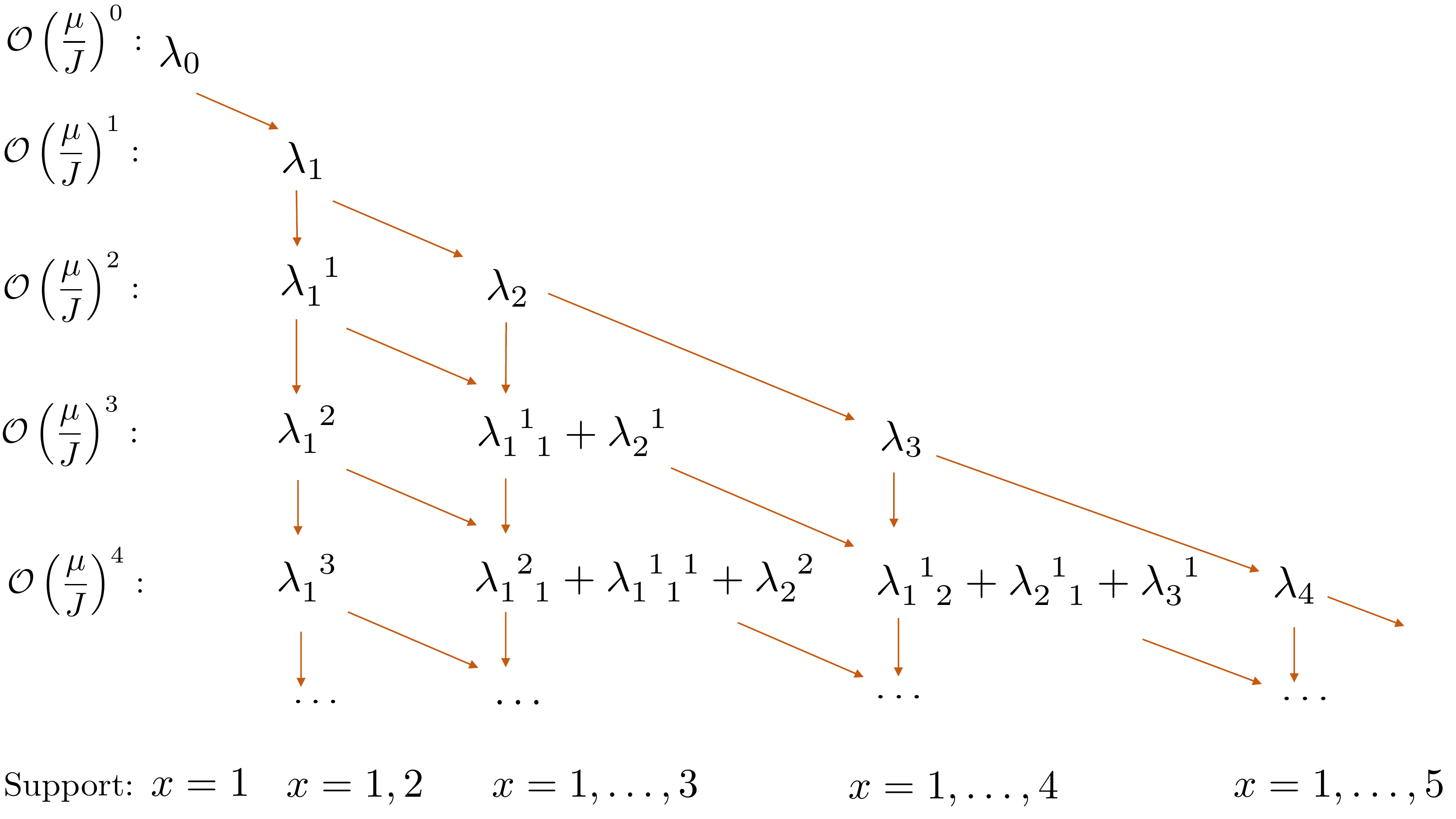}
\caption{Diagram of the structure of the terms at each order in $\frac{\mu}{J}$. For a given term, its commutator with $H_\mu$ is canceled by the subsequent terms' commutator with $H_J$. The notation keeps track of which term is derived from this. At each successive order the support may be extended compared to the previous step, in which case a lower index is added. If the support is unchanged, an upper index is added instead. The sum of all the indices gives the order of the term in $\frac{\mu}{J}$, and the support of a given term is given by the sum of lower indices plus one.}
\label{fig:InOutModes}
\end{figure}

In Sec. \ref{diverge}, we discussed the resonances appearing in the definition of the outer modes $\Lambda_{{\rm out},n}$. Here we investigate the behavior of the inner modes. To this purpose it is necessary to refine our definition of the inner part of the commutators $C_n$ and of the inner modes $\Lambda_{{\rm in},n}$. 

We introduce the notation $c_{a_1\phantom{a_2}a_3\phantom{a_4}\ldots}^{\phantom{a_1}a_2\phantom{a_3}\ldots}$ to label all the terms of the commutator $C_n$ appearing at level $n = \sum_i a_i$ in the iteration process. The set $a_1,a_2,\ldots a_n$ is an ordered partition of $n$ where lower and upper indices refer to the number of consecutive times that the outer or inner operators $\theta$ have been considered in the definition of this contribution of the commutator $C_n$. In particular $c_n \equiv C_{{\rm out},n}$, whereas all the other contributions belong to $C_{{\rm in},n}$.

To define in detail $c_{a_1\phantom{a_2}a_3\phantom{a_4}\ldots}^{\phantom{a_1}a_2\phantom{a_3}\ldots}$, let us consider first the second order of iteration.
The operator $C_2$ can be decomposed into:
\begin{align}
c_2 = -\mu \left[\Lambda_1, \sum_{h_2}\theta_{h_2}(2) \right] =  C_{{\rm out},2}\,,\\
c_1^{\phantom{1}1} = -\mu \left[\Lambda_1, \sum_{k_1}\theta_{k_1}(1) \right] =  C_{{\rm in},2}\,.
\end{align} 
The notation for $c_1^{\phantom{1}1}$ refers to the fact that, in the first order of iteration, we considered the outermost $\theta$ operator available ($\theta_{h_1}(1)$) in this case, whereas in the second order of iteration, we considered the commutator with the inner term $\theta_{k_1}(1)$.

In a similar way, we can define different contributions for the inner part of the strong mode $\Lambda_{{\rm in},n}$. In particular, we build the following operators:
\begin{align}
&\lambda_2 = \Lambda_{{\rm out},2} \quad \text{such that} \quad \left[\lambda_2,H_J\right]=-c_2 \,,\\
&\lambda_1^{\phantom{1}1} = \Lambda_{{\rm in},2} \quad \text{such that} \quad \left[\lambda_1^{\phantom{1}1},H_J\right]=-c_1^{\phantom{1}1} \,.
\end{align}

In the following iteration steps, we can define
\begin{align}
&c_n = -\mu \left[\lambda_{n-1}, \sum_{h_n}\theta_{h_n}(n) \right] =  C_{{\rm out},n}\,,\\
&c_{n-1}^{\phantom{n-1}1}(r) = -\mu \left[\lambda_{n-1}, \sum_{k_1}\theta_{k_1}(r) \right] \,, \\
&c_{n-2}^{\phantom{n-2}2}(r_1,r_2) = -\mu \left[\lambda_{n-2}^{\phantom{n-2}1}(r_1), \sum_{k_2}\theta_{k_2}(r_2) \right] \,.
\end{align} 
More in general, given $\lambda_{a_1\phantom{a_2}a_3\phantom{a_4}\ldots}^{\phantom{a_1}a_2\phantom{a_3}\ldots}$, we will define a set of commutators $c_{a_1\phantom{a_2}a_3\phantom{a_4}\ldots}^{\phantom{a_1}a_2\phantom{a_3}\ldots}$, increasing the last upper index when considering the commutator with an inner $\theta$ operator, and increasing the last lower index when considering the commutator with an outer $\theta$ operator. If the last index is not of the type which is increased, a new index of $1$ is added at that position instead.

The construction of $\lambda_{a_1\phantom{a_2}a_3\phantom{a_4}\ldots}^{\phantom{a_1}a_2\phantom{a_3}\ldots}$ follows accordingly, based on the relation
\begin{equation}
\left[\lambda_{a_1\phantom{a_2}a_3\phantom{a_4}\ldots}^{\phantom{a_1}a_2\phantom{a_3}\ldots},H_J\right] =- \sum_{r_1 \ldots} c_{a_1\phantom{a_2}a_3\phantom{a_4}\ldots}^{\phantom{a_1}a_2\phantom{a_3}\ldots}(r_1,\ldots)\,,
\end{equation}
where we are summing over all the possible position indices of the inner part of the commutator.

This construction implies that the modes $\lambda_{a_1\phantom{a_2}a_3\phantom{a_4}\ldots}^{\phantom{a_1}a_2\phantom{a_3}\ldots}$ have support in the first $a_{\rm in} = a_1 + a_3 + a_5 +\ldots $ sites of the flux-ladder model, and they range from $\alpha(1)$ to $\alpha(2a_{\rm in} +1)$.

This construction is summarized in Fig. \ref{fig:InOutModes}. We observe that the order of the indices matters, so each term in Figure $\ref{fig:InOutModes}$ at any given order are in general not equal.

Because of the factor $U^\dagger(1)$ in $\lambda_n$ there is a difference between $c_n^{\phantom{n}1}(1)$ and $c_n^{\phantom{n}1}(j)$ for $n\geq j>1$. 
To get an idea of the structure of all these many terms it is illustrative to calculate a few of them,  and by using \eqref{GammaOutn} we see 
 \bwt
\begin{align}
c_2^{\phantom{n}1}(1) &= -\frac{\mu^{3}}{J^2} \sum_{h_1,h_2 \neq e}\sum_{k_1\neq e}\chi^A(h_1^{-1})\chi^A(h_2^{-1})\chi^A(k_1^{-1})\left[  \tilde{F}_{1} F_{2} \theta_{h_1}(1)\theta_{h_2}(2) U^\dagger(1) (D^\dagger(h_1)-\id), \theta_{k_1}(1) \right]\nonumber\\
&=-\frac{\mu^{3}}{J^2} \sum_{h_1,h_2 \neq e}\sum_{k_1\neq e}\chi^A(h_1^{-1})\chi^A(h_2^{-1})\chi^A(k_1^{-1})  \bigg( \tilde{F}_{1} F_{2} \theta_{h_1k_1}(1)\theta_{h_2}(2) U^\dagger(1)D^\dagger(k_1) (D^\dagger(h_1)-\id)\nonumber\\
&\phantom{aaaaaaaaaaaaaaaaaaaaaaaaaaaaaa} -\tilde{G}_{1}(k_1,1) G_{2}(k_1,1) \theta_{k_1h_1}(1)\theta_{h_2}(2) U^\dagger(1) (D^\dagger(h_1)-\id)\bigg),\label{c211}
\end{align}
where
\begin{multline}
\tilde{G}_{1}(k_1,1) =  \theta_{k_1}(1) \tilde{F}_1 \theta_{k_1}^\dag(1)  \\
=\Big(\left( \Tr[U(2)CU^\dagger(1)D(k_1)\big(D(h_1)-\id\big) + \hc]\right)^{-1} - \left( \Tr[U(2)CU^\dagger(1)D(k_1)(D(h_1)-\id )D^\dagger(h_2) + \hc]\right)^{-1}\Big),
\end{multline}
and
\begin{multline}
G_{2}(k_1,1) = \theta_{k_1}(1) F_2 \theta_{k_1}^\dag(1) \\
=\bigg(\Tr[U(2)CU^\dagger(1)D(k_1)(D(h_1^{\phantom{1}}h_2^{-1})-\id) + \hc] +\Tr[U(3)CU^\dagger(2)(D(h_2)-\id) + \hc] \bigg).
\end{multline}

The crucial point to notice is that no new conditions are required on the Hamiltonian in order have this term finite. The next order correction $\lambda_{2}^{\phantom{2}1}(1)$ is also finite, since the only difference from \eqref{c211} is that the two terms have an added factor of $\left([\theta_{h_1k_1}(1) \theta_{h_2}(2),H_j] (\theta_{h_1k_1}(1) \theta_{h_2}(2))^{-1}  \right)^{-1}$ and $\left([\theta_{k_1h_1}(1) \theta_{h_2}(2),H_j] (\theta_{k_1h_1}(1) \theta_{h_2}(2))^{-1}  \right)^{-1}$, respectively.
There is a subtlety we should address however. If for instance we look at $c_2^{\phantom{2}2}(1,2)$, there are commutators of the form

\begin{align}
&[\theta_{h_1k_1}(1)\theta_{h_2k_2}(2),H_J]  \nonumber\\
&=\left(\Tr(U(2)CU^\dagger(1) \left(D(h_1k_1(h_2k_2)^{-1} - \id \right) + \hc) + (\Tr(U(3)CU^\dagger(2) \left(D(h_2k_2 - \id \right) + \hc) \right) \theta_{h_1k_1}(1)\theta_{h_2k_2}(2),
\end{align}
\ewt
and the above is zero for $k_2=h_2^{-1}$ and $k_1=h_1^{-1}$. Therefore, when constructing $\lambda_2^{\phantom{2}2}(1,2)$ we would only have to sum over the $k_1$ and $k_2$ such that $c_2^{\phantom{2}2}(1,2)\neq 0$. 

In conclusion, all the inner terms can be expressed as the sum of terms similar to the outer modes, through a redefinition of the domain and the correct conjugations of the $F$ functions generating suitable $G$ functions. As long as $F_n$ and $\tilde{F}_n$ are bounded, their conjugated counterparts $G_n$ and $\tilde{G}_n$ are as well, and all the inner terms are well-defined to all orders. All the $F$ and $G$ operators assume the general form $\left(\sum_i (m_{g_i}-m_{h_i})\right)^{-1}$ in the group element basis, and the only resonances which may appear are the ones described in Sec. \ref{diverge}. Consequently, the inclusion of the inner modes does not qualitatively modify the general behavior of the decay of the strong modes in the bulk.

\end{document}